\documentclass[iop]{emulateapj}
\pdfoutput=1

\usepackage{natbib,aas_macros,amsmath}
\usepackage{multirow,color}

\newcommand{\average}[1]{\ensuremath{\langle#1\rangle} }

\begin{document}
\slugcomment{Accepted for publication in The Astrophysical Journal}
\shorttitle{Faint Galaxies at $z\sim5-10$ in HFF}
\shortauthors{Ishigaki et al.}

\title{
		Hubble Frontier Fields First Complete Cluster Data: Faint Galaxies at $\lowercase{z}\sim 5-10$\\
		for UV Luminosity Functions and Cosmic Reionization
}

\author{Masafumi Ishigaki\altaffilmark{1,2}, Ryota Kawamata\altaffilmark{3}, Masami Ouchi\altaffilmark{1,4}, Masamune Oguri\altaffilmark{2,4,5}, \\ Kazuhiro Shimasaku\altaffilmark{3,5}, and Yoshiaki Ono\altaffilmark{1}}
\email{ishigaki@icrr.u-tokyo.ac.jp}
\altaffiltext{1}{Institute for Cosmic Ray Research, The University of Tokyo, Kashiwa, Chiba 277-8582, Japan}
\altaffiltext{2}{Department of Physics, University of Tokyo, 7-3-1 Hongo, Bunkyo-ku, Tokyo 113-0033, Japan}
\altaffiltext{3}{Department of Astronomy, University of Tokyo, 7-3-1 Hongo, Bunkyo-ku, Tokyo 113-0033, Japan}
\altaffiltext{4}{Kavli Institute for the Physics and Mathematics of the Universe (Kavli IPMU, WPI), University of Tokyo, Kashiwa, Chiba 277-8583, Japan}
\altaffiltext{5}{Research Center for the Early Universe, University of Tokyo, 7-3-1 Hongo, Bunkyo-ku, Tokyo 113-0033, Japan}

\begin{abstract}
We present the comprehensive analyses of faint dropout galaxies up to $z\sim10$ with the first full-depth data set of Abell 2744
lensing cluster and parallel fields observed by the Hubble Frontier Fields (HFF) program.
We identify $54$ dropouts at $z\sim5-10$ in the HFF fields,
and enlarge the size of $z\sim9$ galaxy sample obtained to date.
Although the number of highly magnified ($\mu\sim10$) galaxies is small
due to the tiny survey volume of strong lensing,
our study reaches the galaxies' intrinsic
luminosities comparable to the deepest-field HUDF studies.
We derive UV luminosity functions with these faint dropouts, carefully evaluating
the combination of observational incompleteness and lensing effects in the image plane
by intensive simulations including 
magnification, distortion, and multiplication of images,
with the evaluations of mass model dependences.
Our results confirm that the faint-end slope, $\alpha$, is as steep as $-2$ at $z\sim6-8$,
and strengthen the evidence of the rapid decrease of UV luminosity densities, $\rho_\mathrm{UV}$,
at $z>8$ from the large $z\sim9$ sample. We examine whether the rapid $\rho_\mathrm{UV}$ decrease trend
can reconcile with the large Thomson scattering optical depth, $\tau_\mathrm{e}$, measured by CMB experiments
allowing a large space of free parameters
such as average ionizing photon escape fraction and stellar-population dependent conversion factor.
No parameter set can reproduce both the rapid $\rho_\mathrm{UV}$ decrease and the large $\tau_\mathrm{e}$.
It is possible that the $\rho_\mathrm{UV}$ decrease moderates at $z\gtrsim11$,
that the free parameters significantly evolve towards high-$z$, or 
that there exist additional sources of reionization such as X-ray binaries and faint AGNs.
\end{abstract}


\section{Introduction}

Cosmic reionization history and sources of reionization are open questions in astronomy today.
Studies of QSO Gunn-Peterson absorption indicate that the intergalactic medium (IGM) is rapidly ionized at $z \sim 6$ \citep{2006ARA&A..44..415F}.
A moderately large neutral hydrogen fraction at $z \gtrsim 6$ is implied by the Ly$\alpha$ damping wing
absorption features in the spectra of gamma-ray bursts (GRBs) at $z \sim 6$ \citep{2006PASJ...58..485T,2013arXiv1312.3934T} and 
Ly$\alpha$ emitters at $z \sim 6-7$ \citep{2006ApJ...648....7K, 2010ApJ...723..869O, 2011ApJ...734..119K, 2014arXiv1404.6066K}.
Similarly, there are reports of Ly$\alpha$-emitting galaxy fraction drops 
at $z \sim 7$ probably due to the increase of Ly$\alpha$ damping wing absorption given by the
neutral hydrogen in the IGM
 \citep{2011ApJ...743..132P,2014arXiv1403.5466P,2012ApJ...744...83O,2012ApJ...744..179S,2014arXiv1404.4632S,2013ApJ...775L..29T,2013Natur.502..524F}.
Recent observations of cosmic microwave background (CMB) present the large value of the Thomson scattering optical depth $\tau_e = 0.091^{+0.013}_{-0.014}$ \citep{2013arXiv1303.5076P}. 
The large value of $\tau_e$ indicates that the reionization takes place at $z = 11.1 \pm 1.1$ 
if an instantaneous reionization is assumed.
The combination of these data implies that the reionization process is extended at $z \sim 6-11$.

Star-forming galaxies are thought to be major sources of the cosmic reionization (see reviews of \citealt{2006ARA&A..44..415F}; \citealt{2010Natur.468...49R}).
Recent ultra-deep observations with the Wide Field Camera 3 (WFC3) aboard the {\it Hubble Space Telescope} (HST) have provided improved estimates of the abundances of star-forming galaxies at $z \sim 7 - 10$ \citep{2013ApJ...763L...7E, 2013ApJ...768..196S, 2013MNRAS.432.2696M}.
Combining these results with the {\it WMAP} constraints on the Thomson scattering optical depth \citep{2013ApJS..208...19H} and stellar mass densities, 
\citet{2013ApJ...768...71R} suggest that all these observations can be explained consistently 
if their population of star-forming galaxies extends below the survey limits down to absolute UV magnitudes of $M_{\rm UV} \sim -13$.
However, it is difficult to translate the UV luminosity function measurements into the ionized hydrogen fraction,
because of uncertainties of the following three unknown parameters.
The first is the escape fraction $f_{\rm esc}$, which is the fraction of the numbers of ionizing photons escaping into the IGM
to those produced by star-formation in a galaxy.
The second is the conversion factor $\xi_{\rm ion}$, which converts a UV luminosity density 
to the ionizing photon emission rate in a star-forming galaxy.
The third is a clumping factor $C_{\rm H_{II}} \equiv \left< n_{\rm H_{II}}^2 \right> / \left< n_{\rm H_{II}} \right>^2$, 
where $n_{\rm H_{II}}$ are the local number density of ionized hydrogen and the brackets indicate spatial average.
It is critically important to take into account the uncertainties of these parameters 
to estimate the contribution of galaxies to reionization.

Moreover, the abundance of faint galaxies at high redshift is unknown. 
Some theoretical studies indicate that the star formation is suppressed in low-mass halos.
\citet{2014arXiv1405.1040B} suggest that the star formation is suppressed in halos smaller than $\sim 10^9 M_\odot$ at high redshift, 
corresponding to $M_{\rm UV} \simeq -14$.
Cosmological hydrodynamical simulations of \citet{2013ApJ...766...94J} exhibit a turnover of the $z = 8$ UV luminosity function at $M_{\rm UV} \sim -17$.
Thus it is not obvious if the UV luminosity function of star-forming galaxies indeed extends down to $M_{\rm UV} \sim -13$, as assumed in \citet{2013ApJ...768...71R}.
Recent observations of nearby dwarf galaxies find that the star formation in dwarf galaxies is suppressed 
at the epoch of reionization \citep{2014arXiv1405.5540B,2014arXiv1405.3281W}.
The faint-end slope $\alpha$ of the UV luminosity function is also not well known at high redshift. 
The steepening of UV luminosity functions towards high-$z$ is a general agreement of observational studies.
\citet{2014arXiv1403.4295B} conclude that
the value of $\alpha$ evolves from $\alpha \sim -1.6$ at $z\sim4$ to $\alpha \sim -2.0$ at $z\sim7$.
However, the determination of $\alpha$ includes a large uncertainty at $z \gtrsim 9$, 
due to the poor statistics of the $z \gtrsim 9$ luminosity function measurement.

Gravitational lensing by massive clusters is an effective tool to reveal properties of faint galaxies at high redshift.
Lensing magnifications of background sources enable us to observe intrinsically faint sources that are not detected without lensing magnifications.
For example, 
Cluster Lensing And Supernova survey with Hubble (CLASH) studies properties of faint star-forming galaxies at $z \sim 6-9$ 
using the lensing technique \citep{2012arXiv1211.2230B,2013arXiv1308.1692B}.
Recently, HST has started revolutionary deep imaging on the six massive clusters
with parallel observations, 
the Hubble Frontier Fields (HFF; PI: J. Lotz) project whose data are $\sim 1$ mag deeper than those of CLASH.
The HFF project identifies faint sources reaching $\sim$ 29 AB mag,
allowing us to detect background sources 
with intrinsic magnitudes of $\gtrsim 30$ mag by lensing magnification \citep{2014arXiv1405.0011C}.
The HFF first targets the Abell 2744 cluster, followed by other five clusters: 
MACSJ0416.1-2403, MACSJ0717.5+3745, MACSJ1149.5+2223, Abell S1063 (RXCJ2248.7-4431), and Abell 370.
The observations of Abell 2744 were just completed in July 2014, which provide the first full-depth data set 
on an HFF target.

In this paper, we identify star-forming galaxies at $z \sim 5-10$ magnified by gravitational lensing in the Abell 2744 cluster 
and its parallel fields.
We refer to the former as the cluster field and the latter as the parallel field in the remainder of this paper.
This work serves as a precursor study that uses the first one sixth of the full-depth HFF data set.
We construct the mass model of Abell 2744, and derive the UV luminosity functions 
with the star-forming galaxies at $z \sim 5-10$.
Calculating the UV luminosity densities from the UV luminosity functions of
our and previous studies,
we discuss cosmic reionization based on the UV luminosity density measurements and 
Thomson scattering optical depths from CMB observations
with the ionization equation that allows a large free parameter space.

We present details of the observational data in Section \ref{sec:Data}.
The photometric catalog and dropout selection methods are described in Section \ref{sec:Samples}, 
and our mass model of Abell 2744 is presented in Section \ref{sec:Mass model}.
Using these data, we derive the parameters of UV luminosity functions in Section \ref{sec:uvlf}.
In Section \ref{sec:Discussion}, we discuss cosmic reionization with 
the UV luminosity densities and Thomson scattering optical depths.
Finally, we summarize our results in Section \ref{sec:Summary}.
We adopt a cosmology with $\Omega_{\rm m} = 0.3$, $\Omega_\Lambda = 0.7$, $\Omega_{\rm b} = 0.04$, and $H_0 = 70$ km s$^{-1}$ Mpc$^{-1}$.

\section{Data} \label{sec:Data}

The Abell 2744 cluster and the parallel fields were observed with WFC3-IR and Advanced Camera for Survey (ACS)
in the HFF project.
These data were reduced and released to the public through the HFF official website.
\footnote{http://archive.stsci.edu/pub/hlsp/frontier/abell2744/\\images/hst/v1.0/} 
They provide drizzled science images and inverse variance weight images 
in four WFC3-IR bands, F105W ($Y_{105}$), F125W ($J_{125}$), F140W ($JH_{140}$), and F160W ($H_{160}$), 
and 
in three ACS bands, F435W ($B_{435}$), F606W ($V_{606}$), and F814W ($i_{814}$). 
We use version 1.0 of the public images with a pixel scale of $0 \farcs 03$ pixel$^{-1}$. 
For the measurements of object colors,
we homogenize the point spread functions (PSFs) of the WFC3 images with {\sc iraf} \citep{1986SPIE..627..733T,1993ASPC...52..173T} \verb|imfilter| package. 
A summary of the HST data is shown in Table \ref{summary_of_data}.
We measure limiting magnitudes in a $0\farcs4$-diameter circular 
aperture using {\sc sdfred} \citep{2002AJ....123...66Y,2004ApJ...611..660O}. 
We find that the $5 \sigma$ limits are 
$28.5-29.1$ mag in the cluster field and $28.6-29.2$ mag in the parallel field. 
The cluster field contains the bright intracluster light \citep{2014arXiv1405.2070M}, 
which makes the depths of the cluster field shallower than those of the parallel field.

\begin{figure}
	\plotone{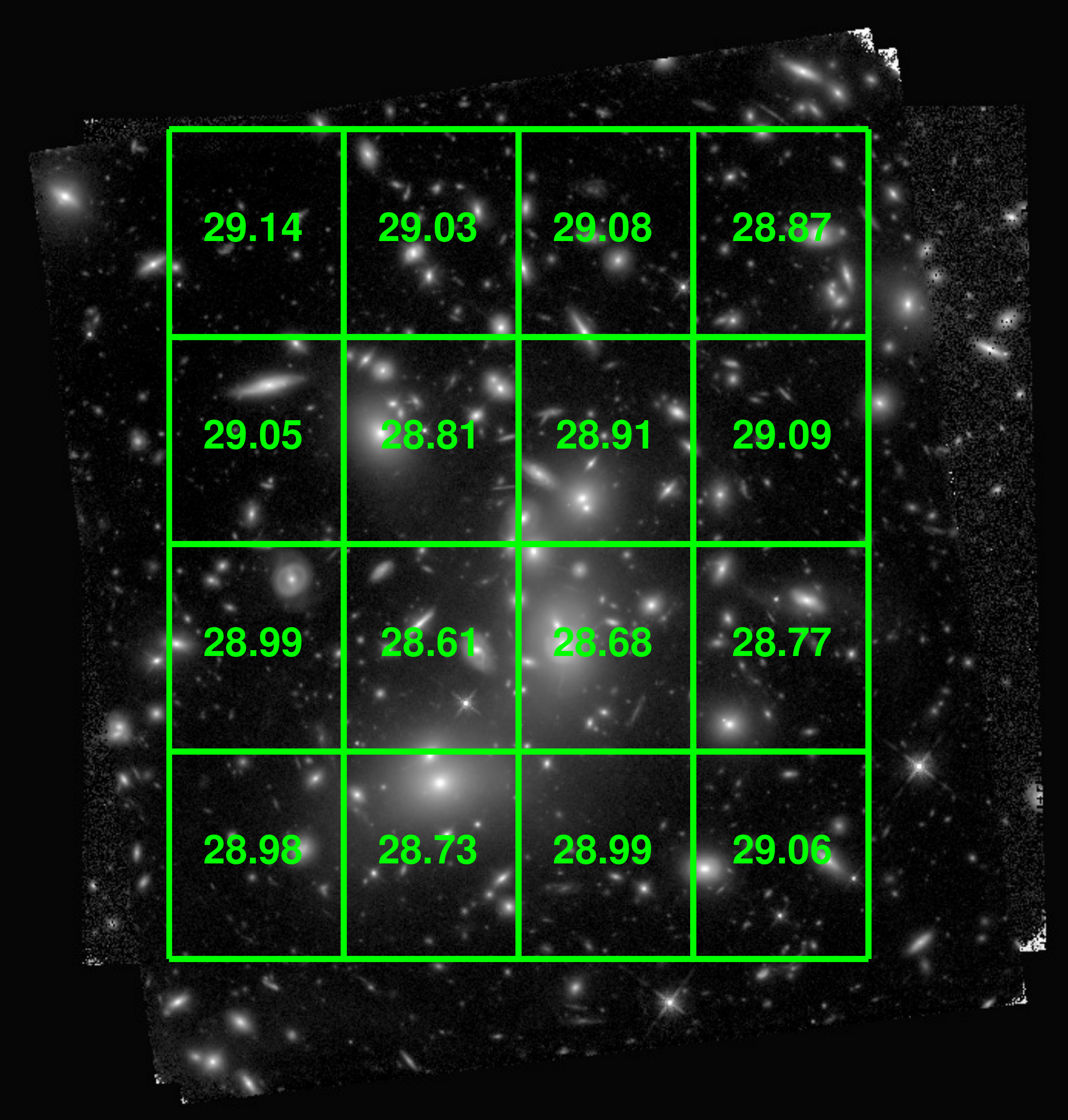}
	\caption{
	\textit{HST}/WFC3-IR $H_{160}$ image of the Abell 2744 cluster field. 
	The numbers denote $5 \sigma$ limiting magnitudes defined in a $0\farcs4$-diameter aperture, measured in the $4 \times 4$ grid cells.  
	}
	\label{limitmag_cut}
\end{figure}

The $5 \sigma$ limiting magnitudes presented in Table \ref{summary_of_data} 
are measured in the entire field.  
However, 
the intracluster light is brighter in the cluster center than in the outskirts, 
which causes spatial variations of the depth in the HST images. 
To evaluate the spatial variations, 
we measure the limiting magnitudes in each of $4 \times 4$ grid cells
defined in Figure \ref{limitmag_cut}.
We present the $H_{160}$-band limiting magnitudes in the cells in 
Figure \ref{limitmag_cut}.
In Figure \ref{limitmag_cut}, we find that 
the depth in the third-row second-column cell ($28.61$ mag) 
is about $0.5$ mag shallower than those of the cluster outskirts ($\sim 29$ mag) due to the bright intracluster light.
The peak-to-peak magnitudes of spatial variations of the depths are also
$\sim 0.5$ mag 
in the rest of WFC3-IR images and the ACS data.
In the cluster field, we adopt the space-dependent limiting magnitudes as illustrated 
in Figure \ref{limitmag_cut}.
For sources in the outside of the cell, we apply a limiting magnitude of the nearest cell.

\setlength{\tabcolsep}{4pt}
\begin{deluxetable}{lccc}
\tabletypesize{\scriptsize}
\tablecaption{Summary of the HFF Abell 2744 Data\label{summary_of_data}}
\tablewidth{0pt}
\tablehead{
		\colhead{Filter} & \colhead{Orbits}  & \colhead{Detection Limits\tablenotemark{a}} & \colhead{PSF FWHM}\\
		\colhead{} & \colhead{} & \colhead{$5\sigma$} & \colhead{arcsec}  
}
\startdata
\sidehead{Cluster Field}
		$B_{435}$ & $24$ & $28.51$ & $0.10$ \\
		$V_{606}$ & $15$  & $28.67$ & $0.09$ \\
		$i_{814}$ & $49$ & $28.72$ ($28.93$)\tablenotemark{b} & $0.09$ ($0.18$)\tablenotemark{b} \\
		$Y_{105}$ & $26$ & $29.05$ ($29.03$)\tablenotemark{b} & $0.16$ ($0.18$)\tablenotemark{b}\\
		$J_{125}$ & $13.5$ & $28.67$ ($28.72$)\tablenotemark{b} & $0.17$ ($0.18$)\tablenotemark{b}\\
		$JH_{140}$ &$10$ & $28.74$ ($28.74$)\tablenotemark{b} & $0.17$ ($0.18$)\tablenotemark{b}\\
		$H_{160}$ & $27$ & $28.75$ ($28.75$)\tablenotemark{b} & $0.18$ ($0.18$)\tablenotemark{b}\\ 
		$J_{125} + JH_{140} + H_{160}$\tablenotemark{c} &  \nodata  & $28.92$ & $0.17$ \\
		$JH_{140} + H_{160}$\tablenotemark{d} &  \nodata  & $28.79$ & $0.17$ \\
\tableline
\sidehead{Parallel Field}
		$B_{435}$ & $28$ & $28.66$ & $0.10$ \\
		$V_{606}$ & $16.5$ & $28.98$ & $0.10$ \\
		$i_{814}$ & $42.5$ & $28.90$ ($28.97$)\tablenotemark{b} & $0.10$ ($0.19$)\tablenotemark{b} \\
		$Y_{105}$ & $24$ & $29.24$ ($29.20$)\tablenotemark{b} & $0.19$ ($0.19$)\tablenotemark{b}\\
		$J_{125}$ & $12$ & $28.88$ ($28.87$)\tablenotemark{b} & $0.18$ ($0.19$)\tablenotemark{b} \\
		$JH_{140}$ &$10$ & $28.82$ ($28.93$)\tablenotemark{b} & $0.18$ ($0.19$)\tablenotemark{b} \\
		$H_{160}$ & $24$ & $28.99$ ($28.99$)\tablenotemark{b} & $0.19$ ($0.19$)\tablenotemark{b} \\ 
		$J_{125} + JH_{140} + H_{160}$\tablenotemark{c} &  \nodata  & $29.11$ & $0.19$ \\
		$JH_{140} + H_{160}$\tablenotemark{d} &  \nodata  & $29.03$ & $0.19$ 
\enddata
\tablenotetext{a}{Measured in a $0\farcs4$-diameter circular aperture.}
\tablenotetext{b}{The values in the parenthesis correspond to the detection limits and the PSF FWHMs of PSF-homogenized images.}
\tablenotetext{c}{The detection image for $i$- and $Y$-dropout selections.}
\tablenotetext{d}{The detection image for $YJ$-dropout selection.}
\end{deluxetable}

\section{Samples} \label{sec:Samples}

In this section, 
we select the $i$-, $Y$-, and $YJ-$dropout candidates in the cluster and the parallel fields
with the color criteria from our source catalogs.
We also compare our dropout samples 
with those obtained in previous studies.

\subsection{Photometric catalog}\label{sec:photometric catalog}

Using {\sc SWarp} \citep{2002ASPC..281..228B},
we make two detection images that are co-added data of
$(J_{125} + JH_{140} + H_{160})$ and 
$(JH_{140} + H_{160})$
for our $i$- and $Y$-dropout candidates and $YJ$-dropout candidates, respectively. 
We match the PSFs of these band images in the same manner
as the WFC3-IR images (Section \ref{sec:Data}),
and produce the detection images.
To apply the criteria of no blue-continuum detections for our dropout selections,
we do not match the PSFs of the blue bands whose wavelengths are shorter 
than the redshifted Ly$\alpha$-break feature of our dropout candidates.
Because the PSF-unmatched images in the blue bands provide
upper limits on the flux densities that are stronger than PSF-homogenized images 
for point-like sources of high-$z$ galaxies,
we use the PSF-unmatched data of the individual ACS images to obtain the upper limits.

We construct our source catalogs from the HFF images using SExtractor \citep{1996A&AS..117..393B}
in a total of $9.5$ arcmin$^2$ area where all the WFC3 and ACS images are available.   
We run SExtractor in dual-image mode for each set of the images. 
In the cluster field, we set {\tt DEBLEND\_NTHRESH} to $16$  
and  
{\tt DEBLEND\_MINCONT} to a small value of $0.0005$ in order to detect objects even in highly crowded regions. 
In the parallel field, we use more conservative values, ${\tt DEBLEND\_NTHRESH} = 32$  
and  
${\tt DEBLEND\_MINCONT} = 0.005$, because the parallel field are not crowded. 
\footnote{Although we use the different deblending parameter sets in the cluster and the parallel fields,
		this difference does not affect our final results of the UV luminosity functions.
                This is because we use the same deblending parameter sets, each for the cluster and parallel fields, 
                self-consistently in our simulations to derive our UV luminosity functions (Section \ref{sec:uvlf}).}
The number of objects identified in the detection images is $\sim 4300$ in total.
The colors of the objects are measured with magnitudes of {\tt MAG\_APER} ($m_{\rm AP}$),
which are estimated from the flux density within a fixed circular aperture.
The aperture diameters used for $m_{\rm AP}$ are two times of the FWHMs of the PSFs. 
We adopt 
the diameters of $0\farcs36$ ($0\farcs38$) and $\sim 0\farcs2$ for the PSF-matched images and for the PSF-unmatched blue-band images in the cluster (parallel) field, respectively.
The detection limits are also defined with $0\farcs36$ ($0\farcs38$) diameter apertures for the PSF-matched images, and $\sim 0\farcs2$ diameter apertures for the PSF-unmatched images in the cluster (parallel) field.

We apply an aperture correction that is defined by the following procedure.
We create a median stacked $J_{125}$-band image of our dropout candidates selected in Section \ref{sec:dropout selection},
and measure the aperture flux of the stacked dropout candidate as a function of aperture size.
Because the flux almost levels off at around a $1\farcs2$ diameter, 
we regard the flux within a $1\farcs2$-diameter aperture as the total flux
corresponding to the total magnitude $m_{\rm tot}$.
In the stacked image, $m_{\rm AP}$ is fainter than $m_{\rm tot}$ by $0.82$ mag.
We thus estimate the total magnitudes with $m_{\rm tot} = m_{\rm AP} - c_{\rm AP}$,
where $c_{\rm AP}$ is the aperture correction factor of $0.82$ mag.
We also make median stacked images for bright and faint subsamples of our dropout candidates,
and obtain $c_{\rm AP}$ values. We confirm that the values of $c_{\rm AP}$ do not
depend on luminosity beyond the statistical uncertainties 
in the magnitude range of our dropout candidates. 
Thus we apply one aperture correction factor of $c_{\rm AP}=0.82$ 
for all our dropout candidates.

\begin{figure}
	\plotone{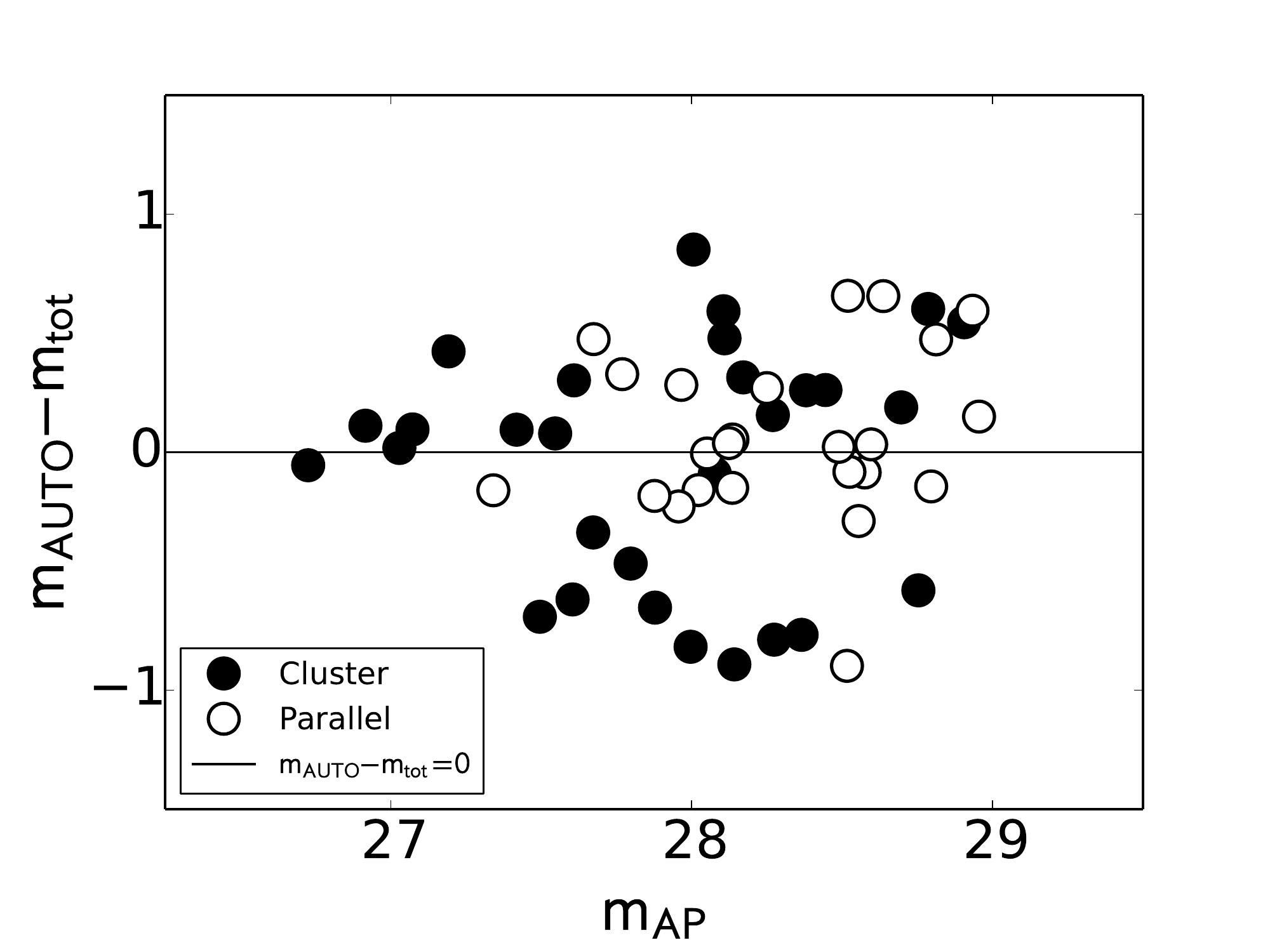}
	\caption{
	Difference between the aperture-corrected total magnitudes ($m_{\rm tot}$) and the SExtractor's AUTO magnitudes ($m_{\rm AUTO}$) as a function of the aperture magnitude $m_{\rm AP}$
	for our dropout candidates in the cluster (filled circles) and parallel (open circles) fields .
	The horizontal line corresponds to the case that 
	$m_{\rm AUTO}$ is equal to $m_{\rm tot}$. 
	}
	\label{mag_test}
\end{figure}

To check the accuracy of our aperture correction, 
we compare $m_{\rm tot}$ with the magnitude of {\tt MAG\_AUTO} ($m_{\rm AUTO}$),
which is calculated with the Kron elliptical aperture \citep{1980apjs...43..305k}. 
Figure \ref{mag_test} presents $m_{\rm AUTO}-m_{\rm tot}$ as a function of $m_{\rm AP}$,
and indicates that $m_{\rm tot}$ is comparable to $m_{\rm AUTO}$ for bright dropout candidates with $m_{\rm AP} < 27$ mag.  
The values of $m_{\rm AUTO}-m_{\rm tot}$ have significant scatters at the faint magnitudes, 
which is mainly due to uncertainties in determining the Kron elliptical apertures of faint sources. 
We adopt $m_{\rm tot}$ for our estimates of total magnitudes, as we expect that
the $m_{\rm tot}$ values are more reliable than $m_{\rm AUTO}$ for faint sources.

\subsection{Dropout Selection}\label{sec:dropout selection}

For the selection of $i$-dropouts at $z \sim 6-7$,
we use the color criteria defined by \citet{2014ApJ...786...60A}:
\begin{eqnarray}
		i_{814} - Y_{105} > 0.8, \label{eq:i-Y}\\  
	i_{814} - Y_{105} > 0.6 + 2( Y_{105} - J_{125}), \\ 
	Y_{105} - J_{125} < 0.8. \label{eq:Y-J}
\end{eqnarray}
For objects fainter than the $3\sigma$ limiting magnitude in $i_{814}$,
$i_{814}$ $3\sigma$ upper limiting magnitude is replaced with the 
$i_{814}$ magnitude (see \citealt{2014ApJ...786...60A}). 
For secure source detection, we apply the source identification thresholds
of the $>5\sigma$ significance levels both in the $Y_{105}$ and $J_{125}$ bands.
From our $i$-dropout candidate catalog, we remove
sources detected at the $>2\sigma$ level either 
in the $B_{435}$ or $V_{606}$ band.

For $Y$-dropouts at $z \sim 8$, 
we adopt the following criteria \citep{2013ApJ...768..196S}: 
\begin{eqnarray}
		Y_{105} - J_{125} > 0.5, \label{eq:Y-J2}\\
	J_{125} - H_{160} < 0.4. \label{eq:J-H}
\end{eqnarray}
As described in \citet{2013ApJ...768..196S}, we use the $1 \sigma$ upper limiting magnitude in the $Y_{105}$ band.
In this selection, sources with the $>3.5\sigma$ levels both in the $J_{125}$ and $JH_{140}$ bands
are regarded as real objects.
We reject sources detected at $2\sigma$ in the optical bands.
Additionally, we apply a criterion that no more than one of the optical bands shows a detection above the $1.5\sigma$ level. 
We use a collective $\chi_{{\rm opt}}^2$ value to eliminate contamination, the details of which are described in Section 3.3 of \citet{2011ApJ...737...90B} and Section 3.2 of \citet{2013ApJ...768..196S}.
The corrective $\chi_{{\rm opt}}^2$ is defined by $\chi_{{\rm opt}}^2\equiv \sum_j {\rm SGN}(f_j)({\rm SNR}_j)^2$, 
where $f_j$ is the flux density in the $j$-th band, ${\rm SNR}_j$ is the signal-to-noise ratio of the source in the $j$-th band, and ${\rm SGN}(f_j)$ is a sign function;
${\rm SGN}(f_j) = 1$ if $f_j > 0$ and $-1$ if $f_j < 0$. 
The $j$ index runs across $B_{435}$, $V_{606}$, and $i_{814}$.
We remove objects with $\chi_{{\rm opt}}^2 > 5.0$ from our dropout candidates if they are brighter than the $10\sigma$ limit in the $JH_{140}$ band,
and remove ones with $\chi_{{\rm opt}}^2 > 2.5$ if they are fainter than the $5\sigma$ limit.
A linear interpolation is used for objects with $JH_{140}$ between the $5\sigma$ and $10\sigma$ limit.

For $YJ$-dropouts at $z \sim 9$, 
we use the following criteria:
\begin{eqnarray}
		(Y_{105} + J_{125})/2 - JH_{140} > 0.75, \label{eq:YJ-JH}\\
		\begin{split}
		(Y_{105} + J_{125})/2 - JH_{140} > \\ 
		0.75 + 0.8 \times (JH_{140} - H_{160}),
\end{split}\\
J_{125} - H_{160} < 1.15, \\
JH_{140} - H_{160} < 0.6. \label{eq:JH-H}
\end{eqnarray}
We replace the $Y_{105}$ or $J_{125}$ magnitude with the $1 \sigma$ upper limiting magnitude if an object is fainter than the
$1\sigma$ magnitude in $Y_{105}$ or $J_{125}$, following \citet{2013ApJ...773...75O}. 
For the $YJ$-dropouts, 
we require detection significance levels beyond $3\sigma$ in the $JH_{140}$ and $H_{160}$ bands, 
and $3.5\sigma$ in at least one of the $JH_{140}$ and $H_{160}$ bands.
From our $YJ$-dropout sample, we remove sources detected at the $2\sigma$ level
in, at least, one of the optical bands and sources with $\chi^2_{{\rm opt}} > 2.8$.
These criteria are similar to those defined by \citet{2013ApJ...773...75O},
but we slightly relax the criteria to include dropout candidates at $z \sim 9.5$.

\begin{figure}
	\plotone{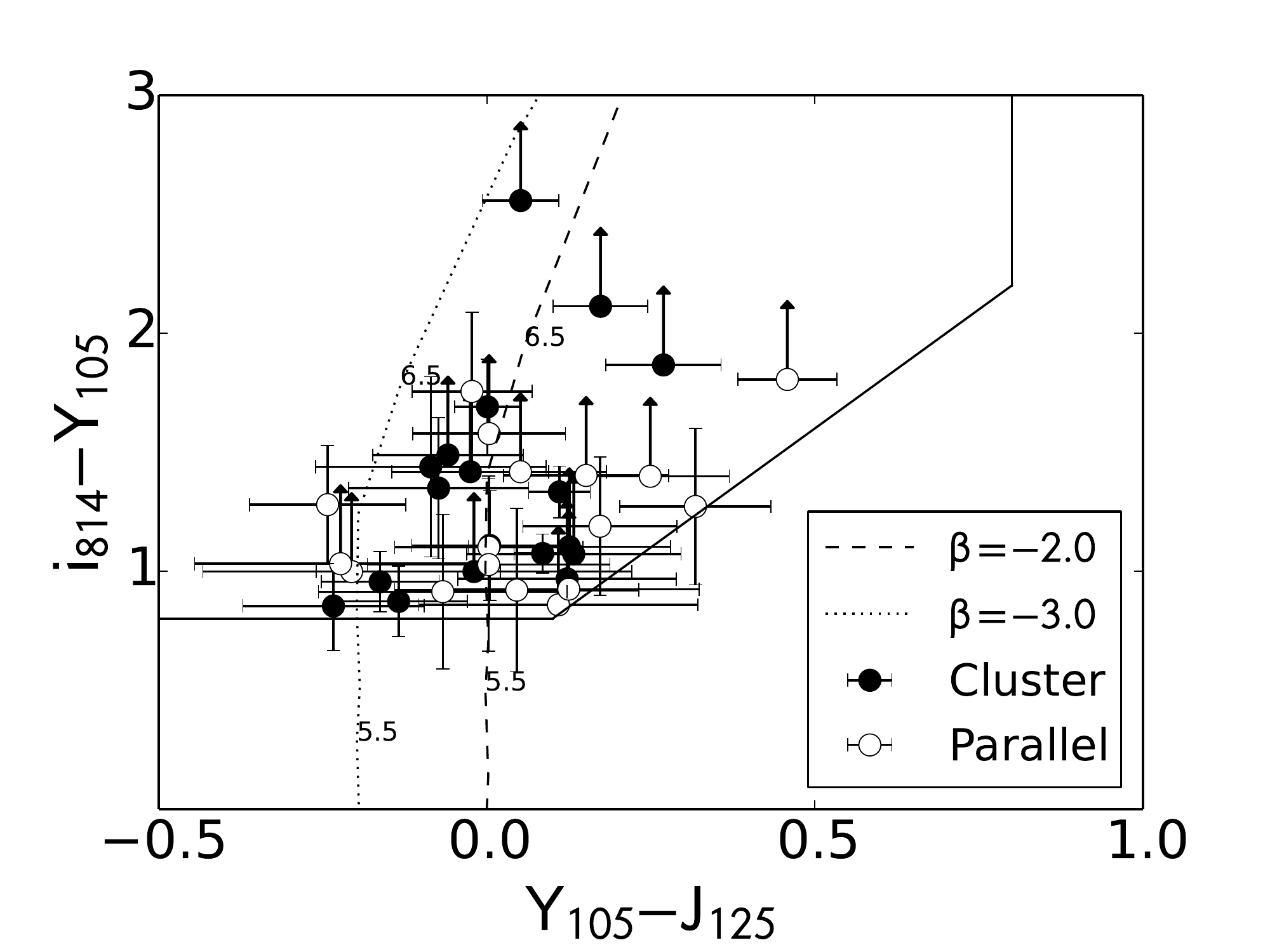}
	\plotone{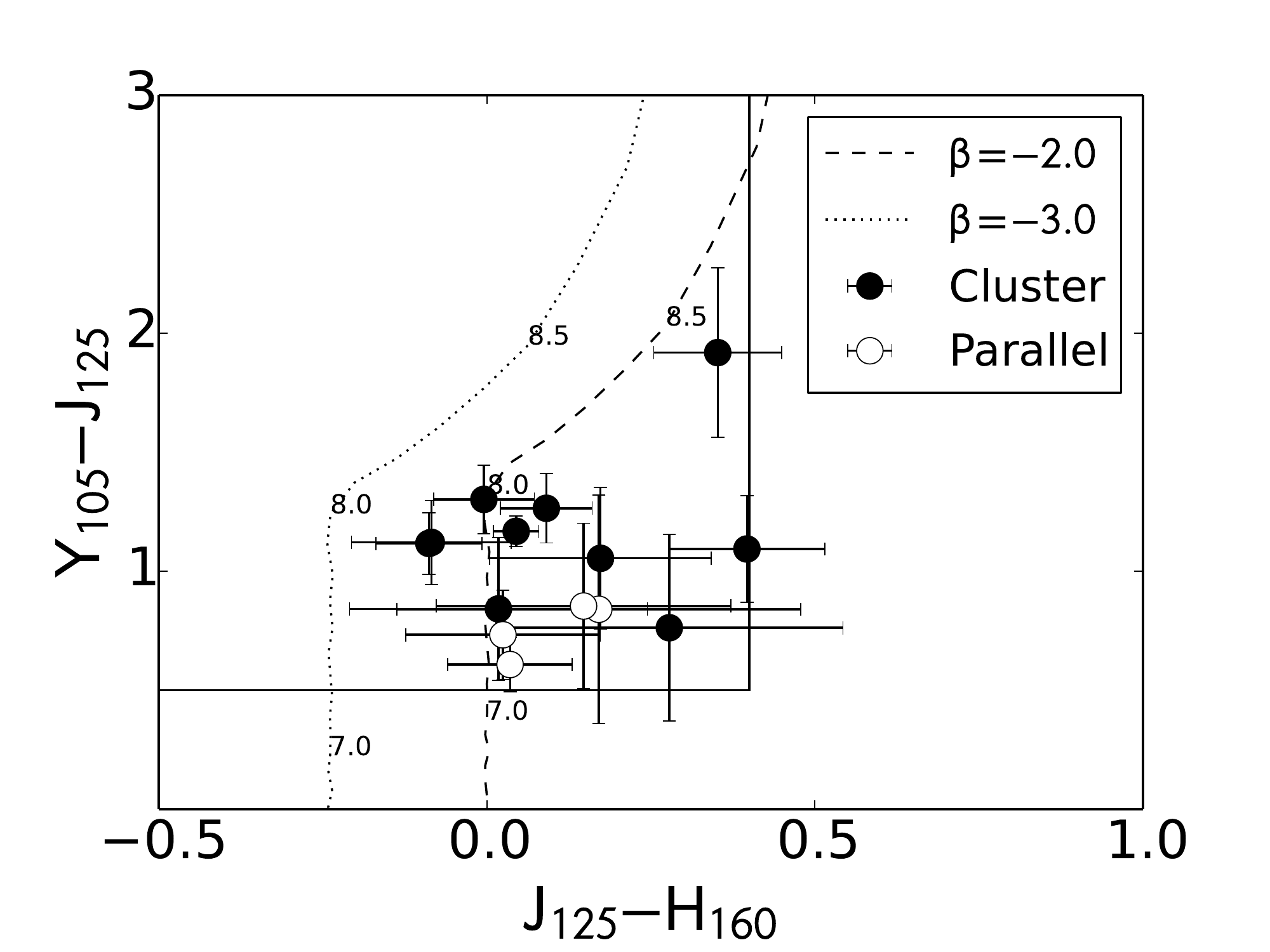}
	\plotone{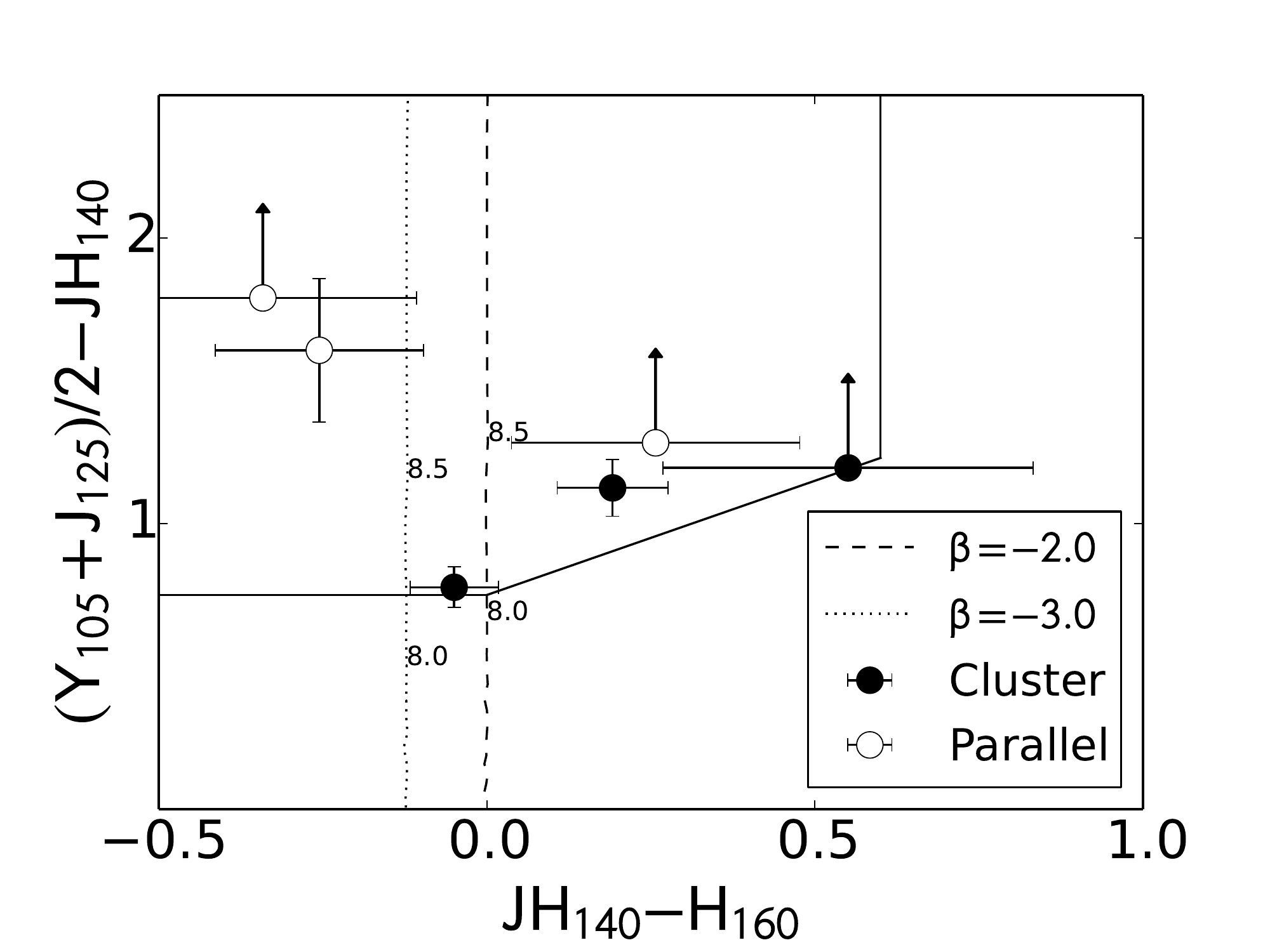}
	\caption{
	Two-color diagrams for $i$-dropout (top), $Y$-dropout (middle), and $YJ$-dropout candidates (bottom). 
	The dropout selection windows are indicated with the solid lines. 
	The filled circles (open circles) denote our dropout candidates in the cluster (parallel) field.
	The arrows indicate that the $i_{814}$ ($Y_{105}$ and $J_{125}$) magnitude is replaced 
	with the $3\sigma$ ($1\sigma$) limiting magnitude
	for the $i$-dropout ($Y$-dropout and $YJ$-dropout) candidates.
	The dashed and dotted lines present the expected colors of 
	star-forming galaxies with UV-continuum slopes of $\beta = -2$ and $-3$, respectively. 
	}
	\label{color_color}
\end{figure}

\begin{figure*}
		\epsscale{1.2}
	\plotone{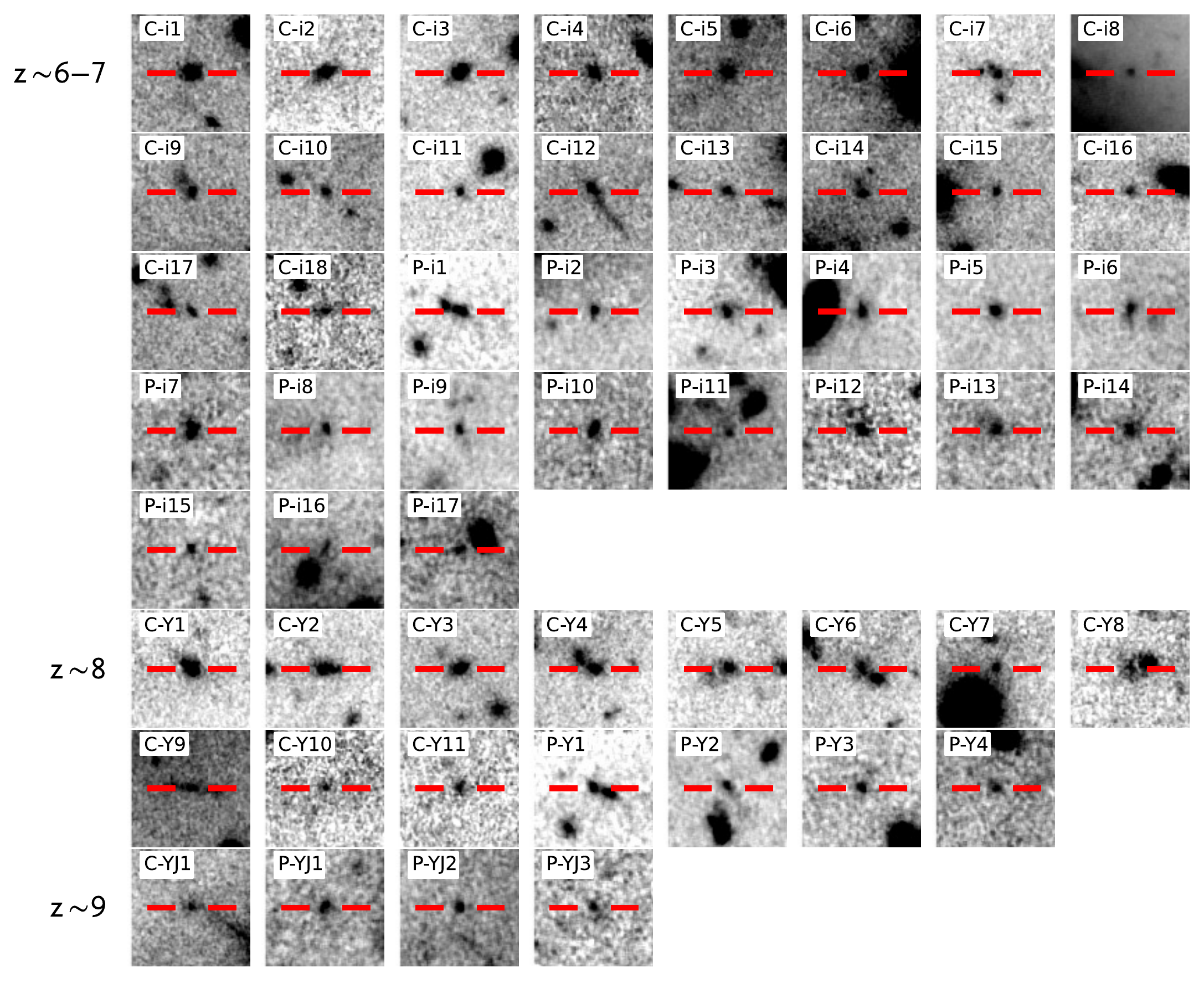}
	\caption{
			Cutouts of the detection images of our dropout candidates.
			The size of each cutout is $3'' \times 3''$.
			Each candidate is placed at the image center, and marked with the red lines.
			North is up, and east is to the left.
	}
	\label{cutout}
\end{figure*}

We select $i$-, $Y$-, and $YJ$-dropouts with the selection criteria 
shown above.
Figure \ref{color_color} shows the two-color diagrams for our dropout candidates,
together with the expected tracks of high-redshift star-forming galaxies with UV slopes of $\beta = -2$ and $-3$  
(see \citealt{1999ApJ...521...64M} for the definition of $\beta$).
Our dropout samples consist of $35$ $i$-dropout, $15$ $Y$-dropout, and $6$ $YJ$-dropout candidates.
These dropout candidates are listed
in Tables \ref{candidates7}-\ref{candidates9}. 
Figure \ref{cutout} shows cutout images of our dropout candidates.
Note that two out of $6$ $YJ$-dropout candidates are the objects which are also selected as $Y$-dropout candidates.
The numbers of dropout candidates in the cluster field are comparable to those in the parallel field, 
although the cluster field is subject to the strong lensing effects.
The numbers of dropout candidates are affected by two effects of the lensing magnification.
One is the magnification of surface brightness that enhances the observed brightness of the dropout candidates.
The other is the magnification of the observed area, which reduces the effective survey area on the source plane.
In the numbers of dropout candidates, these two effects compensate \citep{2014arXiv1405.0011C}.
Albeit the small statistics of a single HFF pointing, this would 
be one of the reasons why the numbers of dropout candidates 
are similar in the cluster and the parallel fields. More quantitative
arguments of the lensing effects are presented in
Section \ref{sec:uvlf}.

We estimate photometric redshifts of our dropout candidates using the Bayesian photometric redshift code {\sc bpz} \citep{2000ApJ...536..571B}. 
Tables \ref{candidates7}, \ref{candidates8}, and \ref{candidates9} include the photometric redshifts.
The redshift ranges are $5.7$-$7.3$, $7.2$-$8.4$, and $8.1-9.6$ for the $i$-, $Y$, and $YJ$-dropout candidates, 
except a candidate with the ID of HFF1P-i12 at $z = 4.5$. The three samples of dropout candidates cover $z\sim 5-10$.
We confirm that the ranges of the photometric redshifts are consistent 
with the redshifts defined by the dropout selections, $z \sim 6-7$, $8$, and $9$.
In the remainder of this paper, we refer to $i$-, $Y$-, and $YJ$-dropout candidates as 
$z \sim 6-7$, $z \sim 8$, $z\sim9$ dropouts, respectively.

Recently, 
\citet{2014ApJ...786...60A}, \citet{2014arXiv1402.6743Z}, \citet{2014arXiv1405.0011C}, and \citet{2014arXiv1409.0512A} have identified a total of 16, 18, 7, and 58 dropouts at $z>6$ in the Abell 2744 cluster field, respectively. 
We recover $10$, $12$, $6$, and $25$ dropouts of their samples. 
\citet{2014A&A...562L...8L} analyze the spectral energy distribution of a $z \sim 8$ dropout,
which is identified in our selection with the ID of HFF1C-Y1 and in all the other three studies.
Our dropout with the ID of HFF1C-Y9 is also found in \citet{2014arXiv1402.6743Z} as one of the three multiple images.
Although we find HFF1C-Y9 by the $Y$-dropout selection,
\citet{2014arXiv1402.6743Z} identify the multiply-imaged object as an $i$-dropout.
The different selections of $Y$- and $i$-dropouts are explained by the fact that 
the photometric redshift of HFF1C-Y9 is $7.3$ that is the border value 
of the $i$-dropout and $Y$-dropout redshift ranges.
\citet{2014arXiv1407.3769Z} have reported a triply-imaged $z \sim 10$ dropout in the cluster field (see also \citealt{2014arXiv1409.1228O}).
We recover one of the multiple images as a $YJ$-dropout with the ID of HFF1C-YJ1. 
In Section \ref{sec:optimization},
we discuss the multiple images including HFF1C-Y9 and HFF1C-YJ1 with our mass model.

We do not recover $7$, $6$, $1$, and $33$ dropouts found in \citet{2014ApJ...786...60A}, \citet{2014arXiv1402.6743Z}, \citet{2014arXiv1405.0011C}, and \citet{2014arXiv1409.0512A}, respectively.
The difference of our and their samples can be explained by the following three reasons.
First, we use the full-depth images of HFF Abell 2744 observations, while 
\citet{2014ApJ...786...60A}, \citet{2014arXiv1402.6743Z}, and \citet{2014arXiv1405.0011C} only use relatively shallow ACS images observed in HST Cycle 17 (GO 11689, PI: Dupke).
Second, the limiting magnitude definitions are different.
We take account of the space-dependent limiting magnitudes in the cluster field as described in Section \ref{sec:Data}.
Third, there are differences in galaxy selection techniques.
\citet{2014arXiv1402.6743Z} and \citet{2014arXiv1405.0011C} do not use
the well-tested color selections, but sophisticated photometric redshifts for the selections.
Although we concur with all 9 confident $z>8$ dropouts from \citet{2014arXiv1402.6743Z}, we recover only one $z\sim7-8$ dropout from \citet{2014arXiv1402.6743Z}.
Similarly, we recover only 19 out of 50 dropouts at $z\sim7$ from \citet{2014arXiv1409.0512A}.
This is possibly because our selection criteria for our $z<8$ dropouts are more conservative than those of \citet{2014arXiv1402.6743Z} and \citet{2014arXiv1409.0512A}.

\setlength{\tabcolsep}{6pt}
\begin{deluxetable*}{lcccccccc}
\tabletypesize{\scriptsize}
\tablecaption{Dropout candidates at $z\sim 6-7$\label{candidates7} in the HFF Abell 2744 Fields}
\tablewidth{0pt}
\tablehead{
		\colhead{ID} & \colhead{R.A. (J2000)} & \colhead{Dec (J2000)} & \colhead{$i_{814}-Y_{105}$} & \colhead{$Y_{105}-J_{125}$} & \colhead{$J_{125}$\tablenotemark{a}} & \colhead{Magnification\tablenotemark{b}} & \colhead{Photo-$z$} & \colhead{Reference\tablenotemark{c}}
}
\startdata
\sidehead{Cluster field}
HFF1C-i1 & $ 3.593804 $ & $ -30.415447 $ &$ > 2.37 $&$ 0.05 \pm 0.07 $&$ 26.10 \pm 0.05 $& $3.73^{+0.24}_{-0.21}$ &                  $6.6\pm0.8$    & $1$, $3$\\                     
HFF1C-i2 & $ 3.570654 $ & $ -30.414659 $ &$ 1.33 \pm 0.12 $&$ 0.11 \pm 0.05 $&$ 26.21 \pm 0.03 $& $1.62 \pm 0.06$ &         $6.0\pm0.7$     & $1$, $3$\\                      
HFF1C-i3 & $ 3.606222 $ & $ -30.386644 $ &$ 1.07 \pm 0.09 $&$ 0.09 \pm 0.04 $&$ 26.25 \pm 0.04 $ & $1.69\pm0.05$ &         $5.8\pm0.7$            & $1$, $3$\\                
HFF1C-i4 & $ 3.606385 $ & $ -30.407282 $ &$ 1.69 \pm 0.21 $&$ 0.00 \pm 0.05 $&$ 26.37 \pm 0.04 $& $2.25^{+0.12}_{-0.10}$ &            $6.3\pm0.7$        & $1$, $3$\\                   
HFF1C-i5 & $ 3.580452 $ & $ -30.405043 $ &$ > 2.10 $&$ 0.17 \pm 0.07 $&$ 26.60 \pm 0.05 $& $5.64\pm0.39$ &           $6.8\pm0.8$         & $1$, $3$\\                 
HFF1C-i6 & $ 3.597834 $ & $ -30.395961 $ &$ > 1.71 $&$ 0.27 \pm 0.09 $&$ 26.79 \pm 0.07 $& $2.87\pm0.19$ &                        $7.0\pm0.8$              & $1$, $3$\\              
HFF1C-i7 & $ 3.590761 $ & $ -30.379408 $ &$ 0.95 \pm 0.13 $&$ -0.16 \pm 0.09 $&$ 27.06 \pm 0.08 $& $1.87^{+0.06}_{-0.05}$ &                $5.9\pm0.7$       & \nodata\\                 
HFF1C-i8  & $ 3.585321 $ & $ -30.397958 $ &$ > 1.10 $&$ 0.13 \pm 0.15 $&$ 27.19 \pm 0.10 $ & $4.60^{+0.43}_{-0.48}$   &                        $6.8\pm0.8$  & $1$, $3$  \\    
HFF1C-i9 & $ 3.601072 $ & $ -30.403991 $ &$ 1.11 \pm 0.27 $&$ 0.00 \pm 0.11 $&$ 27.26 \pm 0.09 $ & $3.56^{+0.26}_{-0.23}$ &  $5.9\pm0.7$         &   $1$, $3$\\                
HFF1C-i10 & $ 3.600619 $ & $ -30.410296 $ &$ > 1.43 $&$ -0.06 \pm 0.11 $&$ 27.29 \pm 0.08 $& $11.43^{+1.60}_{-1.20}$ &       $6.4\pm0.7$        &  $3$\\              
HFF1C-i11 & $ 3.603426 $ & $ -30.383219 $ &$ 0.87 \pm 0.16 $&$ -0.13 \pm 0.11 $&$ 27.29 \pm 0.09 $ & $1.71^{+0.04}_{-0.05}$ &        $5.8\pm0.7$          & $3$\\              
HFF1C-i12 & $ 3.603214 $ & $ -30.410350 $ &$ > 1.36 $&$ -0.03 \pm 0.12 $&$ 27.32 \pm 0.09 $& $3.88^{+0.29}_{-0.21}$ &         $6.3\pm0.7$      & $1$, $2$, $3$ \\                
HFF1C-i13 & $ 3.592944 $ & $ -30.413328 $ &$ > 1.25 $&$ -0.09 \pm 0.20 $&$ 27.35 \pm 0.15 $& $6.85^{+0.60}_{-0.54}$ &                         $6.1\pm0.7$     & $3$\\                  
HFF1C-i14 & $ 3.585016 $ & $ -30.413084 $ &$ 0.85 \pm 0.19 $&$ -0.23 \pm 0.14 $&$ 27.45 \pm 0.12 $ & $2.94^{+0.18}_{-0.17}$ &                $5.7^{+0.7}_{-1.1}$       & \nodata\\       
HFF1C-i15 & $ 3.576889 $ & $ -30.386329 $ &$ > 0.96 $&$ 0.13 \pm 0.18 $&$ 27.45 \pm 0.16 $& $2.77^{+0.15}_{-0.13}$ &                        $6.1^{+0.8}_{-0.7}$            & $3$\\   
HFF1C-i16 & $ 3.609003 $ & $ -30.385283 $ &$ 1.35 \pm 0.33 $&$ -0.07 \pm 0.14 $&$ 27.56 \pm 0.12 $ & $1.59\pm0.04$ &               $6.1\pm0.7$           & $3$\\             
HFF1C-i17 & $ 3.604563 $ & $ -30.409364 $ &$ > 0.91 $&$ 0.12 \pm 0.16 $&$ 27.62 \pm 0.11 $& $2.94^{+0.19}_{-0.16}$ &                         $6.1\pm0.8$         &  $3$\\             
HFF1C-i18 & $ 3.590518 $ & $ -30.379763 $ &$ > 0.95 $&$ -0.02 \pm 0.25 $&$ 28.09 \pm 0.19 $ & $1.94^{+0.06}_{-0.05}$      &                 $6.1^{+0.9}_{-1.4}$         &$3$\\       
\sidehead{Parallel field}
HFF1P-i1& $ 3.474802 $ & $ -30.362578 $ &$ > 1.80 $&$ 0.46 \pm 0.07 $&$ 26.52 \pm 0.05 $ & $1.04$& $7.3\pm0.8$ &\nodata\\               
HFF1P-i2& $ 3.480642 $ & $ -30.371175 $ &$ 1.76 \pm 0.34 $&$ -0.02 \pm 0.09 $&$ 26.95 \pm 0.07 $ & $1.05$& $6.3\pm0.7$&\nodata\\    
HFF1P-i3& $ 3.487575 $ & $ -30.364380 $ &$ 1.27 \pm 0.33 $&$ 0.32 \pm 0.11 $&$ 27.06 \pm 0.08 $ & $1.05$& $5.8\pm0.7$&\nodata\\      
HFF1P-i4& $ 3.488924 $ & $ -30.394630 $ &$ > 1.39 $&$ 0.25 \pm 0.11 $&$ 27.14 \pm 0.08 $ & $1.05$& $6.7\pm0.8$&\nodata\\              
HFF1P-i5& $ 3.482550 $ & $ -30.371559 $ &$ 1.19 \pm 0.29 $&$ 0.17 \pm 0.11 $&$ 27.15 \pm 0.09 $ & $ 1.05$& $5.8^{+0.7}_{-1.4}$&\nodata\\     
HFF1P-i6& $ 3.483960 $ & $ -30.397152 $ &$ > 1.57 $&$ 0.00 \pm 0.11 $&$ 27.20 \pm 0.09 $ & $1.05$& $6.3\pm0.7$&\nodata\\                
HFF1P-i7& $ 3.467582 $ & $ -30.396908 $ &$ > 1.39 $&$ 0.15 \pm 0.12 $&$ 27.23 \pm 0.09 $ & $1.04$& $6.8\pm0.8$&\nodata\\             
HFF1P-i8& $ 3.467097 $ & $ -30.387686 $ &$ 1.28 \pm 0.25 $&$ -0.24 \pm 0.11 $&$ 27.30 \pm 0.10 $ & $ 1.04$& $6.0\pm0.7$&\nodata\\      
HFF1P-i9& $ 3.489520 $ & $ -30.399528 $ &$ > 1.41 $&$ 0.05 \pm 0.13 $&$ 27.32 \pm 0.10 $ & $1.05$& $6.6\pm0.8$&\nodata\\               
HFF1P-i10& $ 3.466056 $ & $ -30.394409 $ &$ 1.10 \pm 0.30 $&$ 0.00 \pm 0.14 $&$ 27.43 \pm 0.11 $ & $1.04$& $6.0\pm0.7$&\nodata\\         
HFF1P-i11& $ 3.460587 $ & $ -30.366320 $ &$ 0.92 \pm 0.34 $&$ 0.05 \pm 0.18 $&$ 27.70 \pm 0.14 $ & $1.04$& $5.8^{+0.7}_{-1.1}$&\nodata\\       
HFF1P-i12& $ 3.455844 $ & $ -30.366359 $ &$ 1.03 \pm 0.36 $&$ 0.00 \pm 0.18 $&$ 27.70 \pm 0.14 $ & $1.03$& $4.5^{+1.6}_{-3.9}$&\nodata\\       
HFF1P-i13& $ 3.488139 $ & $ -30.367864 $ &$ > 0.91 $&$ 0.12 \pm 0.19 $&$ 27.73 \pm 0.15 $ & $1.06$& $5.8^{+1.0}_{-5.2}$&\nodata\\            
HFF1P-i14& $ 3.486988 $ & $ -30.399579 $ &$ 0.91 \pm 0.33 $&$ -0.07 \pm 0.18 $&$ 27.75 \pm 0.15 $ & $1.05$& $5.9^{+0.7}_{-1.1}$&\nodata\\     
HFF1P-i15& $ 3.484920 $ & $ -30.376917 $ &$ > 0.85 $&$ 0.11 \pm 0.20 $&$ 27.82 \pm 0.16 $ & $ 1.05$& $5.9^{+0.8}_{-0.7}$&\nodata\\             
HFF1P-i16& $ 3.477238 $ & $ -30.385998 $ &$ > 1.02 $&$ -0.22 \pm 0.21 $&$ 27.98 \pm 0.18 $ & $1.05$& $6.5\pm0.7$&\nodata\\            
HFF1P-i17& $ 3.481928 $ & $ -30.389557 $ &$ > 0.99 $&$ -0.21 \pm 0.22 $&$ 27.99 \pm 0.19 $ & $1.05$& $6.2^{+0.7}_{-0.8}$&\nodata            
\enddata
\tablenotetext{a}{Total magnitudes estimated with the aperture correction.}
\tablenotetext{b}{The magnification errors in the parallel field are less than $1\%$ based on our model extrapolation estimates. 
		Note that the errors in the parallel field would be underestimated because Abell 2744 is a complex merging cluster.
}
\tablenotetext{c}{References: (1) \citet{2014ApJ...786...60A}; (2) \citet{2014arXiv1402.6743Z}; (3) \citet{2014arXiv1409.0512A}}.
\end{deluxetable*}

\begin{deluxetable*}{lcccccccc}
\tabletypesize{\scriptsize}
\tablecaption{Dropout candidates at $z \sim 8$\label{candidates8} in the HFF Abell 2744 Fields}
\tablewidth{0pt}
\tablehead{
		\colhead{ID} & \colhead{R.A. (J2000)} & \colhead{Dec (J2000)} & \colhead{$Y_{105}-J_{125}$} & \colhead{$J_{125}-H_{160}$} & \colhead{$JH_{140}$\tablenotemark{a}} & \colhead{Magnification\tablenotemark{b}} & \colhead{Photo-$z$} & \colhead{Reference\tablenotemark{c}}
}
\startdata
\sidehead{Cluster field}
HFF1C-Y1 & $ 3.604518 $ & $ -30.380467 $ &$ 1.17 \pm 0.06 $&$ 0.04 \pm 0.04 $&$ 25.91 \pm 0.02 $ & $1.49\pm0.04$      &       $8.0\pm 0.9 $      & $1$, $2$, $4$, $5$, $6$\\ 
HFF1C-Y2 & $ 3.603378 $ & $ -30.382254 $ &$ 1.26 \pm 0.14 $&$ 0.09 \pm 0.07 $&$ 26.62 \pm 0.05 $ & $1.61\pm0.05$    &         $8.2\pm 0.9 $      & $2$, $4$, $6$\\           
HFF1C-Y3 & $ 3.596091 $ & $ -30.385833 $ &$ 1.30 \pm 0.16 $&$ -0.00 \pm 0.08 $&$ 26.67 \pm 0.05 $ & $2.25\pm0.09$&     $8.2\pm 0.9 $      & $2$, $4$, $6$\\           
HFF1C-Y4 & $ 3.606461 $ & $ -30.380996 $ &$ 1.12 \pm 0.13 $&$ -0.09 \pm 0.08 $&$ 26.94 \pm 0.06 $ & $1.49\pm0.04$    &        $8.0\pm 0.9 $      & $2$, $4$, $6$\\           
HFF1C-Y5 & $ 3.603859 $ & $ -30.382263 $ &$ 1.92 \pm 0.34 $&$ 0.35 \pm 0.10 $&$ 26.98 \pm 0.07 $ & $1.60\pm0.05$     &         $8.4\pm 0.9 $      & $2$, $4$, $6$\\           
HFF1C-Y6 & $ 3.606577 $ & $ -30.380924 $ &$ 1.09 \pm 0.22 $&$ 0.40 \pm 0.12 $&$ 27.15 \pm 0.08 $ & $1.48\pm0.04$      &        $7.9^{+0.9}_{-6.1}   $      & $2$, $6$\\       
HFF1C-Y7 & $ 3.588980 $ & $ -30.378668 $ &$ 1.12 \pm 0.19 $&$ -0.08 \pm 0.11 $&$ 27.28 \pm 0.09 $ & $1.82^{+0.06}_{-0.05}$      &       $7.9\pm 0.9  $     & $2$, $4$, $6$\\           
HFF1C-Y8 & $ 3.603997 $ & $ -30.382304 $ &$ 1.05 \pm 0.29 $&$ 0.17 \pm 0.17 $&$ 27.59 \pm 0.12 $ & $1.60\pm0.05$    &         $7.9^{+0.9}_{-6.1}  $      & $2$, $6$\\       
HFF1C-Y9 & $ 3.592349 $ & $ -30.409892 $ &$ 0.60 \pm 0.39 $&$ -0.24 \pm 0.32 $&$ 28.00 \pm 0.24 $ & $9.82^{+0.65}_{-0.57}$   & $7.3^{+0.8}_{-1.9}  $       & $2$, $6$\\    
HFF1C-Y10 & $ 3.605263 $ & $ -30.380604 $ &$ 0.76 \pm 0.39 $&$ 0.28 \pm 0.27 $&$ 28.03 \pm 0.17 $ & $1.48\pm0.04$      &      $7.6^{+0.8}_{-6.4}   $     & $3$\\       
HFF1C-Y11 & $ 3.605062 $ & $ -30.381463 $ &$ 0.84 \pm 0.30 $&$ 0.02 \pm 0.22 $&$ 28.06 \pm 0.18 $  & $1.53^{+0.04}_{-0.05}$    &        $7.7^{+0.8}_{-6.4}  $      & $2$\\       
\sidehead{Parallel field}
HFF1P-Y1 & $ 3.474918 $ & $ -30.362542 $ &$ 0.61 \pm 0.11 $&$ 0.04 \pm 0.08 $&$ 26.93 \pm 0.07 $ & $1.05$& $7.5^{+0.8}_{-1.5}$& \nodata\\    
HFF1P-Y2 & $ 3.459245 $ & $ -30.367360 $ &$ 0.73 \pm 0.18 $&$ 0.02 \pm 0.13 $&$ 27.44 \pm 0.11 $ & $1.04$& $7.6^{+0.8}_{-1.7}$& \nodata\\   
HFF1P-Y3 & $ 3.479684 $ & $ -30.366359 $ &$ 0.85 \pm 0.33 $&$ 0.15 \pm 0.21 $&$ 27.71 \pm 0.14 $ & $1.05$& $7.8^{+0.9}_{-6.4}$& \nodata\\  
HFF1P-Y4 & $ 3.457192 $ & $ -30.379281 $ &$ 0.84 \pm 0.45 $&$ 0.17 \pm 0.29 $&$ 28.02 \pm 0.19 $ & $1.03$& $7.8^{+0.9}_{-6.6}$& \nodata  
\enddata
\tablenotetext{a}{Total magnitudes estimated with the aperture correction.}
\tablenotetext{b}{The magnification errors in the parallel field are less than $1\%$}.
\tablenotetext{c}{References: (1) \citet{2014ApJ...786...60A}; (2) \citet{2014arXiv1402.6743Z}; (3) \citet{2014arXiv1402.6743Z} possible candidates; (4) \citet{2014arXiv1405.0011C}; (5) \citet{2014A&A...562L...8L}; (6) \citet{2014arXiv1409.0512A}.}
\end{deluxetable*}

\setlength{\tabcolsep}{3pt}
\begin{deluxetable*}{lcccccccc}
\tabletypesize{\scriptsize}
\tablecaption{Dropout candidates at $z \sim 9$\label{candidates9} in the HFF Abell 2744 Fields}
\tablewidth{0pt}
\tablehead{
		\colhead{ID} & \colhead{R.A. (J2000)} & \colhead{Dec (J2000)} & \colhead{$(Y_{105}+J_{125})/2-JH_{140}$} & \colhead{$JH_{140}-H_{160}$} & \colhead{$H_{160}$\tablenotemark{a}} & \colhead{Magnification\tablenotemark{b}} & \colhead{Photo-$z$} & \colhead{Reference\tablenotemark{c}}
}
\startdata
\sidehead{Cluster field}
HFF1C-YJ1   & $ 3.592512 $ & $ -30.401486 $ &$ >1.22 $&$ 0.55 \pm 0.30 $&$ 27.37 \pm 0.16 $       & $14.40^{+1.20}_{-1.06}$ &   $9.6^{+1.0}_{-7.1}$       & $1$, $2$\\  
HFF1C-Y2\tablenotemark{d} & $ 3.603380 $ & $ -30.382255 $ &$ 0.78 \pm 0.07 $&$ -0.05 \pm 0.07 $&$ 26.67 \pm 0.05 $      & $1.61\pm0.05$ &  $8.2\pm 0.9 $        & $3$,$4$,$5$\\   
HFF1C-Y5\tablenotemark{d} & $ 3.603859 $ & $ -30.382262 $ &$ 1.13 \pm 0.10 $&$ 0.19 \pm 0.08 $&$ 26.78 \pm 0.05 $       & $1.60\pm0.05$ &  $8.4\pm 0.9 $        & $3$,$4$,$5$\\   
\sidehead{Parallel field}
HFF1P-YJ1 & $ 3.488893 $ & $ -30.396183 $ &$ 1.61 \pm 0.25 $&$ -0.26 \pm 0.16 $&$ 27.67 \pm 0.11 $      & $1.05$ &   $8.7\pm1.0$       & \nodata \\  
HFF1P-YJ2 & $ 3.473522 $ & $ -30.384024 $ &$ > 1.28 $&$ 0.26 \pm 0.22 $&$ 27.70 \pm 0.11 $               & $1.04$ &   $8.8^{+1.0}_{-1.7}$         & \nodata \\  
HFF1P-YJ3 & $ 3.474445 $ & $ -30.368728 $ &$ > 1.79 $&$ -0.34 \pm 0.23 $&$ 28.14 \pm 0.17 $            & $1.04$ &    $8.9^{+1.0}_{-6.9}$       & \nodata     
\enddata
\tablenotetext{a}{Total magnitudes estimated with the aperture correction.}
\tablenotetext{b}{The magnification errors in the parallel field are less than $1\%$}.
\tablenotetext{c}{References: (1) \citet{2014arXiv1407.3769Z}; (2) \citet{2014arXiv1409.1228O}; (3) \citet{2014arXiv1402.6743Z}; (4) \citet{2014arXiv1405.0011C}; (5) \citet{2014arXiv1409.0512A}.}
\tablenotetext{d}{Identified by our two selections for $Y$-dropouts and $YJ$-dropouts.}
\end{deluxetable*}

\section{Mass Model} \label{sec:Mass model}

In this section, 
we construct a mass model of Abell 2744 at $z=0.308$
using the parametric gravitational lensing package {\sc glafic} \citep{2010PASJ...62.1017O}. 
\footnote{we name the mass model in this work 'glafic model version 1.0'. This mass model will be released on the STScI website (http://archive.stsci.edu/prepds/frontier/lensmodels/). The mass model version 2.0 is being developed,
in which we include new multiple images.}
Our mass model includes 
three types of mass distributions:
cluster-scale halos,  cluster member galaxy halos, and external perturbation. 
With the positions of multiple images provided in the literature, 
we optimize free parameters of the mass profiles based on a standard $\chi^2$ minimization to determine the best-fit mass model
whose parameters are summarized in Table \ref{glaficresult}. 
We then calculate magnification factors $\mu$ of our dropouts and positions of the multiple images
using the best-fit mass model.

\subsection{Cluster-Scale Halos} \label{subsec:halos}

We place three cluster-scale halos at the positions of three brightest galaxies in the core of the cluster.
We adopt the Navarro-Frenk-White (NFW) profiles \citep{1997ApJ...490..493N} for the mass distributions of the cluster-scale halos.
The radial profiles of NFW are described as 
\begin{eqnarray}
\rho(r) = \frac{\rho_s}{(r/r_s)(1+r/r_s)^2},
\end{eqnarray}
where $\rho_s$ is the characteristic density and $r_s$ is the scale radius.
The scale radius is defined by 
\begin{eqnarray}
		r_s = \frac{r_{\rm vir}}{c_{\rm vir}}, 
\end{eqnarray}
where 
$r_{\rm vir}$ is the virial radius of the cluster-scale halo
and $c_{\rm vir}$ is the concentration parameter. 
The scale radius and the characteristic density 
are related to the virial mass $M_{\rm vir}$ and the concentration parameter 
with the equations,
\begin{align}
		r_s 
			&= \dfrac{1}{c_{\rm vir}} \left( \dfrac{3 M_{\rm vir}}{4 \pi \Delta(z) \bar \rho(z)} \right)^{1/3}, \\
		\rho_s 
		&= \dfrac{\Delta (z) \bar \rho(z) c_{\rm vir}^3}{3 m_{\rm nfw}(c_{\rm vir})}, \\
		m_{\rm nfw} (c_{\rm vir}) 
			&= \displaystyle \int^{c_{\rm vir}}_0 \dfrac{r}{(1+r)^2} dr, 
\end{align}
where $\Delta (z)$ is the nonlinear overdensity \citep[e.g.,][]{1997PThPh..97...49N}
and 
$\bar \rho (z)$ is the mean matter density of the universe at a redshift $z$. 
		The surface mass density $\Sigma(r')$ is obtained as a function of a radius $r' \equiv \sqrt{x^2 + y^2}$ by integrating $\rho(r)$ along the line of sight:
		\begin{eqnarray}
				\Sigma(r') = \int^{+\infty}_{-\infty} \rho(r',z)dz
		\end{eqnarray}
		In the obove discussion, we assumed spherical halos.
		We then introduce an ellipticity $e$ in the isodensity contour by replacing $r'$ in $\Sigma(r')$ \citep{2010PASJ...62.1017O}:
		\begin{eqnarray}
				\Sigma(r'): r' \rightarrow \sqrt{\frac{\tilde{x}^2}{(1-e)}+(1-e)\tilde{y}^2},
		\end{eqnarray}

where $\tilde{x}$ and $\tilde{y}$ are defined by the following equations with the position angle $\theta_e$ (measured east of north) of the isodensity contours:
\begin{eqnarray}
		\tilde{x} = x\cos \theta_e + y \sin \theta_e \\
		\tilde{y} = -x\sin \theta_e + y \cos \theta_e.
\end{eqnarray}
We use the ellipsoidal halos in this work.
Each cluster-scale halo has four free parameters: $M_{\rm vir}$, $c_{\rm vir}$,
$e$, and $\theta_e$.

\setlength{\tabcolsep}{6pt}
\begin{deluxetable*}{lccccccc}
\tabletypesize{\scriptsize}
\tablecaption{Best-fit mass model parameters\label{glaficresult}}
\tablewidth{0pt}
\tablehead{
		\colhead{Component} & 
		\colhead{model} & 
		\colhead{Mass ($h^{-1} M_\odot$)} &  
		\colhead{$e$} & 
		\colhead{$\theta_e$ ($^\circ$)} & 
		\colhead{$c$} & 
		\colhead{R.A. (J2000)} & 
		\colhead{Dec (J2000)}
}
\startdata
Cluster halo 1  & NFW & $3.5 \times 10^{14}$ & $0.20$ & $30.0 $& $3.41 $& $3.585972 $& $-30.400122 $\\
Cluster halo 2  & NFW & $2.5 \times 10^{14}$ & $0.49$ & $-41.9$ &$ 8.01$ &$ 3.592074$ &$ -30.405165$ \\
Cluster halo 3  & NFW & $1.3 \times 10^{13}$ & $0.60$ & $72.2 $& $28.2 $& $3.583417 $& $-30.392069 $\\ 
\tableline
& & $\sigma_*$ (km s$^{-1}$) & $r_{{\rm trun,*}}$ ($''$) & $\eta$ &  &  &  \\[2pt]
\cline{3-8}\\
Member galaxies & PJE & $2.0 \times 10^2$ & $5.09 \times 10$ & $1.22$ &  & &  \\[2pt]
\tableline
& & $z_{s,{\rm fid}}$ & $\gamma$ & $\theta_\gamma$ ($^\circ$) & $\kappa$ &  &  \\[2pt]
\cline{3-8}\\
Perturbation & PRT & $2.0$(fix) & $6.21 \times 10^{-2}$ & $15.2$ & $0.0$(fix) & &
\enddata
\end{deluxetable*}

\subsection{Cluster Member Galaxy Halos} \label{subsec:membergalaxy}

\begin{figure}
		\epsscale{1.2}
	\plotone{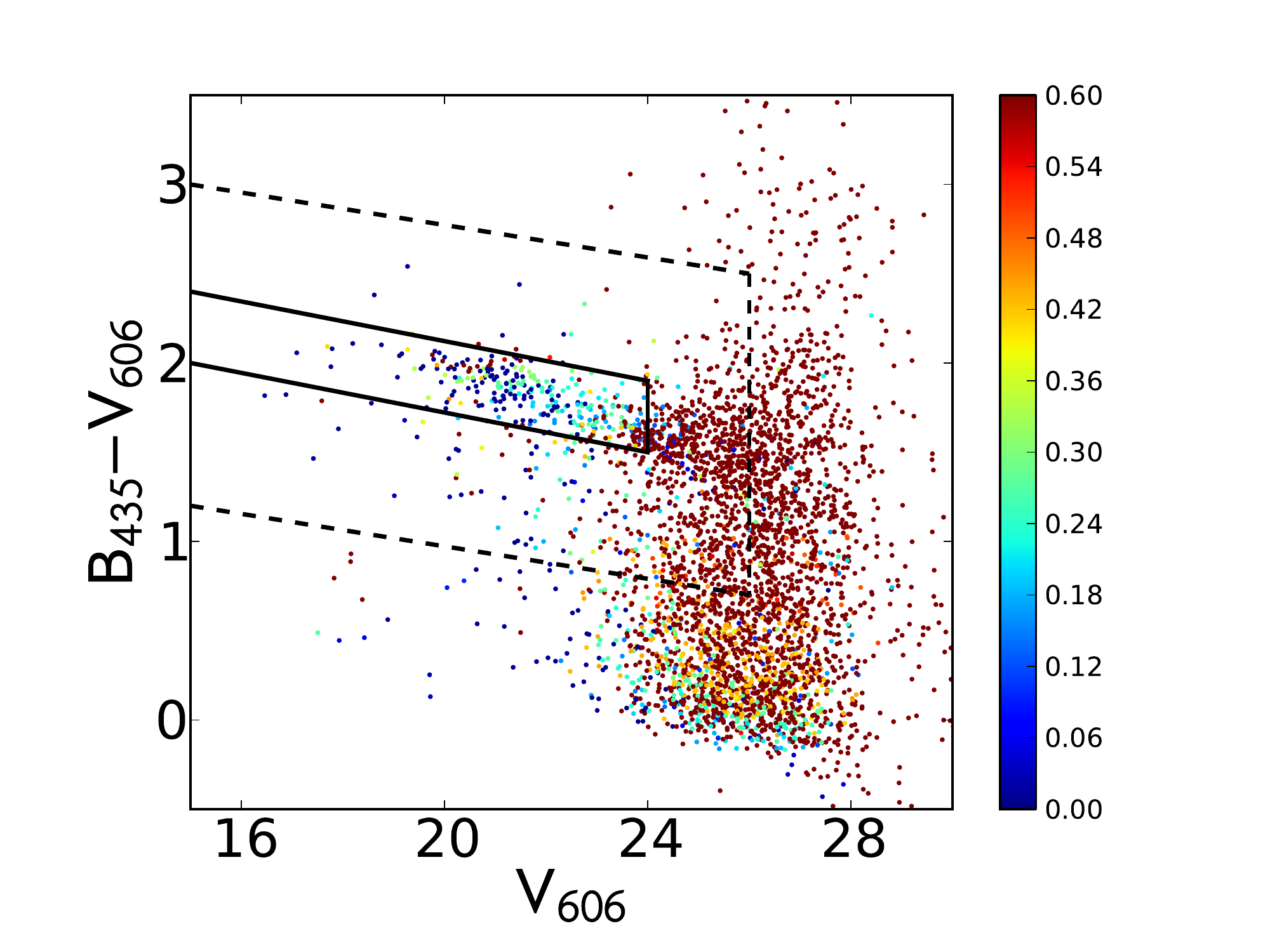}
	\caption{
	Color-magnitude diagram for the objects (circles) detected in the optical images. 
	The color code for the circles indicates photometric redshifts estimated with {\sc bpz}. 
	The solid line box indicates the color selection window for the cluster members of the red-sequence galaxies at the cluster redshift.
	The dashed line box represents the relaxed selection window that is applied with the photometric redshift criterion.
	}
	\label{colormag}
\end{figure}

To estimate contributions from cluster member galaxy halos, 
we identify cluster member galaxies with spectroscopic redshifts $z_{\rm spec}$, 
photometric redshifts $z_{\rm photo}$, and $B_{435} - V_{606}$ colors. 
First, we use $z_{\rm spec}$ 
presented in Table 5 of \citet{2011ApJ...728...27O}. 
Galaxies at $0.28 < z_{\rm spec} < 0.34$ are regarded as cluster member galaxies. 
For objects with no $z_{\rm spec}$, we apply color criteria,
\begin{align}
		-\frac{1}{18}(B_{435} - V_{606}) + 2 &< V_{606}, \\
		-\frac{1}{18}(B_{435} - V_{606}) + 2.4 &> V_{606}, \\
		V_{606} &< 24, 
\end{align}
and select galaxies on the red-sequence of the cluster redshift.
Figure \ref{colormag} presents the $B_{435} - V_{606}$ versus $V_{606}$ color-magnitude diagram
of our objects, together with the color criteria of the red-sequence galaxies
shown with the solid-line box.
Finally, for objects that are selected neither by $z_{\rm spec}$ nor the red-sequence criteria, 
we refer $z_{\rm photo}$ estimated with {\sc bpz} (Section \ref{sec:Samples}). 
We select member galaxies with the $z_{\rm photo}$ criterion and 
the relaxed color-magnitude criteria, 
\begin{align}
		0.09 < z_{\rm photo} &< 0.4, \\
		-\frac{1}{22}(B_{435} - V_{606}) + 1.2 &< V_{606}, \\
		-\frac{1}{22}(B_{435} - V_{606}) + 3 &> V_{606}, \\
		V_{606} &< 26.
\end{align}
The dashed-line box in Figure \ref{colormag} indicates the boundary 
of the relaxed color-magnitude criteria.

We describe halo mass distributions of these member galaxies by the sum of pseudo-Jaffe ellipsoids (PJE; \citealt{2001astro.ph..2341K}, see also \citealt{1983MNRAS.202..995J}). 
In this model, the mass profile is characterized by 
the velocity dispersion $\sigma$ 
and the truncation radius $r_{\rm trun}$. 
We assume that the parameters of $\sigma$ and $r_{\rm trun}$ are scaled with the galaxy luminosity $L$ in the $i_{814}$ band, 
\begin{align}
		\frac{\sigma}{\sigma_*} &= \left( \frac{L}{L_*} \right)^{1/4}, \\
		\frac{r_{\rm trun}}{r_{\rm trun,*}} &= \left( \frac{L}{L_*} \right)^\eta,
\end{align}
where $L_*$ is the normalization luminosity at the cluster redshift, 
and $\sigma_*$, $r_{\rm trun,*}$, and $\eta$ are free parameters.  
The mass-to-light ratio is constant for $\eta = 0.5$.
The ellipticities and position angles of the member galaxy halos are determined 
from shapes of the member galaxies in the $JH_{140}$ band.

\subsection{External Perturbation} \label{subsec:pert}

Although the mass distribution of Abell 2744 is mostly explained by 
the contributions from the three cluster-scale halos (Section \ref{subsec:halos}) and the member galaxy halos (Section \ref{subsec:membergalaxy}),
we include perturbation induced by external sources  
to improve our mass model. 
If the perturbation is weak, 
its potential can be described by \citep[e.g.,][]{1991ApJ...373..354K} 
\begin{eqnarray}
		\phi = \frac{1}{2} r^2 \kappa + \frac{1}{2} r^2 \gamma \cos 2(\theta - \theta_\gamma), 
\end{eqnarray}
where 
$\kappa$ is the constant convergence 
and 
$\gamma$ is the constant tidal shear. 
The amplitude of the potential $\phi$ is defined for a given fiducial source redshift $z_{\rm s,fid}$. 
We refer to this potential as PRT.
In this paper, 
$\gamma$ and $\theta_\gamma$ are free parameters.
We fix $z_{s,{\rm fid}}$ and $\kappa$; $z_{s,{\rm fid}} \equiv 2.0$ and $\kappa \equiv 0$.

\subsection{Model Optimization} \label{sec:optimization}

To constrain the mass-model parameters of the cluster, we use the positions of multiply imaged systems. 
We identify multiply imaged galaxies based on their colors and morphologies 
while iteratively refining the mass-model parameters. 
In total, we use the positions of $67$ multiple images of $24$ systems summarized in Table \ref{multipleimages}. 
Seventeen out of the $24$ systems, IDs $1.1 - 17.2$ (Table \ref{multipleimages}),
are the same as those listed in Table 1 of the document provided by the HFF map-making team (PI: K. Sharon; see also \citealt{2014arXiv1405.0222J})
\footnote{http://archive.stsci.edu/prepds/frontier/lensmodels/\\hlsp\_frontier\_model\_abell2744\_sharon\_v1\_readme.pdf}.
Similarly, four high-redshift systems from the $24$ systems, IDs $19.1 - 22.2$, are 
listed in Table 3 of \citet{2014ApJ...786...60A},
among which IDs $19.1$, $19.2$, and $19.3$ are identified as $z\sim6-7$ dropouts in Section \ref{sec:dropout selection}. 
The images of IDs $19.1$, $19.2$, and $19.3$ are referred to
as HFF1C-i5, HFF1C-i6, and HFF1C-i8 in Table \ref{candidates7}.
We also find three new sets of multiple images, IDs $18.1-18.3$ and $23.1 - 24.2$.

We search for the best-fit mass model which reproduces the positions of the multiple images.
We optimize the 23 free parameters described in Section \ref{subsec:halos}-\ref{subsec:pert}
based on a $\chi^2$ minimization with the downhill-simplex algorithm 
(See \citealt{2010PASJ...62.1017O} for more details). 
We assign a positional error of $0 \farcs 4$ in the image plane for each multiple image, following \citet{2010PASJ...62.1017O},
and obtain the best-fit parameters as summarized in Table \ref{glaficresult}. 
The best-fit model has $\chi^2 = 52.8$ for $41$ degrees of freedom,
suggesting that our model is reasonable.

In the following sections, we use this best-fit mass model to estimate the
lensing effects both in the cluster and parallel fields. 
Because the parallel field has no multiple images to constrain 
the parameters of the mass model, it should be noted that 
the lensing effects in the parallel field are estimated 
from the extrapolation of the mass model determined at the cluster field.
However, the extrapolation gives negligibly small uncertainties in
our final results, due to the very small magnification factors (see Section \ref{sec:magnification_factors}).

\setlength{\tabcolsep}{6pt}
\begin{deluxetable*}{lcccccc}
\tabletypesize{\scriptsize}
\tablecaption{Multiple images\label{multipleimages}}
\tablewidth{0pt}
\tablehead{
		\colhead{ID} & \colhead{R.A. (J2000)} & \colhead{Dec (J2000)} & \colhead{$z_{\rm model}$\tablenotemark{a}} & \colhead{$z_{\rm photo}$\tablenotemark{b}} & \colhead{$z_{\rm spec}$\tablenotemark{b}} & \colhead{Reference\tablenotemark{c}}
}
\startdata
$1.1$ & $3.595958$ & $-30.40682$ & $1.61$ & $0.5-2.2$ & \nodata & $1$\\             
$1.2$ & $3.597542$ & $-30.40392$ &      &         &&\\              
$1.3$ & $3.586208$ & $-30.40999$ &      &         &&\\[2pt]         
\tableline                                            
$2.1$ & $3.596417$ & $-30.40612$ & $1.63$ & $0.5-2.2$ & \nodata &$ 1$\\              
$2.2$ & $3.597042$ & $-30.40475$ &      &         &&\\              
$2.3$ & $3.585744$ & $-30.41010 $&       &         &&\\[2pt]          
\tableline                                            
$3.1$ & $3.585417$ & $-30.39990 $&$ 2.00$  & $0.5-2.9$ & \nodata & $1$\\               
$3.2$ & $3.583250$ & $-30.40335 $&       &         &&\\               
$3.3$ & $3.597292$ & $-30.39672$ &      &         &&\\              
$3.4$ & $3.586417$ & $-30.40213$ &      &         &&\\[2pt]         
\tableline                                            
$4.1$ & $3.596750$ & $-30.39630 $& $1.97$   & $0.5-2.9$ & \nodata & $1$\\                
$4.2$ & $3.582542$ & $-30.40227$ &      &         &&\\              
$4.3$ & $3.586250$ & $-30.40085$ &       &         &&\\               
$4.4$ & $3.584500$ & $-30.39929$ &        &         &&\\[2pt]           
\tableline                                            
$5.1$ & $3.589375$ & $-30.39388$ & $2.17$ &\nodata & $3.98$ & $1$\\              
$5.2$ & $3.588792$ & $-30.39380$ &       &        &&\\               
$5.3$ & $3.577500$ & $-30.39957$ &        &        &&\\[2pt]           
\tableline                                            
$6.1$ & $3.592125$ & $-30.40263$ & $3.58$\tablenotemark{d} &\nodata & $3.58$ & $1$\\              
$6.2$ & $3.595625$ & $-30.40162$ &      &    &&\\              
$6.3$ & $3.580417$ & $-30.40892$ &      &    &&\\
$6.4$ & $3.593208$ & $-30.40491$ &      &    &&\\
$6.5$ & $3.593583$ & $-30.40511$ &      &    &&\\[2pt]
\tableline
$7.1$ & $3.583417$ & $-30.39207$ & $1.18$ &\nodata && $1$\\
$7.2$ & $3.585000$ & $-30.39138$ &         &        &&\\
$7.3$ & $3.579958$ & $-30.39476$ &      &        &&\\[2pt]
\tableline
$8.1$ & $3.598535$ & $-30.40180 $& $2.02$\tablenotemark{d} &\nodata & $2.019$ &  $1$\\
$8.2$ & $3.594042$ & $-30.40801$ &      &                        & &\\
$8.3$ & $3.586417$ & $-30.40937$ &      &                        & &\\[2pt]
\tableline
$9.1$ & $3.598261$ & $-30.40232$ & $2.71$ & $0.3-3.4$ & \nodata & $1$\\
$9.2$ & $3.595233$ & $-30.40741$ &      &         && \\
$9.3$ & $3.584601$ & $-30.40982$ &      &         &&\\[2pt]
\tableline
$10.1$ & $3.589708$ & $-30.39434$ & $2.59$ &\nodata & \nodata &$ 1$\\
$10.2$ & $3.588833$ & $-30.39422$ &      &        & & \\[2pt]
\tableline
$11.1$ & $3.588375$ &$ -30.40527$ & $3.76$ & $0.6-2.8$ & \nodata & $1$\\
$11.2$ & $3.587125$ &$ -30.40624$ &      &         &&\\
$11.3$ & $3.600150 $& $-30.39715 $&       &         &&\\[2pt]
\tableline
$12.1$ & $3.588417$ & $-30.40588$ & $5.38$  & $1.8-3.2$ & \nodata & $1$\\
$12.2$ & $3.587375$ & $-30.40648$ &      &         &&\\
$12.3$ & $3.600721$ & $-30.39709$ &      &         &&\\[2pt]
\tableline
$13.1$ & $3.597264$ & $-30.40143$ & $2.73$  & $0.4-2.8$ & \nodata & $1$\\
$13.2$ & $3.582792$ & $-30.40891$ &      &&&\\[2pt]
\tableline
$14.1$ & $3.593239$ & $-30.40325$ &$3.32$  & $1.4-3.1$& \nodata &$1$\\
$14.2$ & $3.594555$ & $-30.40300$ &        &&&\\[2pt]
\tableline
$15.1$ & $3.592375$ & $-30.40256$ & $1.46$  & $0.6-2.6$& \nodata &$1$\\
$15.2$ & $3.593792$ & $-30.40216$ &      &&&\\
$15.3$ & $3.582792$ & $-30.40804$ &      &&&\\[2pt]
\tableline
$16.1$ & $3.589750$ & $-30.39464$ & $1.79$   & $1.8-3.2$& \nodata &$1$\\
$16.2$ & $3.588458$ & $-30.39444$ &      &&&\\[2pt]
\tableline
$17.1$ & $3.590750$ & $-30.39556$ & $1.44$   & $1.5-5.4$&  \nodata &$1$\\
$17.2$ & $3.588375$ & $-30.39564$ &      &&&\\[2pt]
\tableline
$18.1$ & $3.598676$ & $-30.40491$ & $2.19$  &\nodata& \nodata & \nodata \\
$18.2$ & $3.587053$ & $-30.41126$ &      &&&\\
$18.3$ & $3.596818$ & $-30.40783$ &      &&&\\[2pt]
\tableline
$19.1$ & $3.580452$ & $-30.40504$ & $7.94$  &\nodata& \nodata &$2$\\
$19.2$ & $3.597831$ & $-30.39596$ &      &&&\\
$19.3$ & $3.585321$ & $-30.39796$ &      &&&\\[2pt]
\tableline
$20.1$ & $3.596572$ & $-30.40900$ & $7.50$    &\nodata& \nodata &$2$\\
$20.2$ & $3.600058$ & $-30.40440$ &       &&&\\
$20.3$ & $3.585801$ & $-30.41175$ &      &&&\\[2pt]
\tableline
$21.1$ & $3.591432$ & $-30.39669$ & $4.64$  &\nodata& \nodata &$2$\\
$21.2$ & $3.576122$ & $-30.40449$ &      &&&\\[2pt]
\tableline
$22.1$ & $3.593552$ & $-30.40971$ & $3.73$  &\nodata& \nodata &$2$\\
$22.2$ & $3.600541$ & $-30.40182$ &      &&&\\[2pt]
\tableline
$23.1$ & $3.578090$ & $-30.39964$ & $1.75$   &\nodata& \nodata& \nodata\\
$23.2$ & $3.589237$ & $-30.39444$ &      &&&\\[2pt]
\tableline
$24.1$ & $3.584284$ & $-30.40893$ & $2.62$  &\nodata& \nodata & \nodata \\
$24.2$ & $3.598125$ & $-30.40098$ &      &&&

\enddata
\tablenotetext{a}{Redshift from the best-fit mass model.}
\tablenotetext{b}{Numbers quoted from \citet{2014arXiv1405.0222J}.}
\tablenotetext{c}{References: (1) http://archive.stsci.edu/prepds/frontier/lensmodels/hlsp\_frontier\_model\_abell2744\_sharon\_v1\_readme.pdf, see also \citet{2014arXiv1405.0222J}; (2) \citet{2014ApJ...786...60A}.}
\tablenotetext{d}{Fixed to $z_{\rm spec}$.}
\end{deluxetable*}

\subsection{Magnification Factors and Multiple Images} \label{sec:magnification_factors}

\begin{figure*}
	\plotone{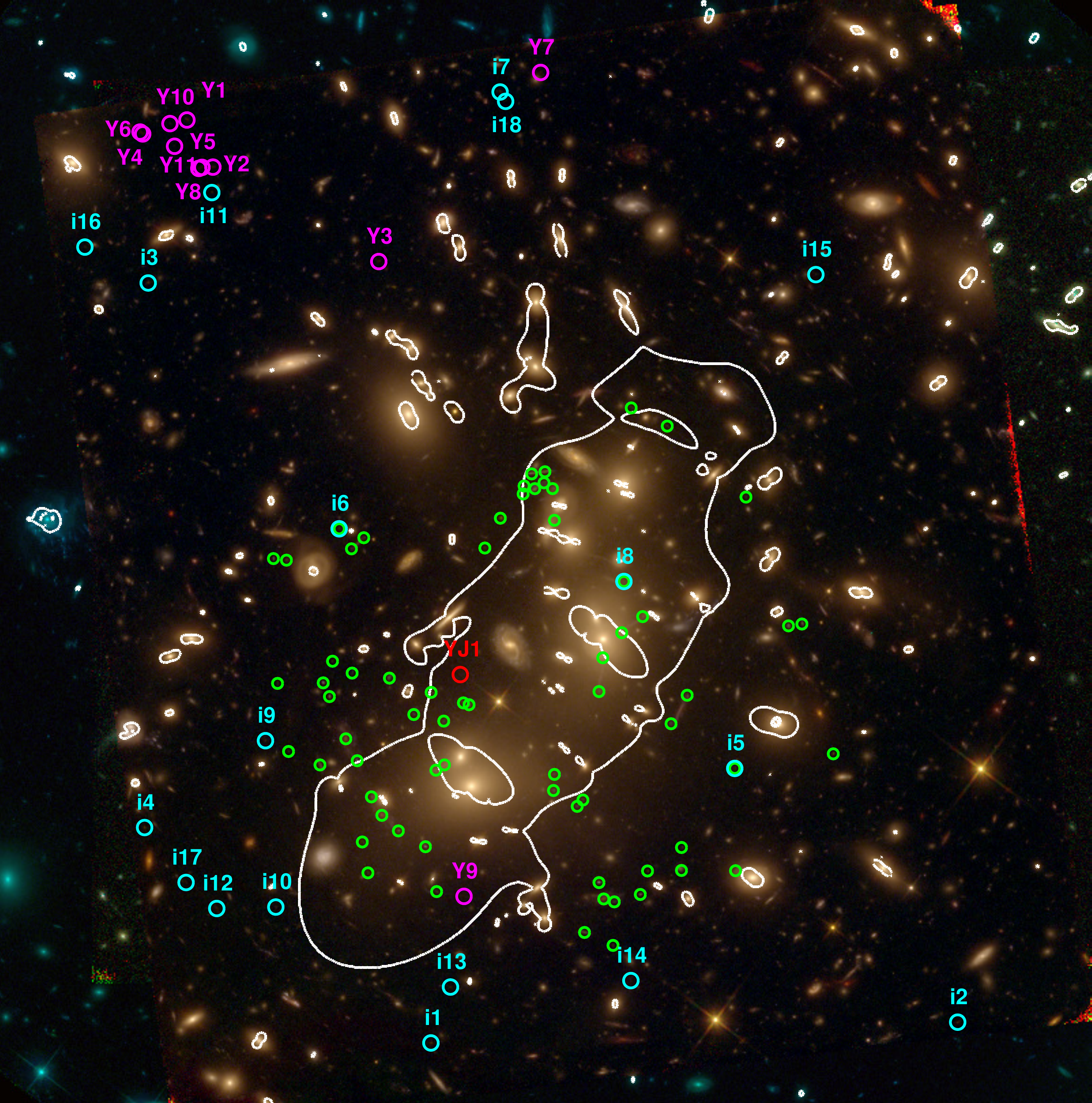}
	\caption{
			False-color image of the Abell 2744 cluster field (red: $J_{125}+JH_{140}+H_{160}$, green: $i_{814}+Y_{105}$, blue: $B_{435}+V_{606}$). 
	The cyan, magenta, and red circles 
	denote the positions of our dropout candidates at $z \sim 6-7$, $z \sim 8$, and $z \sim 9$, respectively. 
	The green circles indicate the multiply imaged systems 
	used for determination of our mass model. 
	The critical lines for background sources at $z = 8$ are shown with the white lines. 
	}
	\label{criticalline}
\end{figure*}

\begin{figure*}
	\plotone{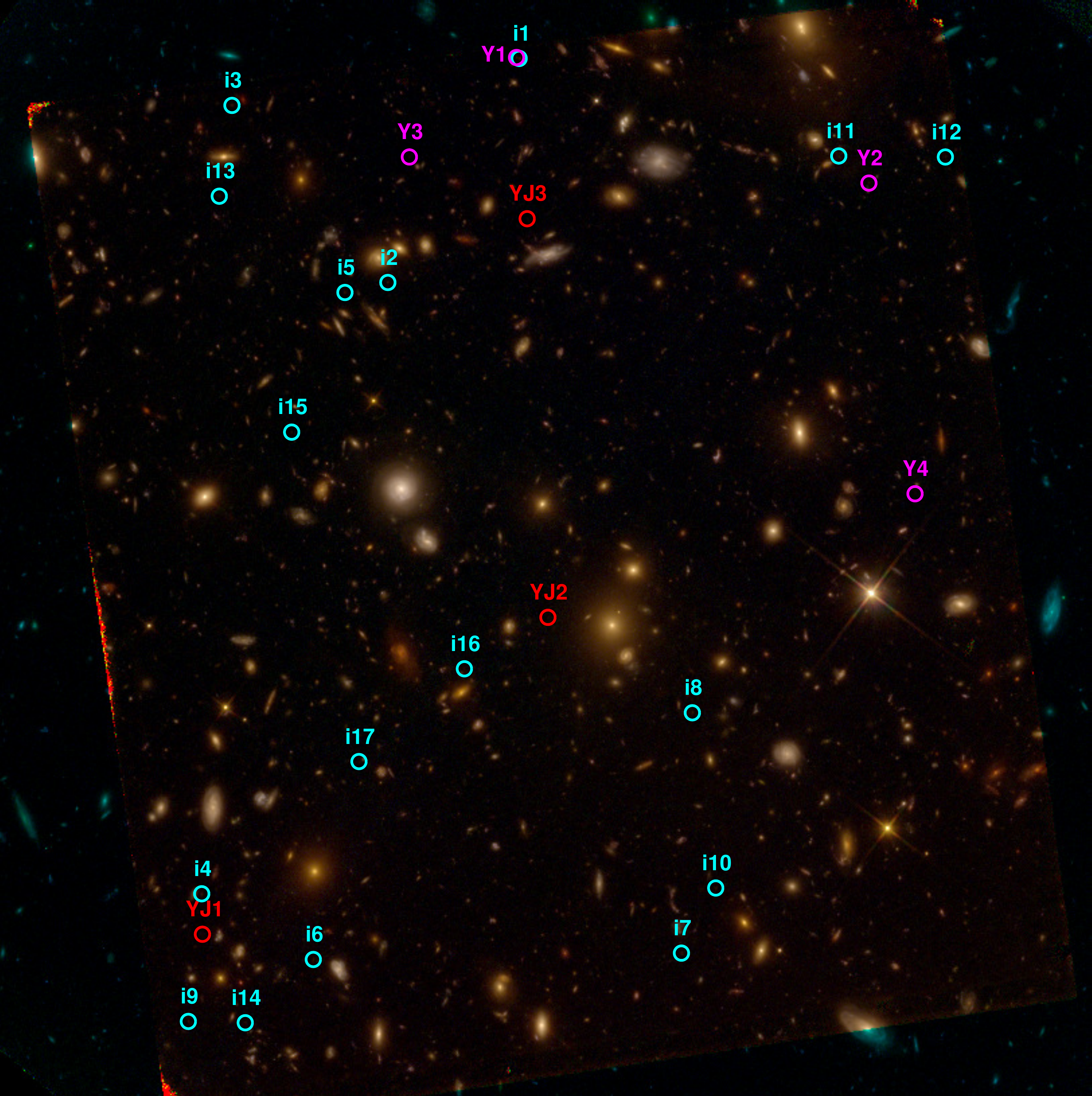}
	\caption{
			Same as Figure \ref{criticalline}, but for the Abell 2744 parallel field.
	}
	\label{drops_parallel}
\end{figure*}

Figure \ref{criticalline} displays  
the critical lines for $z = 8$ sources 
and the positions of our $z \sim 5-10$ dropouts in the cluster field. 
Because most of the dropouts are located far from the critical lines,
the magnification factors are generally small, $\mu \sim1.5-2$, 
as presented in Tables \ref{candidates7}-\ref{candidates9}. 
However, some of the dropouts are placed near the critical line 
and highly magnified. In particular, the magnification factors of three dropouts, 
HFF1C-i10, HFF1C-Y9, and HFF1C-YJ1, are estimated to be $\mu \sim 10-14$. 
The intrinsic absolute magnitudes of these three dropouts are $-17.00$, $-16.66$, and $-17.06$ mag, respectively.
Figure \ref{drops_parallel} shows the positions of our $z \sim 5-10$ dropouts in the parallel field.
The magnification factors in the parallel field are almost the same and near unity, typically $\sim 1.05$.
Thus, the lensing effects in the parallel field are negligibly small, and the parallel field 
may be regarded as a blank field. Although the magnification of the parallel field is very small,
we adopt the lensing magnifications to our dropouts both in the cluster and the parallel fields in our analysis. 
We estimate the errors of the magnification factors with a Markov Chain Monte Carlo (MCMC) method. 
These errors are shown in Tables \ref{candidates7}-\ref{candidates9}.

Our mass model predicts that three systems in our dropout samples have counter images.
We discuss the positions and the redshifts of these multiple images using our mass model.
Because the predicted positions of multiple images depend on the source redshifts,
we estimate the redshifts of multiple images from the positions in the image plane.

\begin{figure}
	\plotone{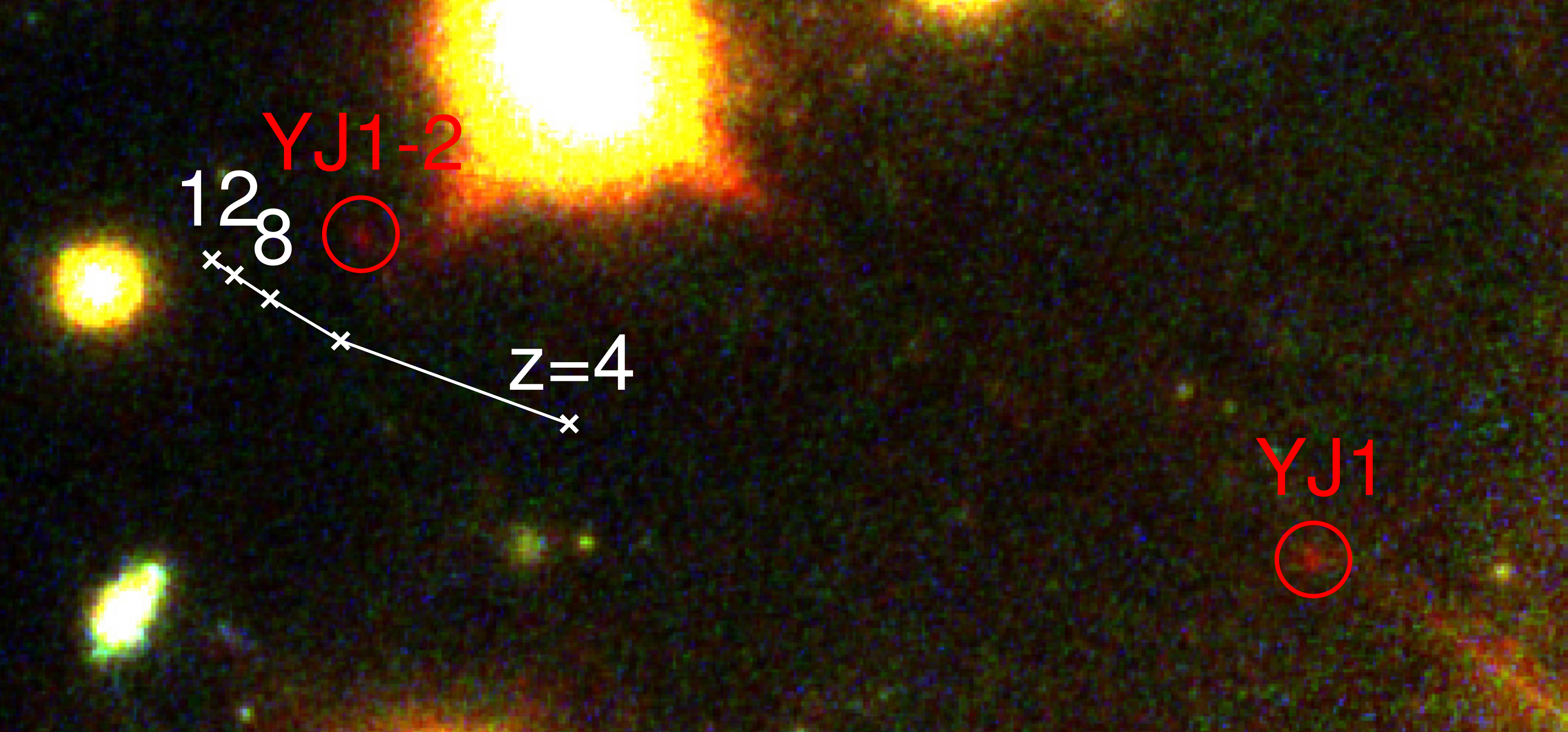}
	\plotone{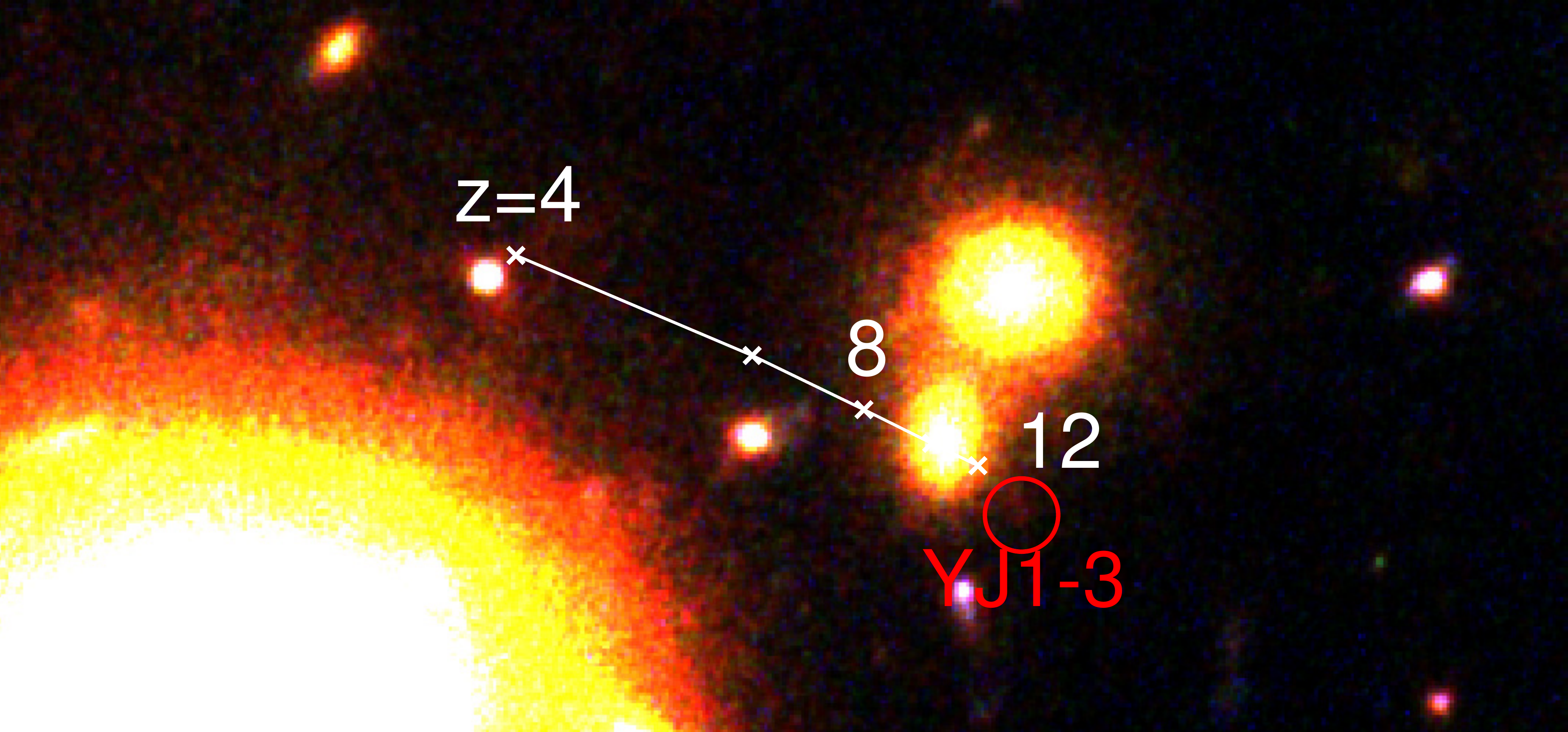}
	\caption{
			Predicted positions of multiple images of HFF1C-YJ1 using our mass model.
			The red circles show the positions of HFF1C-YJ1 and its multiple images with the ID of HFF1C-YJ1-2 and HFF1C-YJ1-3.
			The radius of each circle is $0\farcs3$.
			The white crosses in the upper (lower) panel indicate the predicted position of image HFF1C-YJ1-2 (HFF1C-YJ1-3) at $z=4$, $6$, $8$, $10$, and $12$ with the tracks over a redshift range $4 < z < 12$.
			Our mass model predicts that the redshift of YJ1 is $z > 6$, as discussed in \citet{2014arXiv1407.3769Z}.
	}
	\label{fig:YJ1}
\end{figure}

\paragraph{HFF1C-YJ1 at $z\sim9$}
HFF1C-YJ1 is a highly magnified dropout at $z \sim 9$ which is also reported in \citet{2014arXiv1407.3769Z}.
\citet{2014arXiv1407.3769Z} claim two counter images of this dropout, which we refer to as HFF1C-YJ1-2 and HFF1C-YJ1-3.
In Figure \ref{fig:YJ1}, we show the predicted positions of the counter images of HFF1C-YJ1 whose redshifts are
assumed to be $z = 4-12$.
The positions of the counter images are consistent with our estimates, if the redshift of HFF1C-YJ1 is $z > 6$.
In fact, the predicted positions of the counter images are largely separated, $>2''$, if HFF1C-YJ1 resides at $z=4$.
Thus our mass model predicts that the redshift of HFF1C-YJ1 is $z > 6$,
which is consistent with the findings of \citet{2014arXiv1407.3769Z}.
The multiple image positions rule out the possibility that HFF1C-YJ1 is 
a low redshift galaxy at $z<4$, and imply that the HFF1C-YJ1 system
is a strong candidate of a high redshift galaxy. Combining the result of photometric redshift,
we find that the system of HFF1C-YJ1, HFF1C-YJ1-2 and HFF1C-YJ1-3 is located at $z\simeq 9.6$.

\begin{figure}
	\plotone{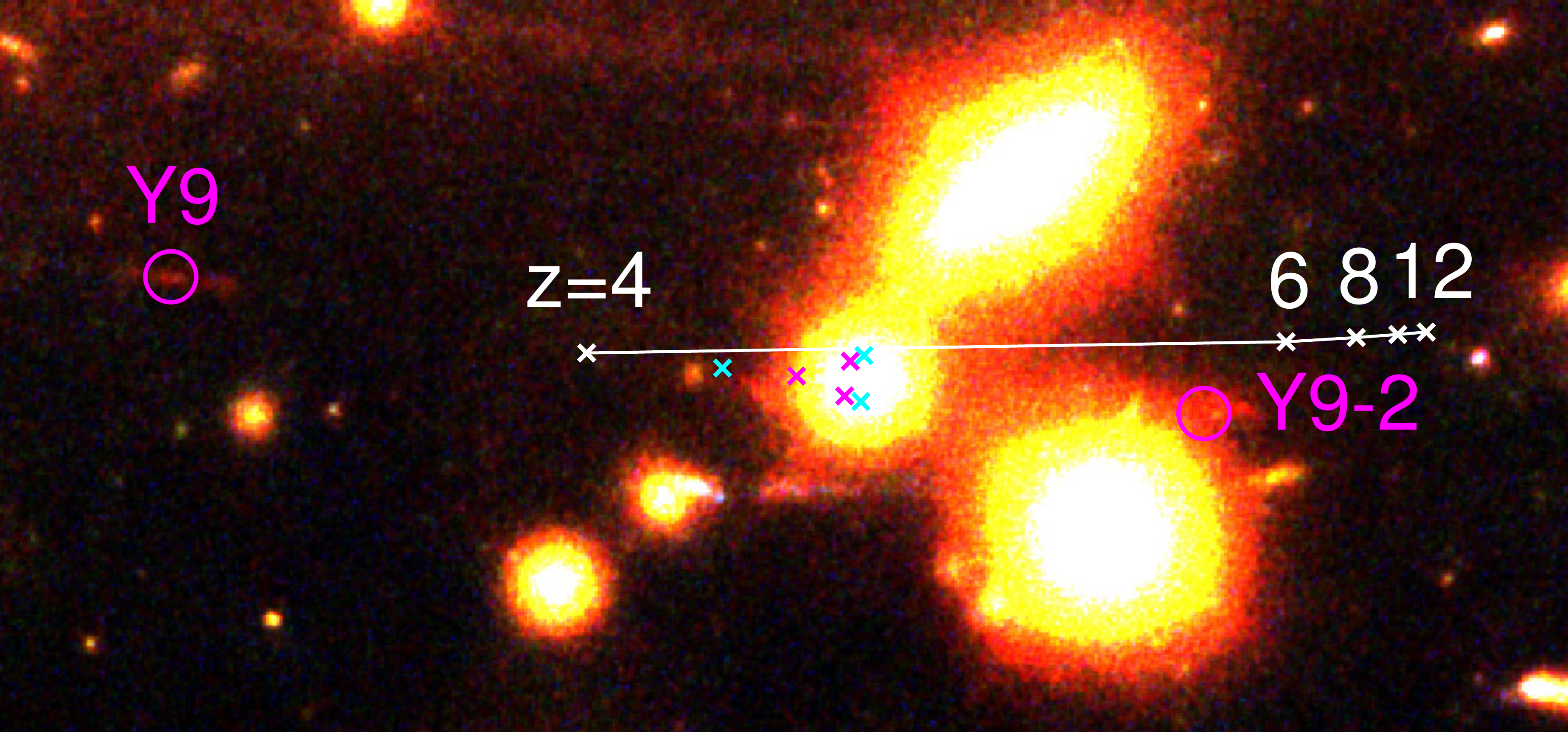}
	\plotone{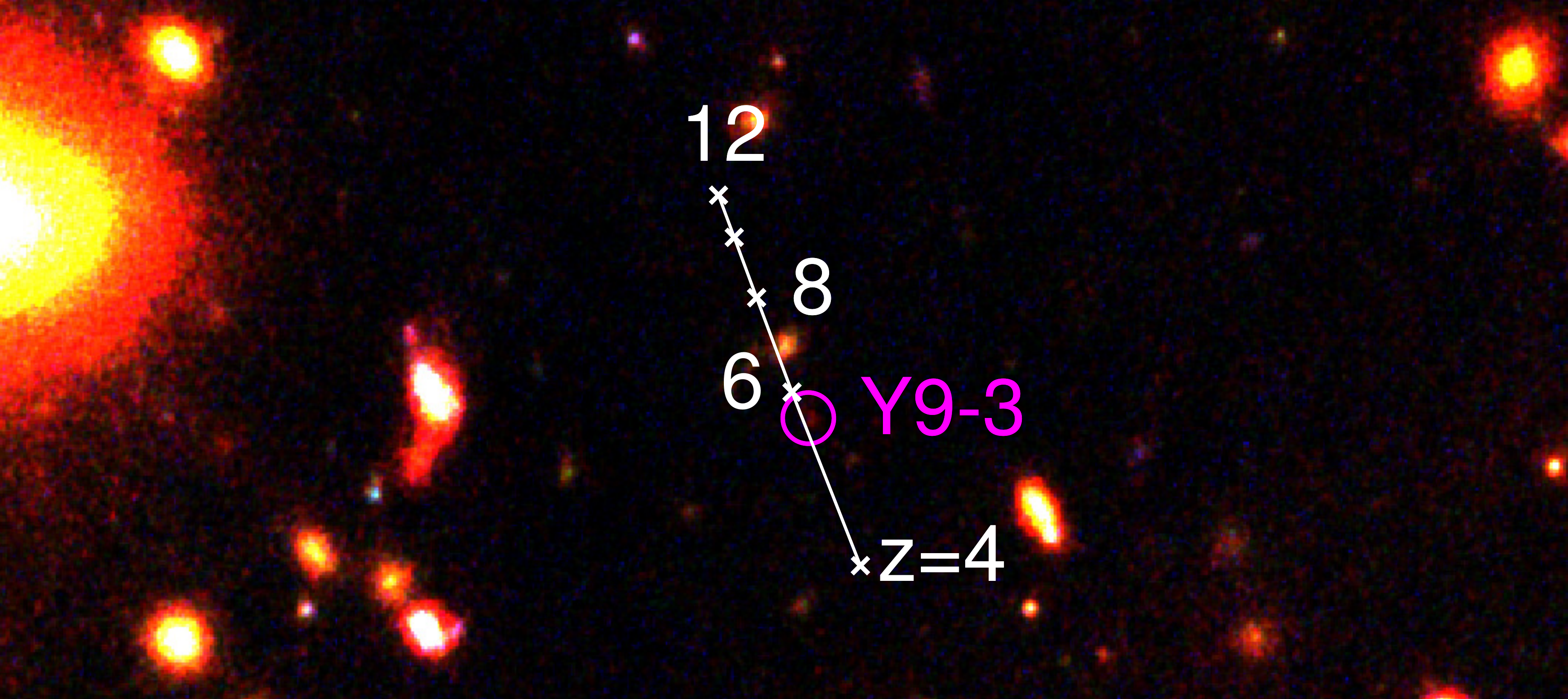}
	\caption{
			Same as Figure \ref{fig:YJ1}, but for HFF1C-Y9.
The magenta circles show the positions of HFF1C-Y9 and its multiple images of HFF1C-Y9-2 and HFF1C-Y9-3.
The cyan (magenta) crosses represent the positions of the other multiple images at $z=6$ ($z = 8$) near a bright foreground galaxy, respectively.
	}
	\label{fig:Y9}
\end{figure}

\paragraph{HFF1C-Y9 at $z\sim8$}
HFF1C-Y9 is a highly magnified dropout at $z \sim 8$.
\citet{2014arXiv1402.6743Z} find two counter images of this dropout, which are referred to as
HFF1C-Y9-2 and HFF1C-Y9-3 in this paper.
Figure \ref{fig:Y9} presents the predicted positions of the counter images of HFF1C-Y9.
Our mass model predicts that HFF1C-Y9 has more than two counter images, if HFF1C-Y9 resides at $z = 6-8$.
However, we cannot investigate the positions of these additional counter images 
due to the bright galaxies along the lines of sight.
We compare the observed images of HFF1C-Y9-2 and HFF1C-Y9-3 with the predicted positions.
The observed images fall in the predicted positions for the system at $z = 4-6$,
implying that HFF1C-Y9 would be a source at a redshift slightly lower than that estimated 
in the dropout selection.

\begin{figure}
	\plotone{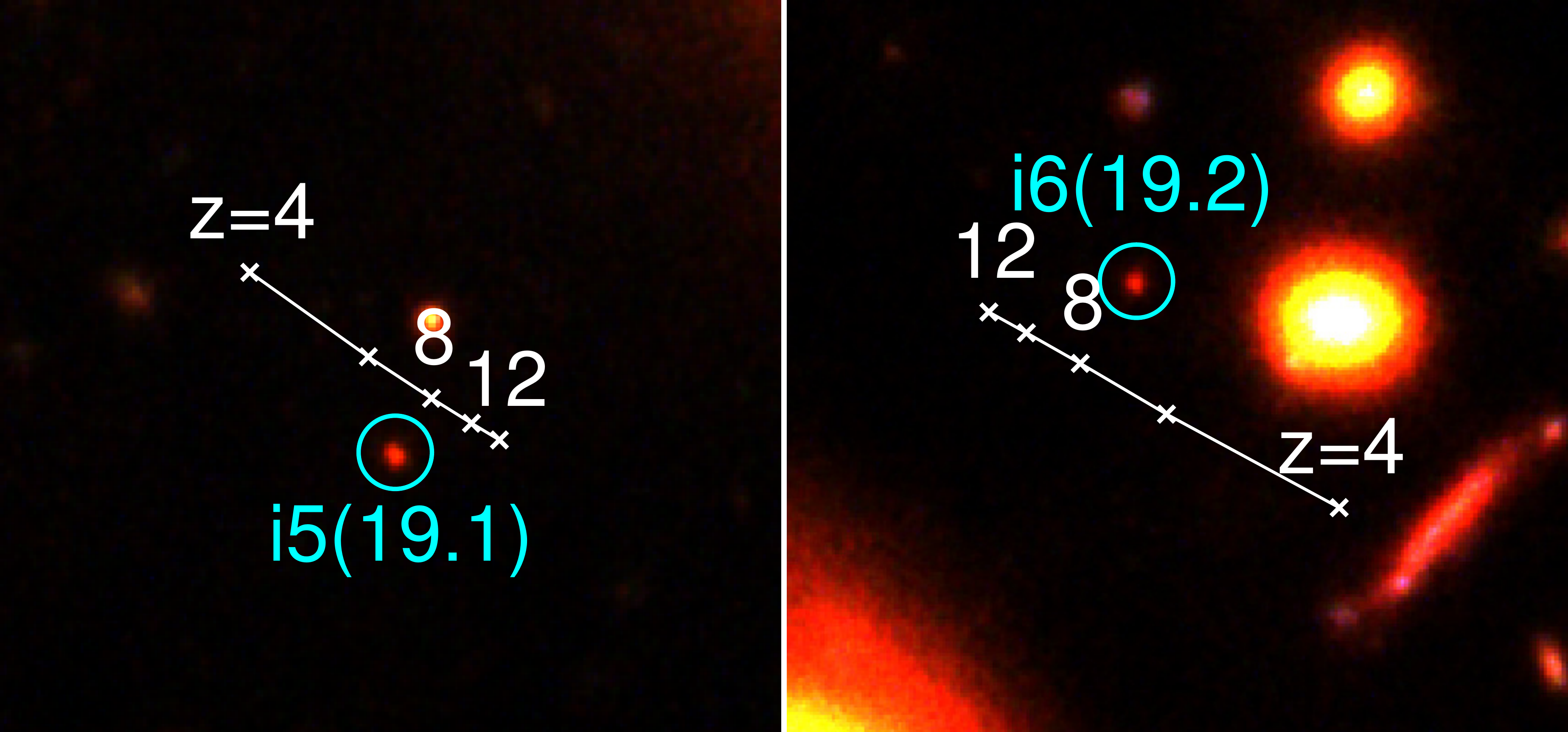}
	\plotone{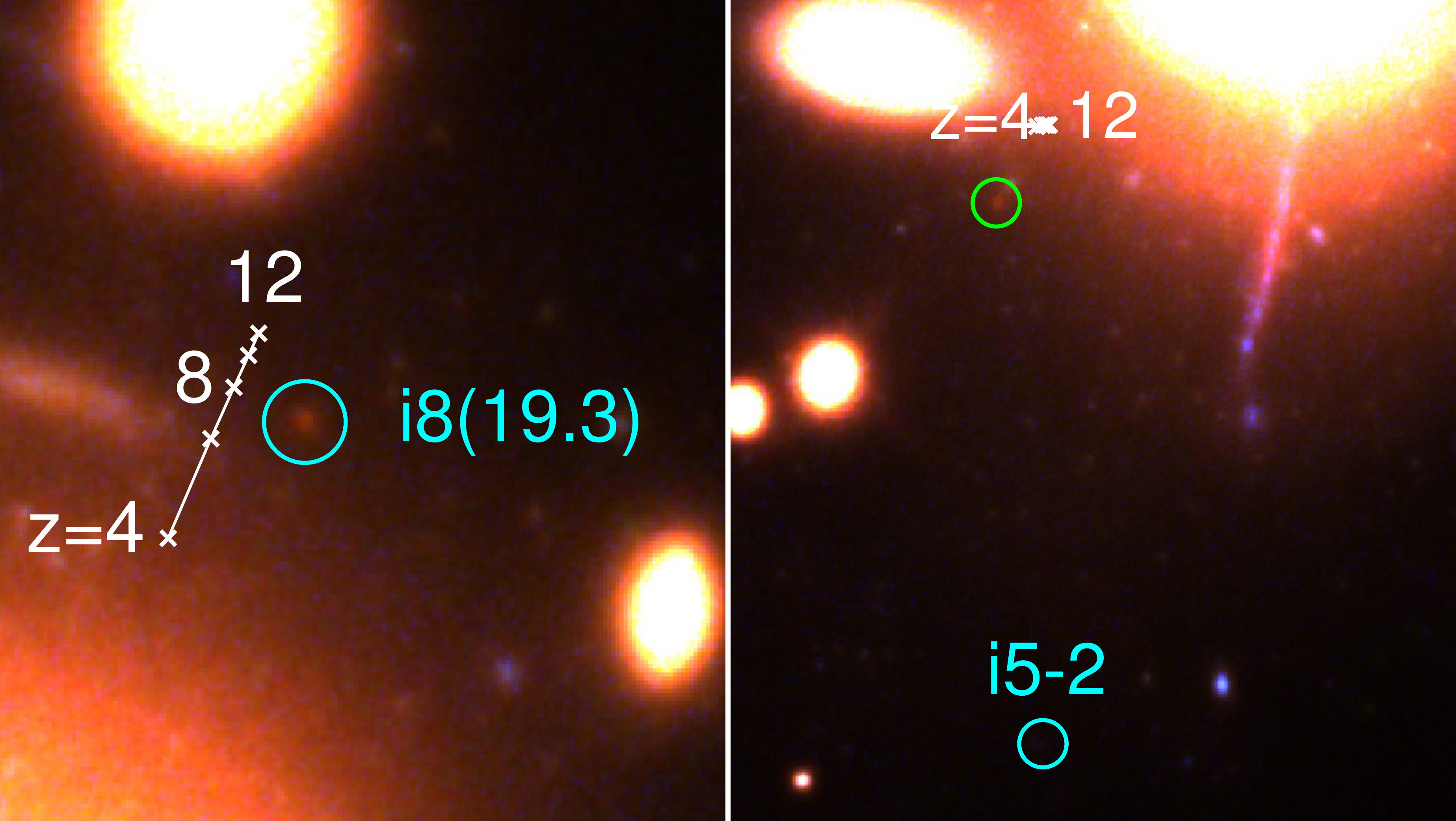}
	\caption{
			Same as Figure \ref{fig:YJ1}, but for HFF1C-i5, HFF1C-i6, and HFF1C-i8.
The cyan circles show the positions of HFF1C-i5, HFF1C-i6, HFF1C-i8, and their multiple images.
The green circle presents the position of HFF1C-i5-2 predicted by \citet{2014arXiv1409.8663J}.
	}
	\label{fig:i5}
\end{figure}

\paragraph{HFF1C-i5, -i6, and -i8 at $z\sim6-7$}
We find three multiple images of an $i$-dropout, which are HFF1C-i5, HFF1C-i6, and HFF1C-i8.
We use the positions of these three multiple images for the construction of our mass model.
Their IDs in Table \ref{multipleimages} are 19.1, 19.2, and 19.3, respectively.
In addition to these three multiple images, \citet{2014ApJ...786...60A} report another counter images,
which is named Image 5.4 in Table 3 of \citet{2014ApJ...786...60A}.
In our paper, we refer to Image 5.4 as HFF1C-i5-2. In Figure \ref{fig:i5}, we plot the predicted positions of these four multiple images.
The predicted position of HFF1C-i5-2 is about $8''$ away from the images reported by \citet{2014ApJ...786...60A}.
Instead, it is close to the position predicted by \citet{2014arXiv1409.8663J} shown with the green circle in Figure \ref{fig:i5}.
The observed images of the other three multiple images lie near the predicted positions at $z = 6$ and $8$.
Our mass model predicts that the best-fit value of their redshift is $z=7.94$, as shown in Table \ref{multipleimages}.

\subsection{Comparisons with the Public Mass Models}

\begin{figure*}
		\epsscale{1.2}
	\plotone{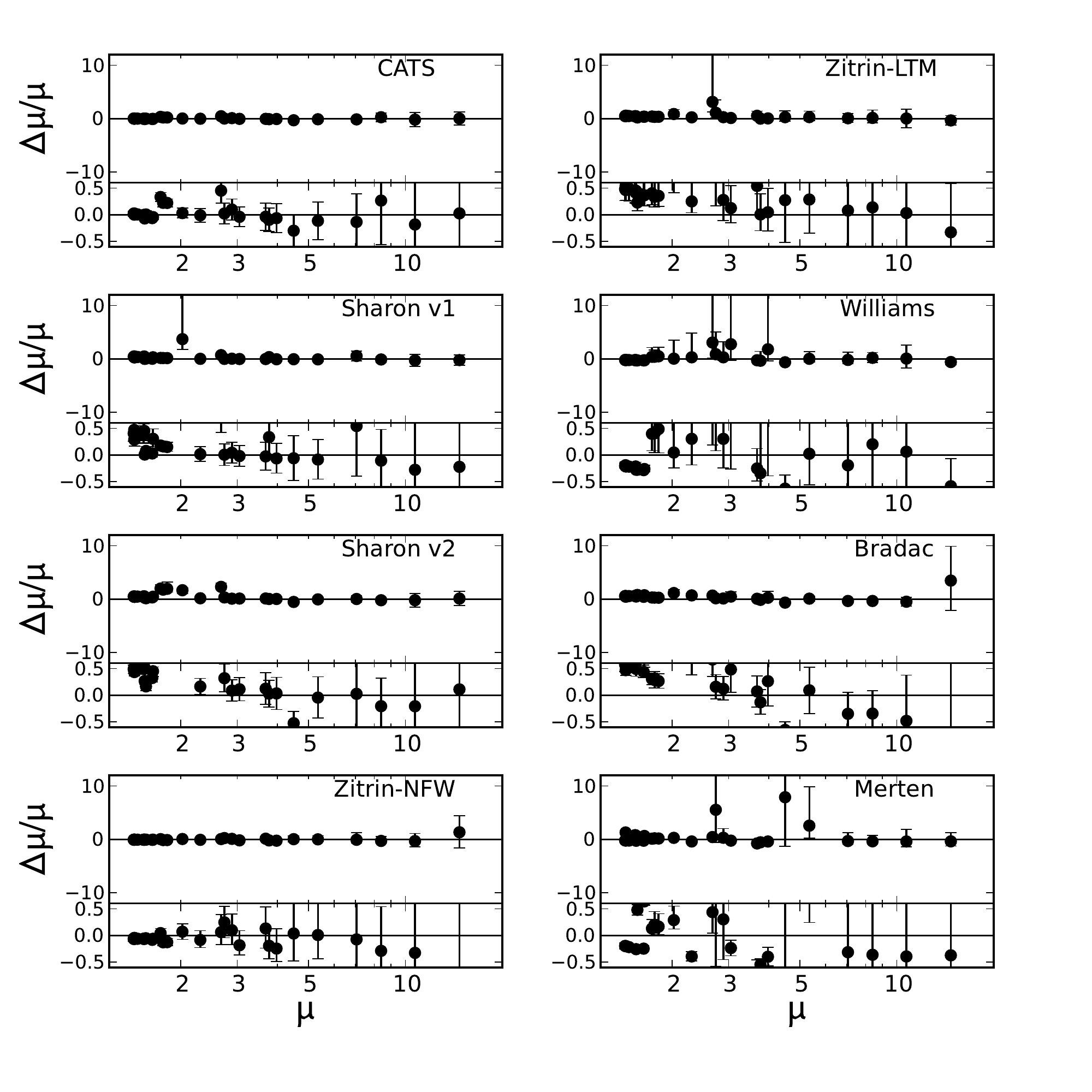}
	\caption{
	Comparison of the magnification factors of our dropouts for the different mass models in the cluster field.
		The horizontal axes show the magnification factor $\mu$ from our mass model,
			and the vertical axes present the difference $\Delta \mu$ between magnification factors of our model and a public model normalized by $\mu$.
			Eight panels present the public models of CATS, Sharon version 1, Sharon version 2, Zitrin-NFW, Zitrin-LTM, Williams, Brada\v{c}, and Merten.
			The upper and lower segments of each panel represent $\Delta \mu / \mu$ in the wide and narrow dynamic ranges, respectively. 
	}
	\label{mag_comp}
\end{figure*}

Mass models of Abell 2744 are also made by other groups (e.g., \citealt{2014arXiv1406.2702L}). 
Eight public mass models are accessible through the STScI website\footnote{http://archive.stsci.edu/prepds/frontier/lensmodels/}
which are made by the five independent groups,
M. Brada\v{c} (PI), The Clusters As TelescopeS (CATS) team (Co-PI's J.P Kneib, P. Natarajan; see \citealt{2014arXiv1405.3303R}), J. Merten \& A. Zitrin (Co-PI's), K. Sharon (PI; see \citealt{2014arXiv1405.0222J}), and L. Williams (PI).
Figure \ref{mag_comp} compares the magnification factors of our mass model and these public mass models at the positions of our dropout candidates in the cluster field.
The vertical axes show $\Delta \mu / \mu$, where $\mu$ is the magnification factor from our mass model 
and $\Delta \mu \equiv \mu_{\rm other} - \mu$ is the difference between magnification factors of our model and a public mass model ($\mu_{\rm other}$).
In the cluster field, the magnification factors from our mass model are broadly consistent with those from the public mass models. 
We especially find excellent agreements with the CATS and Zitrin-NFW models.
The Merten's group extends their mass model to the parallel field using weak lensing data covering both the cluster and parallel field.
The magnification factors in the parallel field from the Merten model are $\sim 1.08 - 1.22$.
Our mass model estimates the magnifications of the dropouts in the parallel field to be $\sim 1.05$, 
which is consistent with those from Merten model.

\section{UV Luminosity Functions}
\label{sec:uvlf}

In this section,  
we derive UV luminosity functions of 
dropout galaxies at $z \sim 6-7$, $8$, and $9$  
based on 
the $z \sim 5-10$ dropouts 
identified by our HFF study and the previous blank field surveys.
In Section \ref{sec:contamination estimates}, 
we estimate the contamination rates
of our dropout samples. 
In Section \ref{sec:completeness}, 
we conduct Monte Carlo simulations with the gravitational lensing effects,
and in Section \ref{sec:best-fit parameters} we obtain simulated number counts of dropouts in the image plane.
Incorporating the contamination estimates, 
we search for the best-fit Schechter parameters 
for the UV luminosity functions of dropout galaxies
at $z \sim 6-7$, $8$, and $9$ .

\subsection{Contamination Estimates}\label{sec:contamination estimates}

One of the major sources of contamination in high-redshift dropout galaxy samples is galaxies at $z \sim 2$  
whose Balmer break mimics a Lyman-$\alpha$ break in the spectra of high-$z$ star-forming galaxies.  
Although bright $z \sim 2$ interlopers are removed by detections of a blue continuum in deep optical images, 
faint interlopers are selected due to the photometric uncertainties.
Here we estimate the expected number of such contaminants 
that meet our dropout selection criteria, 
basically following the method 
described in Section 3.3 of 
\citet[][see also Section 3.1 of \citealt{2004ApJ...611..660O}]{2013ApJ...768..196S}.

To obtain the expected number of contaminants, 
we make use of our catalog 
of bright objects 
detected in the HFF fields. 
We assume that  
bright objects with $22 < H_{160} < 25$ that do not satisfy 
the dropout selection criteria are bright interlopers, 
and that faint interlopers have the color distribution same as that of bright ones. 
We create artificial objects with faint magnitudes of $H_{160} = 25.0-29.5$ 
using the \verb|mkobjects| package in {\sc iraf} \citep{1986SPIE..627..733T,1993ASPC...52..173T} software. 
Their number counts are matched to 
the observed number counts extrapolated from the bright magnitudes.
These artificial objects are placed in random positions of our HFF images.
We conduct the source extraction and the dropout selection in the same manner 
as our dropout galaxy identification in the real HFF data.
The artificial objects selected as dropouts are regarded as contaminants.
We derive the numbers of contamination objects as a function of magnitude,
which is used in our UV luminosity function estimates in Section \ref{sec:best-fit parameters}.
We find that the fraction of total numbers of the contaminants to our dropout candidates 
down to 29.5 magnitude is $\sim 27${\%}.
		This fraction is relatively larger than the contamination rate estimated by some previous studies, which are for examples $\sim7\%$ at $z\sim7-8$ in \citet{2011ApJ...737...90B} and $\sim23\%$ at $z\sim9-10$ in \citet{2012arXiv1211.2230B}.
In Section \ref{sec:Discussion}, we discuss the discrepancy between the UV luminosity densities and Thomson scattering optical depth.
		Because the differences of the number counts given by the contamination estimates are at most 30\% which is smaller than the discrepancy discussed in Section \ref{sec:Discussion}, our conclusion does not change.

\subsection{Completeness Estimates} \label{sec:completeness}

The gravitational lensing effects are important 
to interpret the observational results of our dropouts 
in the HFF fields. The brightness of dropouts are magnified, 
and multiple images appear for some of the dropouts.
Thus, the number counts of our HFF dropout candidates 
are changed from those of the blank field 
by the gravitational lensing effects, especially in the cluster field.
Moreover, to derive UV luminosity functions, one needs 
to correct for the selection incompleteness of dropouts 
that is a function of both magnitude and source redshift.
In our study, we carry out Monte-Carlo simulations in the {\it image plane}
to evaluate the gravitational lensing effects
as well as the selection completeness.  
The simulations contain all lensing effects: magnification,
distortion, and multiplication of images. 
This is called the {\it image plane} technique.

There is another method for the luminosity function estimates
referred to as the {\it source plane} technique \citep{2014ApJ...786...60A}.
The {\it source plane} technique determines an absolute magnitude of 
each dropout candidate with a magnification factor
to derive the number of dropout candidates 
per unit source plane volume. However, 
distortion and multiplication effects are not included
in the {\it source plane} technique.
Note that our method of the {\it image plane} technique is self-consistent and more complete
than the {\it source plane} technique.

We first estimate the completeness of dropouts identified by our selections
in the HFF images, where the completeness depends on redshift and magnitude.
We create a mock catalog of $\sim 1,000,000$ galaxies uniformly distributed at
$z=5.0-10.4$ in the magnitude range of $25.0 - 30.5$ mag.
To define the UV continuum colors of the galaxies,
we assume a spectral UV slope of $\beta = -2.0$, 
which is the same as that used in \citet{2013ApJ...768..196S}. 
IGM attenuation is given with the prescription given by \citet{1996MNRAS.283.1388M}. 
For the galaxies' intrinsic surface brightness profiles, 
we adopt 
a S{\'e}rsic index of $1.0$  
and 
half-light radii of 
$\simeq 0.6$ kpc and $\simeq 0.3$ kpc 
for bright ($M_{\rm UV} \lesssim -19.5$ mag)  
and faint ($M_{\rm UV} \gtrsim -19.5$ mag) dropout candidates, respectively,  
which are motivated by recent 
size measurements for $z \sim 7-8$ dropout candidates
\citep[][see also, \citealt{2010ApJ...709L..21O}]{2013ApJ...777..155O}.  
We assume a uniform distribution of the intrinsic ellipticity 
in the range of $0.0 - 0.9$, 
because the observed ellipticities of $z \sim 3-5$ dropouts
roughly have uniform distributions \citep{2006ApJ...652..963R}.

\begin{figure}
		\epsscale{1.2}
	\plotone{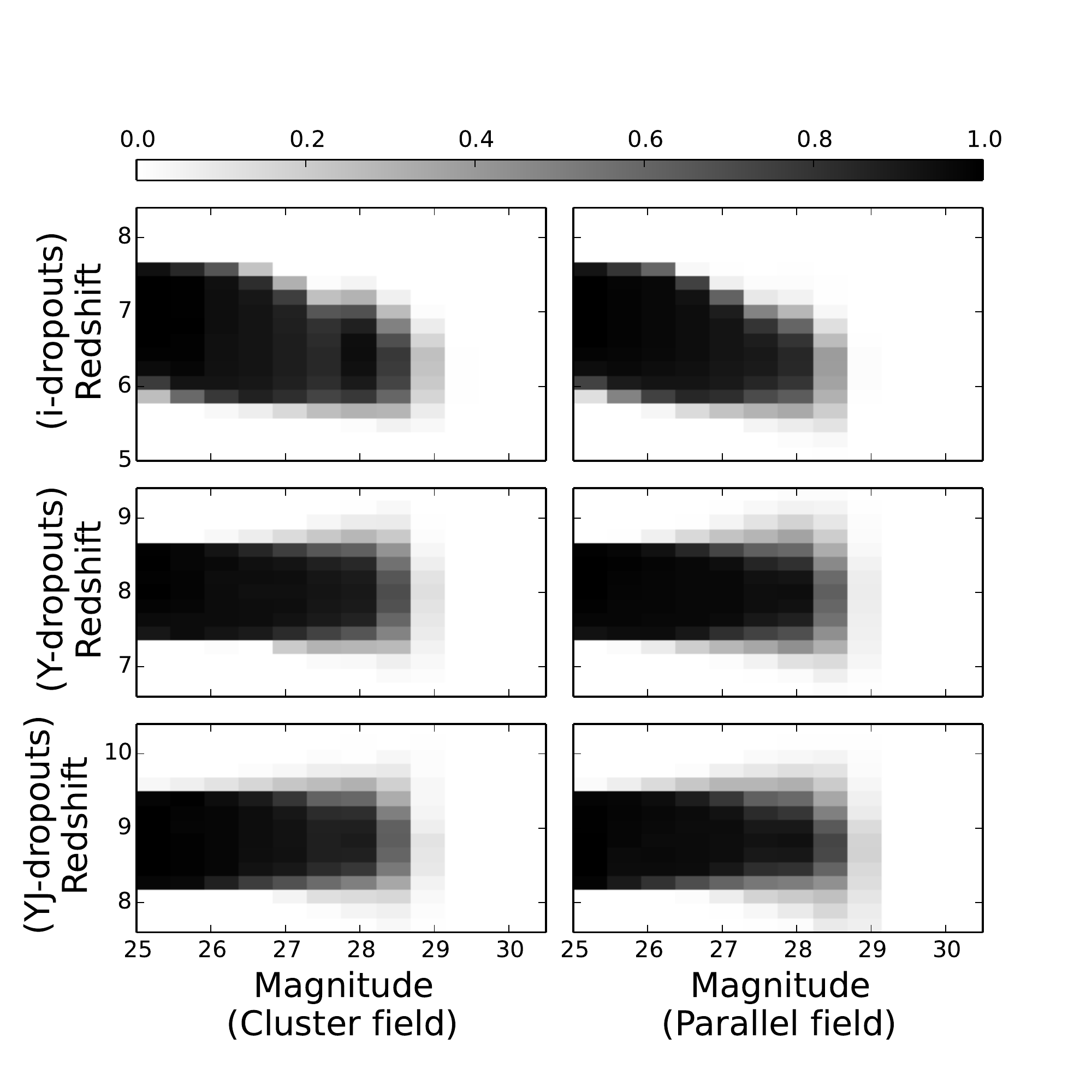}
	\caption{
			Completeness of our dropout selections that depend on redshift and apparent magnitude
			in the cluster (left) and parallel (right) fields.
			The top, middle, and bottom panels are the completeness of $i$-, $Y$, and $YJ$-dropout candidates, respectively.
	The completeness values are normalized to 1.0 at the maximum completeness values in the selection.
	The darker shades indicate the higher completeness.  
	}
	\label{selectfunc}
\end{figure}

Then, 
we produce simulated images of the galaxies that include the HST PSFs and 
the Abell 2744's gravitational lensing effects with \verb|writeimage| command of {\sc glafic}.
We randomly select about 3000 simulated images of galaxies in a magnitude bin of $\Delta m=0.5$,
and place these simulated galaxy images at random positions on the real HFF images
to make simulated HFF images.
In the same manner as the procedure for the identifications
of our real dropouts (Section \ref{sec:photometric catalog}), 
we perform source extractions for the simulated HFF images
with SExtractor, and construct photometric catalogs.
Applying the color selection criteria used in Section \ref{sec:dropout selection}
and 
the magnitude-dependent contamination rates estimated in Section \ref{sec:contamination estimates}, 
we obtain simulated dropout galaxies. 
We make a simulated dropout galaxy sample for the magnitude bin.
We conduct the same simulations over the given magnitude range of $25.0 - 30.5$ mag,
and derive the completeness for our $i$-, $Y$-, and $YJ$-dropout selections
that depend on redshift and magnitude.
Figure \ref{selectfunc} shows the completeness 
derived from our simulations. 
The left (right) panels indicate the selection windows in the cluster (parallel) field.
At bright magnitudes of $\sim 25 - 27$ mag, our $i$-, $Y$-, and $YJ$-dropout selection criteria 
provide a high completeness sample of star-forming galaxies at 
$6.0 < z < 7.4$, $7.4 < z < 8.6$, and $8.2 < z < 9.4$, respectively.

\subsection{Best-fit Schechter Parameters of UV Luminosity Functions} \label{sec:best-fit parameters}

These completeness estimates are then used to predict observed galaxy number counts.
We calculate the predicted number counts from the completeness estimates 
and a UV luminosity function expressed as a Schechter function with a set of parameters, ($M_*$, $\phi_*$, $\alpha$),
and repeat it for various sets of Schechter parameters covering a wide parameter space.
In this way, we obtain the predicted number counts for various sets of Schechter parameters.
Because the completeness values are estimated with all of the observational effects in the {\it image plane},
these predicted number counts include the lensing magnifications,  magnification, distortion, 
and multiplication as well as the corresponding detection incompleteness in the redshift and magnitude
space.

\begin{figure}
	\plotone{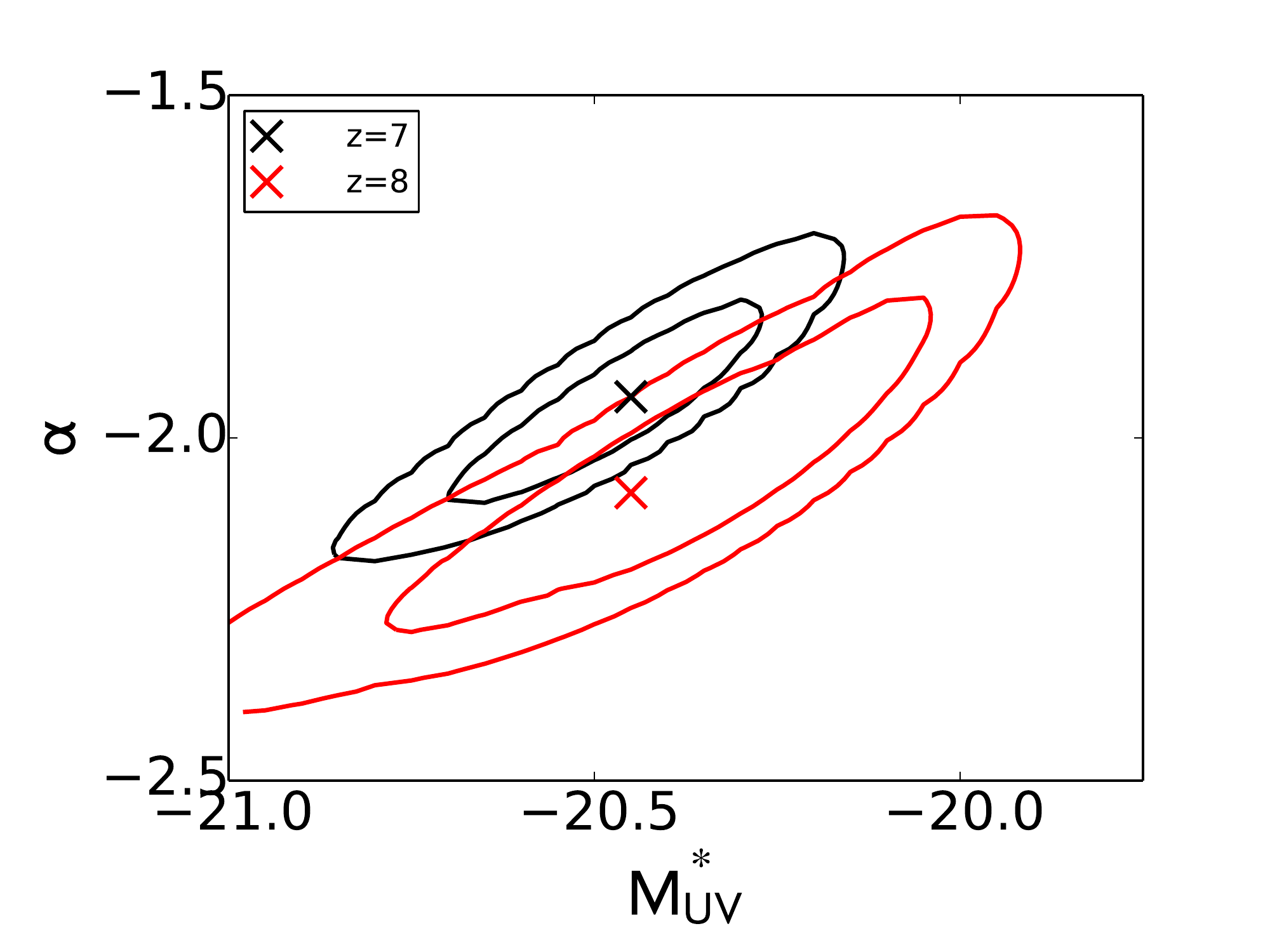}
	\caption{The 68\% and 95\% confidence level contours of Schechter parameters, $M_*$ and $\alpha$, 
			for the UV luminosity functions at $z \sim 6-7$ (black) and $z\sim8$ (red), respectively.
			The contours indicate our best-estimate results with
			our HFF data and all of the previous measurements. 
			The crosses denote the best-fit parameter values.
	}
	\label{fit7_contour}
\end{figure}

Using the predicted number counts, we search for the best-fit Schechter parameters
that reproduce the observed number counts of our dropout candidates.
We adopt a maximum likelihood method, 
assuming a Poisson distribution of the number counts.
The likelihood is written as
\begin{equation}
{\cal L} 
	\propto \prod_{\rm field} \prod_i 
		n_{{\rm exp},i}^{n_{{\rm obs},i}} 
		e^{-n_{{\rm exp},i}},  
\end{equation}
where 
$n_{{\rm exp},i}$ is the expected number counts 
from a given Schechter function in a magnitude interval $i$
and 
$n_{{\rm obs},i}$ is the observed number counts in the magnitude interval.
Constraining the Schechter parameters, 
we simultaneously fit 
both the observed HFF number counts 
and the UV luminosity function data points obtained in the previous studies.
For our dropouts at $z \sim 6-7$,
we compare our number counts of $z \sim 6-7$ with the luminosity function data points of $z \sim 7$ in previous studies,
assuming that the UV luminosity function does not rapidly change in $z \sim 6-7$.
We take the previous blank-field survey results 
from the studies of CANDELS, HUDF09, HUDF12, ERS, and BORG/HIPPIES \citep{2014arXiv1403.4295B}, UltraVISTA+UKIDSS UDS \citep{2014MNRAS.440.2810B}, 
BoRG \citep{2012ApJ...760..108B}, SDF+GOODS-N \citep{2009ApJ...706.1136O}, and HUDF12/XDF+CANDELS \citep{2013ApJ...773...75O}.
We regard $M_\ast$, $\phi_\ast$, and $\alpha$ as free parameters for the fitting of number counts at $z\sim6-7$ and $z\sim8$.
Because the statistics of the $z\sim9$ luminosity function is poor, we 
choose $\phi_\ast$ for a free parameter and $M_\ast$ and $\alpha$
to be fixed to the best-fit values of $z\sim8$.
Maximizing the Poisson likelihood, 
we obtain the best-fit parameters of 
$(M_\ast, \log \phi_\ast {\rm [Mpc^{-3}]}, \alpha)
=(-20.45^{+0.1}_{-0.2},  -3.30^{+0.10}_{-0.20},  -1.94^{+0.09}_{-0.10})$
for the $z \sim 6-7$ dropout candidates, 
$(-20.45^{+0.3}_{-0.2}, -3.65^{+0.15}_{-0.25}, -2.08^{+0.21}_{-0.12})$
for the $z \sim 8$ dropout candidates,  
and 
$(-20.45 {\rm [fixed]}, -4.00\pm0.15, -2.08 {\rm [fixed]})$ 
for the $z \sim 9$ dropout candidates.
Table \ref{lf_parameters} summarizes these parameters, together with those obtained by the previous studies.
We find that 
our results are consistent with the previous results within 
the $1 \sigma$ uncertainties. 
Figures \ref{fit7_contour} and show 
the $1 \sigma$ confidence intervals on the $\alpha $ versus $M_\ast$ plane 
for the UV luminosity functions at $z \sim 6-7$ and $z \sim 8$, respectively. 
To test our results, we also perform Schechter function fittings 
without the results from the HUDF09+ERS data at $z \sim 6-7$ and $z \sim 8$. 
We confirm that the fitting results without the HUDF09+ERS data 
are consistent with the previous results, 
although the uncertainties are substantially large due to the small statistics of the HFF samples.

\setlength{\tabcolsep}{2pt}
\begin{deluxetable}{lccc}
\tabletypesize{\scriptsize}
\tablecaption{Best-fit Schechter parameters of luminosity functions\label{lf_parameters}}
\tablewidth{0pt}
\tablehead{
		\colhead{Reference} & \colhead{$M_*$}  & \colhead{$\log \phi_*$ [Mpc$^{-3}$]} & \colhead{$\alpha$}
}
\startdata
\sidehead{$z \sim 6-7$}
		This Work & $-20.45^{+0.1}_{-0.2}$ & $-3.30^{+0.10}_{-0.20}$ & $-1.94^{+0.09}_{-0.10}$ \\
 \citet{2014arXiv1409.0512A}	& $-20.63^{+0.69}_{-0.56} $ & $-3.34\pm0.36$ & $-1.88^{+0.17}_{-0.20}$ \\
		\citet{2014arXiv1403.4295B} & $-21.04\pm0.26$  & $-3.65^{+0.27}_{-0.17}$ & $-2.06\pm0.12$ \\
		\citet{2011ApJ...737...90B} & $-20.14\pm 0.26$ & $-3.07\pm0.26$ & $-2.01\pm 0.21$ \\
		\citet{2009ApJ...706.1136O} & $-20.10\pm0.76$ & $-3.16\pm0.68$ & $-1.72\pm0.65$ \\
		\citet{2013ApJ...768..196S} & $-20.14^{+0.36}_{-0.48}$ & $-3.19^{+0.27}_{-0.24}$ & $-1.87^{+0.18}_{-0.17}$ \\
\sidehead{$z \sim 8$}
		This Work & $-20.45^{+0.3}_{-0.2}$ & $-3.65^{+0.15}_{-0.25}$ & $-2.08^{+0.21}_{-0.12}$ \\
		\citet{2014arXiv1403.4295B} & $-19.97\pm 0.34$ & $-3.19\pm0.30$ & $-1.86\pm 0.27$ \\
		\citet{2011ApJ...737...90B} & $-20.10\pm 0.52$ & $-3.22\pm0.43$ & $-1.91\pm 0.32$ \\
		\citet{2012ApJ...760..108B} & $-20.26^{+0.29}_{-0.34}$ & $-3.37^{+0.26}_{-0.29}$ & $-1.98^{+0.23}_{-0.22}$ \\
		\citet{2013ApJ...768..196S} & $-20.44^{+0.47}_{-0.35}$ & $-3.50^{+0.35}_{-0.32}$ & $-1.94^{+0.21}_{-0.24}$ \\
\sidehead{$z \sim 9$}
This Work & $-20.45$ (fixed) & $-4.00\pm0.15$ & $-2.08$ (fixed)\\
\citet{2013ApJ...773...75O} & $-18.8\pm0.3$ & $-2.94$ (fixed) & $-1.73$ (fixed) \\
\citet{2012arXiv1211.2230B} & $-20.04$ (fixed)  & $-3.95^{+0.39}_{-0.56}$ & $-2.06$ (fixed)  
\enddata
\end{deluxetable}

\begin{figure}
		\epsscale{1.3}
	\plotone{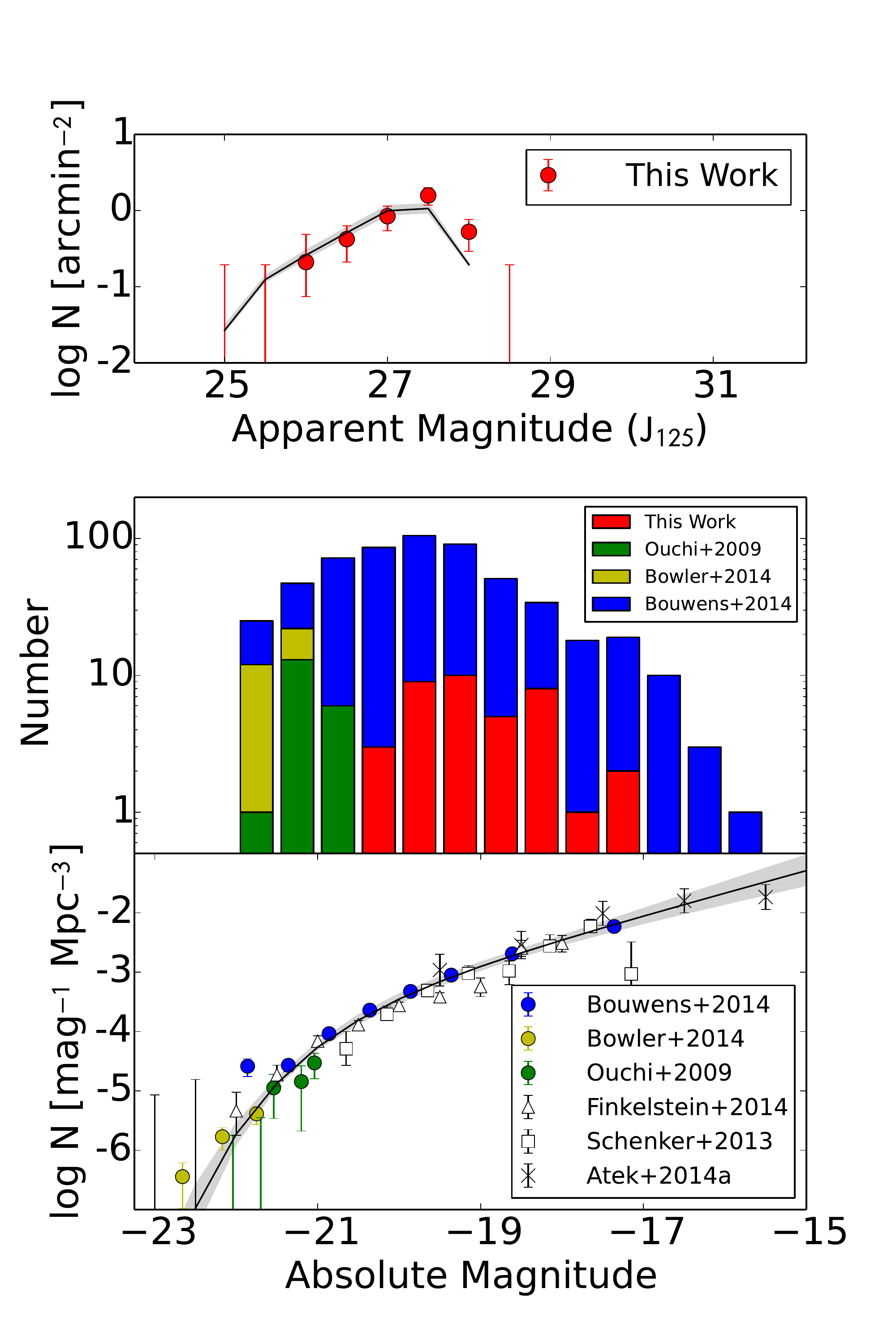}
	\caption{Number counts, histograms, and luminosity functions of $z \sim 6-7$ dropouts.
			Top panel: Our observed number counts in the cluster and parallel fields (red circles) 
			and the simulated number counts of the best-fit Schechter parameters (black line) 
			with the $1\sigma$ uncertainties (gray region).
			The horizontal axis presents observed apparent magnitude in the $J_{125}$ band.
			Middle panel: The histograms of the numbers of dropouts found in our HFF study (red) and the previous work, 
					\citet{2014arXiv1403.4295B} (blue), \citet{2009ApJ...706.1136O} (green), and \citet{2014MNRAS.440.2810B} (yellow).
			Bottom panel: Our best-fit luminosity function (black line) and the $1\sigma$ error (gray region).
			The blue, green, and yellow circles, white triangles, white squares, and black crosses denote luminosity functions derived by \citet{2014arXiv1403.4295B}, \citet{2009ApJ...706.1136O}, \citet{2014MNRAS.440.2810B}, \citet{2014arXiv1410.5439F}, \citet{2013ApJ...768..196S}, and \citet{2014arXiv1409.0512A}, respectively.
			The horizontal axis shows intrinsic absolute magnitude in the $J_{125}$ band.
	}
	\label{fit7_lf}
\end{figure}

\begin{figure}
		\epsscale{1.3}
	\plotone{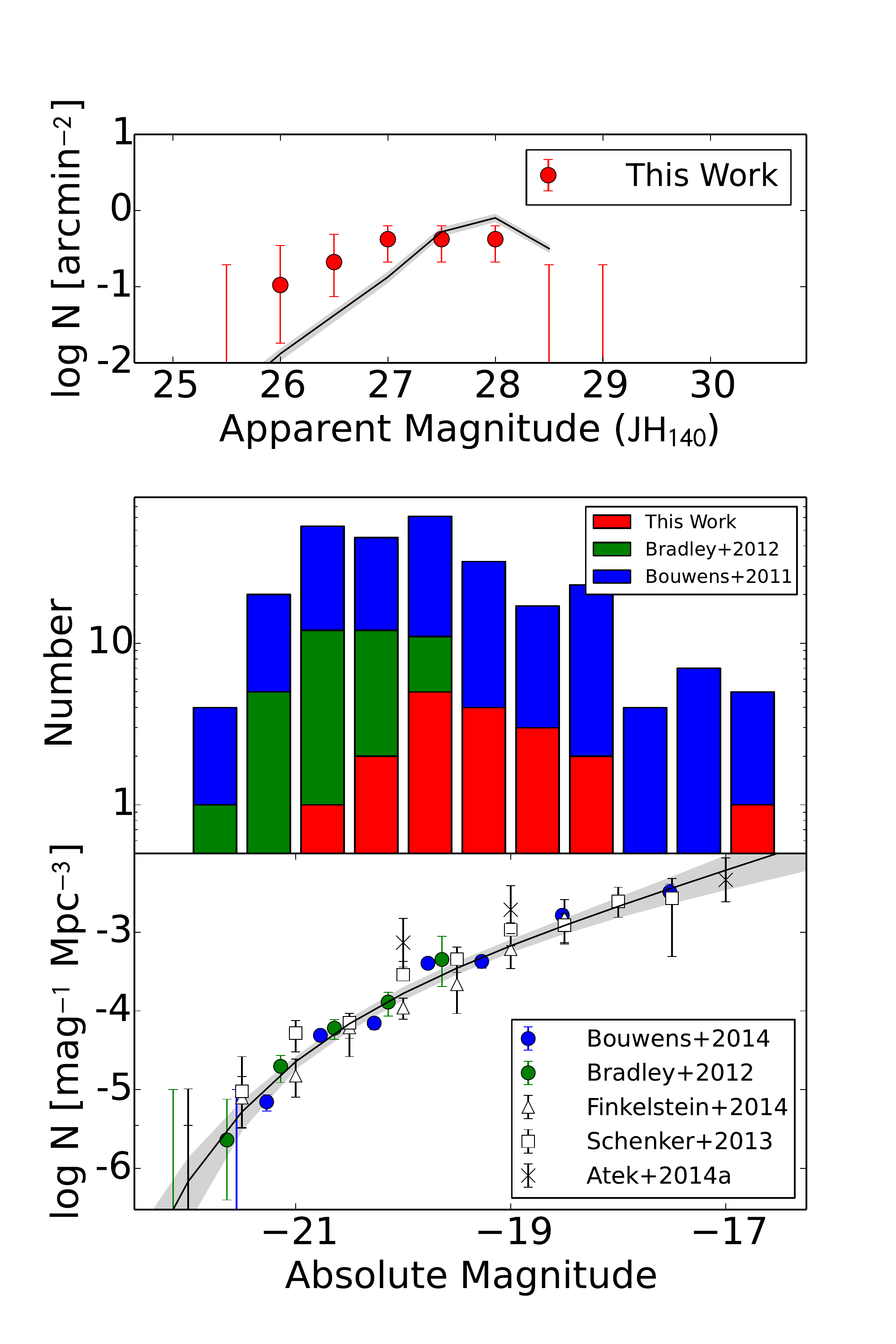}
	\caption{Same as Figure \ref{fit7_lf}, but for $z \sim 8$.
			The horizontal axes in the top and bottom panels present apparent magnitude and intrinsic absolute magnitude in the $JH_{140}$ band, respectively.
	We show the results of \citet{2012ApJ...760..108B} with the green histogram and circles in the middle and bottom panels, respectively.
	}
	\label{fit8_lf}
\end{figure}

\begin{figure}
		\epsscale{1.3}
	\plotone{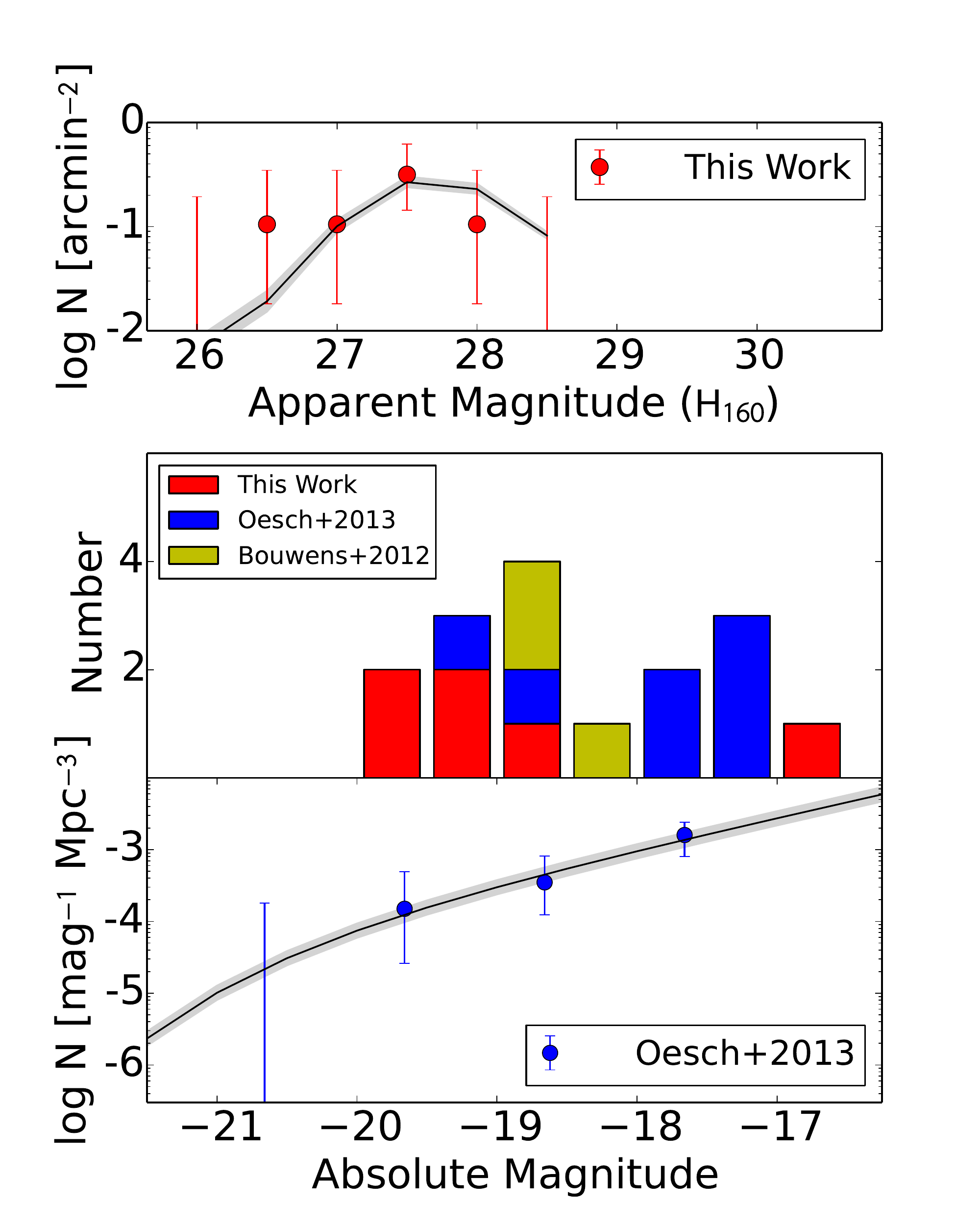}
	\caption{Same as Figure \ref{fit7_lf}, but for $z \sim 9$.
			The horizontal axes in the top and bottom panels present apparent magnitude and intrinsic absolute magnitude in the $H_{160}$ band, respectively.
	The yellow histogram shows the number of dropouts found in \citet{2012arXiv1211.2230B}.
	The blue histogram and circles are the 
	numbers and luminosity functions, respectively, obtained by
	\citet{2013ApJ...773...75O}.
	}
	\label{fit9_lf}
\end{figure}

The top and bottom panels of Figures \ref{fit7_lf}-\ref{fit9_lf} present 
the best-fitting number counts and Schechter functions, respectively.
In the bottom panels, we present recent studies of $z \sim7$ and $8$ luminosity functions, \citet{2013ApJ...768..196S}, \citet{2014arXiv1410.5439F},
and \citet{2014arXiv1409.0512A} for comparison.
The best-fitting results broadly agree with the observed number counts.  
The observed number counts of $z\sim 8$ at the bright end are larger than the best-fit function.
It is probably caused by the field-to-field variance, 
since our effective survey area is only $\simeq 6$ arcmin$^2$ in the source plane.
In fact, 
eight of the $z \sim 8$ dropouts are found 
within a small region with a radius of $6''$ 
(corresponding to a physical length of $\sim 30$ kpc at $z=8$).
This overdensity of $z \sim 8$ dropouts is originally 
claimed by \citet{2014arXiv1402.6743Z} with the
early optical images shallower than our full-depth data by $\sim 1$ magnitudes.
Because one cannot remove a number of foreground interlopers
with the shallow early optical data,
the existence of overdensity is open question (see the discussion
in \citealt{2014arXiv1405.0011C}).
In our study with the full-depth HFF data deep enough to remove such foreground interlopers reliably,
there is the overdensity of $z \sim 8$ dropouts, indicative that the overdensity is real. 
If it is true, the existence of the overdensity
would significantly enhance the source number counts of dropouts at $z\sim 8$ 
in the HFF fields.

The middle panels of Figures \ref{fit7_lf}-\ref{fit9_lf} 
show the histograms of the numbers of dropout galaxies 
used for the UV luminosity function determinations.
We also show the numbers of dropout galaxies newly identified in the observations of the HDF12 at $z\sim7-8$ \citep{2013ApJ...768..196S} and those of CLASH at $z\sim9.2$ \citep{2012arXiv1211.2230B}.
These histograms indicate that 
our HFF samples enable us to probe 
the UV luminosity functions down to a faint UV magnitude of $\simeq -17$, 
which is comparable to survey limits of the deepest blank-field observations of
the HUDF \citep{2013ApJ...763L...7E,2013ApJ...773...75O},
thanks to the gravitational lensing effects.  
In addition to the lensing effects, the HFF Abell 2744 observations 
provide two new ultra-deep imaging regions of the cluster and parallel fields, 
allowing us to significantly increase the number of $z\sim 9$ dropouts
(cf. the $z\sim 9$ dropout samples in UDF12 and CANDELS \citealt{2013ApJ...773...75O}).

\section{Discussion}
\label{sec:Discussion}

In Section \ref{sec:uvlf}, we have derived the UV luminosity functions of  dropout galaxies at 
$z \sim 6-7$, $8$, and $9$ based on our HFF and the previous study data.
We improve the faint-end luminosity function determinations
with the large samples extending the magnitude range.
The UV luminosity functions are tightly connected with the production rates of 
ionizing photons escaping to the IGM, which are important observational quantities
to understand the process of cosmic reionization, 
In this section, we carry out the joint analysis of 
the UV luminosity functions and 
the electron scattering optical depth $\tau_e$ measured by the CMB observations
to discuss the ionizing sources of the IGM.

\subsection{Evolution of the UV Luminosity Density}
\label{subsec:evolution_of_the_uv_ld}

To investigate the ionizing sources for the cosmic reionization, 
we first estimate UV luminosity densities $\rho_{\rm UV}$ 
from our UV luminosity functions.
$\rho_{\rm UV}$ is calculated by
\begin{eqnarray}
		\rho_{{\rm UV}}(z) = \int^{M_{\rm trunc}}_{-\infty} \Phi(M_{\rm UV}) L(M_{\rm UV}) dM_{\rm UV}, \label{eq:rho}
\end{eqnarray}
where $M_{\rm trunc}$ is the truncation magnitude of the UV luminosity function
where no galaxies exist beyond this magnitude. Because the $M_{\rm trunc}$ parameter is not
constrained by observations, in this study we assume two $M_{\rm trunc}$ values 
that bracket the plausible range of the parameter;
$M_{\rm trunc} = -17$ mag corresponding to the current observational limit and
$M_{\rm trunc} = -10$ mag being the predicted magnitude of minimum-mass halos 
that can host star-forming galaxies \citep{2011MNRAS.417.2982F}.

\begin{figure*}
		\epsscale{1.1}
	\plotone{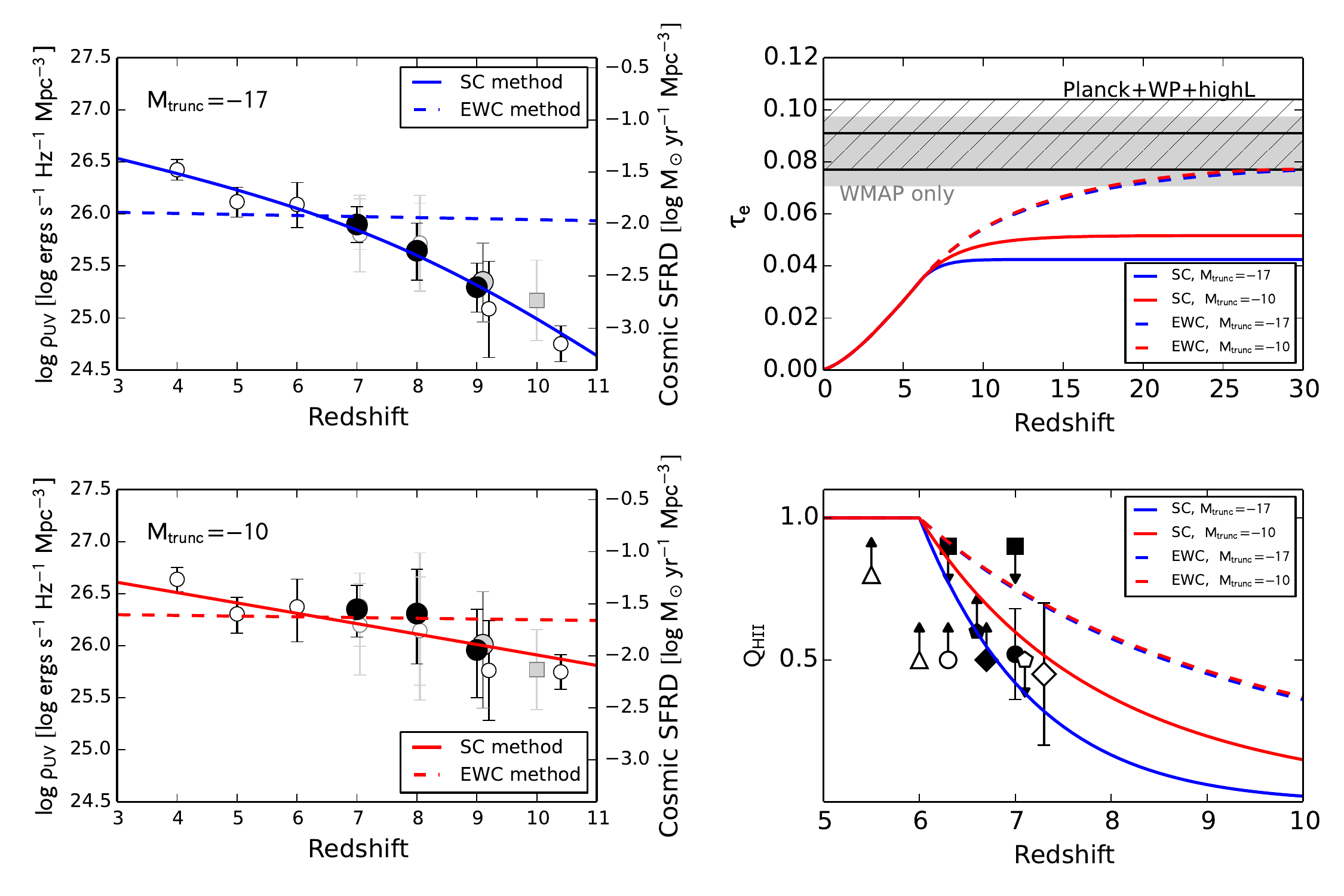}
	\caption{Upper left panel: The UV luminosity densities calculated with $M_{\rm trunc} = -17$.
	The filled and open circles represent the UV luminosity densities from this work and other studies \citep{2007ApJ...670..928B,2013ApJ...768..196S,2013MNRAS.432.2696M,2012arXiv1211.2230B,2014arXiv1403.4295B}, respectively.
	The gray circles and squares denote the UV luminosity densities from the HFF data only in this work and in \citet{2014arXiv1409.1228O}, respectively.
	The solid and dashed lines present our best-fit functions of $\rho_{\rm UV}$ with the SC and the EWC method, respectively.
	The right axes show cosmic SFR densities at a given UV luminosity density estimated with the Equation (2) of \citet{1998ApJ...498..106M}.
	Bottom left panel: Same as the top left panel, but for $M_{\rm trunc} = -10$.
			Upper right panel: Electron scattering optical depth integrating from $z \sim 0$ to a redshift, $z$, for our best-fit parameters by the SC method with $M_{\rm trunc} = -17$ (blue solid line), the EWC method with $M_{\rm trunc} = -10$ (red solid line), the EWC method with $M_{\rm trunc} = -17$ (blue dashed line), and the EWC method with $M_{\rm trunc} = -10$ (red dashed line), respectively.
			The hatched and gray regions indicate the $1 \sigma$ range of $\tau_e$ obtained by {\it WMAP+Planck+highL} \citep{2013arXiv1303.5076P} and nine-year {\it WMAP} \citep{2013ApJS..208...19H,2013ApJS..208...20B}, respectively.
	Bottom right panel: The evolution of ionized hydrogen fraction $Q_{\rm H_{II}}$ of IGM for our best-fit parameters
	with four lines, whose notations are the same as the right upper panel.
	Each symbol represents the observational limit of \citet{2011MNRAS.416L..70B}, \citet{2010ApJ...714..834C} (filled squares), \citet{2006PASJ...58..485T}, \citet{2008MNRAS.388.1101M} (open circle), \citet{2007MNRAS.381...75M}, \citet{2010ApJ...723..869O} (filled diamond), \citet{2010ApJ...723..869O}, \citet{2008ApJ...677...12O} (open pentagon), \citet{2010MNRAS.407.1328M}, \citet{2011MNRAS.415.3237M} (open triangles), \citet{2007MNRAS.381...75M}, \citet{2008MNRAS.386.1990M}, \citet{2011MNRAS.414.2139D} (filled pentagon), and \citet{2014arXiv1404.6066K} (open diamond) (see also \citealt{2013ApJ...768...71R}).
	}
	\label{fig:plot_tile}
\end{figure*}

We use the best-fit Schechter functions shown in Section \ref{sec:uvlf} and those in the literature 
for $z \sim 4-6$ \citep{2007ApJ...670..928B}, 
$z \sim 9.2$ \citep{2012arXiv1211.2230B}, and
$z \sim 10.4$ \citep{2014arXiv1403.4295B}.
For comparison purpose, we plot the data for $z \sim 7-8$ taken from \citet{2013ApJ...768..196S} and \citet{2013MNRAS.432.2696M}. 
Because the data of \citet{2013ApJ...768..196S} and \citet{2013MNRAS.432.2696M} are included in our luminosity function estimates 
via \citet{2014arXiv1403.4295B} data points,
we do not use these data points of \citet{2013ApJ...768..196S} and \citet{2013MNRAS.432.2696M} for fitting analyses carried out in Section \ref{subsec:rho_tau}.
The top and bottom left panels of Figure \ref{fig:plot_tile} present the $\rho_{{\rm UV}}$ as a function of redshift 
under the assumptions of $M_{\rm trunc} = -17$ and $-10$, respectively.
The solid and dashed lines show the best-fit functions of $\rho_{{\rm UV}}$ with two fitting methods detailed in Section \ref{subsec:rho_tau}.
We confirm that our $\rho_{\rm UV}$ at $z \sim 6-9$ are broadly consistent with the previous results,
and that there is a rapid decrease of $\rho_{{\rm UV}}$ 
from $z \sim 8$ towards high redshifts, which is claimed 
by \citet{2013ApJ...773...75O} and \citet{2014arXiv1403.4295B}.
With the improved measurements of $\rho_{\rm UV}$ in our study, 
this trend of the rapid decrease is strengthened.
To test whether the rapid decrease is confirmed with the HFF data alone,
we derive the luminosity function at $z\sim9$ with the HFF data alone,
and estimate $\rho_{\rm UV}$ at $z\sim9$.
The gray circles in the left panels of Figure \ref{fig:plot_tile} indicate $\rho_{\rm UV}$ obtained with our HFF data alone.
Although the uncertainty is large, the HFF data independently support the rapid decrease of $\rho_{\rm UV}$ from $z\sim8$.
Similar analysis is found in \citet{2014arXiv1409.1228O}. They derive the luminosity function at $z\sim10$ from the HFF cluster data alone.
we plot $\rho_{\rm UV}$ at $z\sim10$ calculated from the luminosity function derived by \citet{2014arXiv1409.1228O} with the gray squares.
These plots are also consistent with the rapid decrease from $z\sim8$.

\citet{2014arXiv1410.0962R} estimate the cosmic variance uncertainties of the high-redshift galaxies in the Abell 2744 cluster field.
The uncertainties are $\sim35\%$ at $z\sim7$ and $\gtrsim65\%$ at $z\sim10$.
Our errors of $\rho_{\rm UV}$ slightly increase by the cosmic variance uncertainties. 
However, our conclusion does not change because 
our $\rho_{\rm UV}$ at $z\sim9$ is smaller than $\rho_{\rm UV}$ at $z\sim8$ by a factor of two,
which is significantly larger than the uncertainties of the cosmic variance.

\subsection{Properties of the Ionizing Sources Revealed from the $\rho_{\rm UV}$ and $\tau_e$ Measurements}
\label{subsec:rho_tau}

The evolution of the ionized hydrogen fraction in the IGM, $Q_{{\rm H_{II}}}$, is described by the following 
ionization equation (e.g., \citealt{2013ApJ...768...71R}),
\begin{eqnarray}
		\dot{Q}_{{\rm H_{II}}} = \frac{\dot{n}_{{\rm ion}}}{\average{n_{\rm H}}} - \frac{Q_{{\rm H_{II}}}}{t_{{\rm rec}}},
\label{eq:Q}
\end{eqnarray}
where the dots denote time derivatives.

The first term in the right-hand side of Equation (\ref{eq:Q})
is a source term proportional to the ionizing photon emissivity.
$\dot{n}_{{\rm ion}}$ and $\average{n_{\rm H}}$ are the production rate of ionizing photons and 
the mean hydrogen number density, respectively.
They are defined by 
\begin{eqnarray}
		\dot{n}_{{\rm ion}} &=& \int^{M_{\rm trunc}}_{-\infty} f_{\rm esc}(M_{\rm UV}) \xi_{\rm ion}(M_{\rm UV}) \Phi(M_{\rm UV}) L(M_{\rm UV}) dM_{\rm UV}\nonumber \\
		&\equiv & \left< f_{\rm esc} \xi_{\rm ion} \right> \rho_{\rm UV}, \label{eq:nion}\\
		\average{n_{\rm H}} &=& \frac{X_{\rm p} \Omega_{\rm b} \rho_{\rm c}}{m_{\rm H}}. \label{eq:nH}
\end{eqnarray}
$X_{\rm p}$ is the primordial mass fraction of hydrogen, 
$\rho_{\rm c}$ is the critical density,
and $m_{\rm H}$ is the mass of the hydrogen atom.
Note that $f_{\rm esc}$ and $\xi_{\rm ion}$ are parameters that appear in the product form for our analysis.
If one assumes that $f_{\rm esc}$ and $\xi_{\rm ion}$ depend on $M_{\rm UV}$,
$\left<f_{\rm esc}\xi_{\rm ion}\right>$ is a magnitude-averaged value defined in Equation (\ref{eq:nion}).

The second term in the right-hand side of Equation (\ref{eq:Q}) is a sink term due to recombinations;
$t_{{\rm rec}}$ is the averaged gas recombination time,
\begin{eqnarray}
		t_{{\rm rec}} = \frac{1}{C_{\rm H_{II}} \alpha_{\rm B}(T) (1 + Y_{\rm p} / 4X_{\rm p}) \average{n_{\rm H}} (1+z)^3}, \label{eq:trec}
\end{eqnarray}
where $\alpha_{\rm B}$ is the case B hydrogen recombination coefficient, 
and $T$ is the IGM temperature at a mean density.
$Y_{\rm p}$ is the primordial helium mass fraction.
Substituting Equations (\ref{eq:nion})-(\ref{eq:trec}) into Equation (\ref{eq:Q}), we obtain 
\begin{eqnarray}
		\dot{Q}_{{\rm H_{II}}} &=& A \left( \frac{\rho_{UV}}{10^{26}\ {\rm ergs\ s^{-1}\ Hz^{-1}\ Mpc^{-3}}} \right) - \frac{Q_{{\rm H_{II}}}}{t_{{\rm rec}}} \label{eq:Q2}, \\
		A &=& 2.06\ {\rm Gyr^{-1}} \left( \frac{ \left< f_{{\rm esc}}\xi_{{\rm ion}} \right>}{0.2 \times 10^{25.2}\ {\rm erg\ Hz^{-1}}} \right) \label{eq:A}, \\
		t_{\rm rec} &=& 3.19 \times 10^2 \ {\rm Gyr}\ (1+z)^{-3} \left( \frac{C_{\rm H_{II}}}{3} \right)^{-1}. \label{eq:trec2}
\end{eqnarray}
Once the evolution of $Q_{{\rm H_{II}}}$ is determined by these equations, $\tau_e$ at a redshift $z$ is estimated
(e.g., \citealt{2012MNRAS.423..862K}) from
\begin{eqnarray}
		\begin{aligned}
		\tau_e(z) = \int^z_0 \frac{c(1+z')^2}{H(z')}Q_{{\rm H_{II}}} \sigma_{\rm T} \average{n_{\rm H}} \\
		\times (1+\eta Y_{\rm p}/4X_{\rm p}) dz'  \label{eq:tau_e},
		\end{aligned}
\end{eqnarray}
where $c$ is the speed of light, $H(z)$ is the Hubble parameter, and $\sigma_{\rm T}$ is the Thomson scattering cross section.
We assume that helium is singly ionized ($\eta = 1$) at $z > 4$ and doubly ionized ($\eta = 2$) at $z < 4$ \citep{2012MNRAS.423..862K}.
The value of $\tau_e$ is measured to be $\tau_e = 0.091^{+0.013}_{-0.014}$ \citep{2013arXiv1303.5076P} 
from the combination of the {\it Planck} temperature power spectrum, the {\it WMAP} polarization low-multipole ($l \leq 23$) likelihood \citep{2013ApJS..208...20B}, and the high-resolution ground-based CMB data (e.g., \citealt{2012ApJ...755...70R}, \citealt{2013ApJ...779...86S}).

We assume that
$\rho_{{\rm UV}}$ is approximated by a logarithmic double power law, 
\begin{eqnarray}
		\rho_{\rm UV}(z) = \frac{2 \rho_{{\rm UV}, z=8}}{10^{a(z-8)} + 10^{b(z-8)}} \label{eq:rho_uv_param},
\end{eqnarray}
where $\rho_{{\rm UV}, z=8}$ is a normalization factor, and 
$a$ and $b$ determine the slopes of $\rho_{\rm UV} (z)$.
This double power-law function recovers the rapid decrease of $\rho_{{\rm UV}}$ from $z \sim 8$ towards high redshifts.

With the analytic reionization models described with Equations (\ref{eq:Q2})-(\ref{eq:rho_uv_param}), 
we carry out $\chi^2$ fitting to the observational data of $\tau_e$ and $\rho_{{\rm UV}}$ 
to search for reionization models allowed by these observational constraints.
There are six free parameters in the fit, $\rho_{{\rm UV}, z=8}$, $a$, $b$, $\left<f_{\rm esc}\xi_{\rm ion}\right>$, 
and $C_{\rm H_{II}}$.
Because there is no observational data point of $\rho_{{\rm UV}}$ at $z>11$,
we extrapolate the best-fit $\rho_{{\rm UV}}$ function of $z<11$ 
to $z=30$. At $z>30$, we assume $\rho_{{\rm UV}}=0$.
In conjunction with this assumption, we regard that $\tau_e (z=30)$ should
agree with the $\tau_e$ value from the CMB measurements.

For the data of $\rho_{\rm UV}$ in the fitting, we use all of the $\rho_{\rm UV}$ data points 
presented in Figure \ref{fig:plot_tile}
(Section \ref{subsec:evolution_of_the_uv_ld}), except for those given
by \citet{2013ApJ...768..196S} and \citet{2013MNRAS.432.2696M} at $z=7$ and $8$.
Note that the data from these two studies are already included in our $\rho_{\rm UV}$ estimates
via our best-fit UV luminosity functions in Section \ref{sec:best-fit parameters}.
We, thus, use a total of 8 $\rho_{\rm UV}$ data points for the fitting.

The fitting ranges of $C_{\rm H_{II}}$ and $\left<f_{\rm esc}\xi_{\rm ion}\right>$ are $1.0-9.9$ and $0-10^{25.2}\ {\rm erg^{-1}\ Hz}$, respectively.
The range of $\left<f_{\rm esc}\xi_{\rm ion}\right>$ is motivated by the estimate of spectral properties of high-redshift galaxies in \citet{2013ApJ...768...71R}.
We calculate the $\chi^2$ value by 
simply summing up the $\chi^2$ value of each data point of $\rho_{\rm UV}$ and $\tau_e$,
and obtain the best-fitting parameters.
We refer to this fitting method as a simple $\chi^2$ (SC) method. 

\begin{figure}
	\plotone{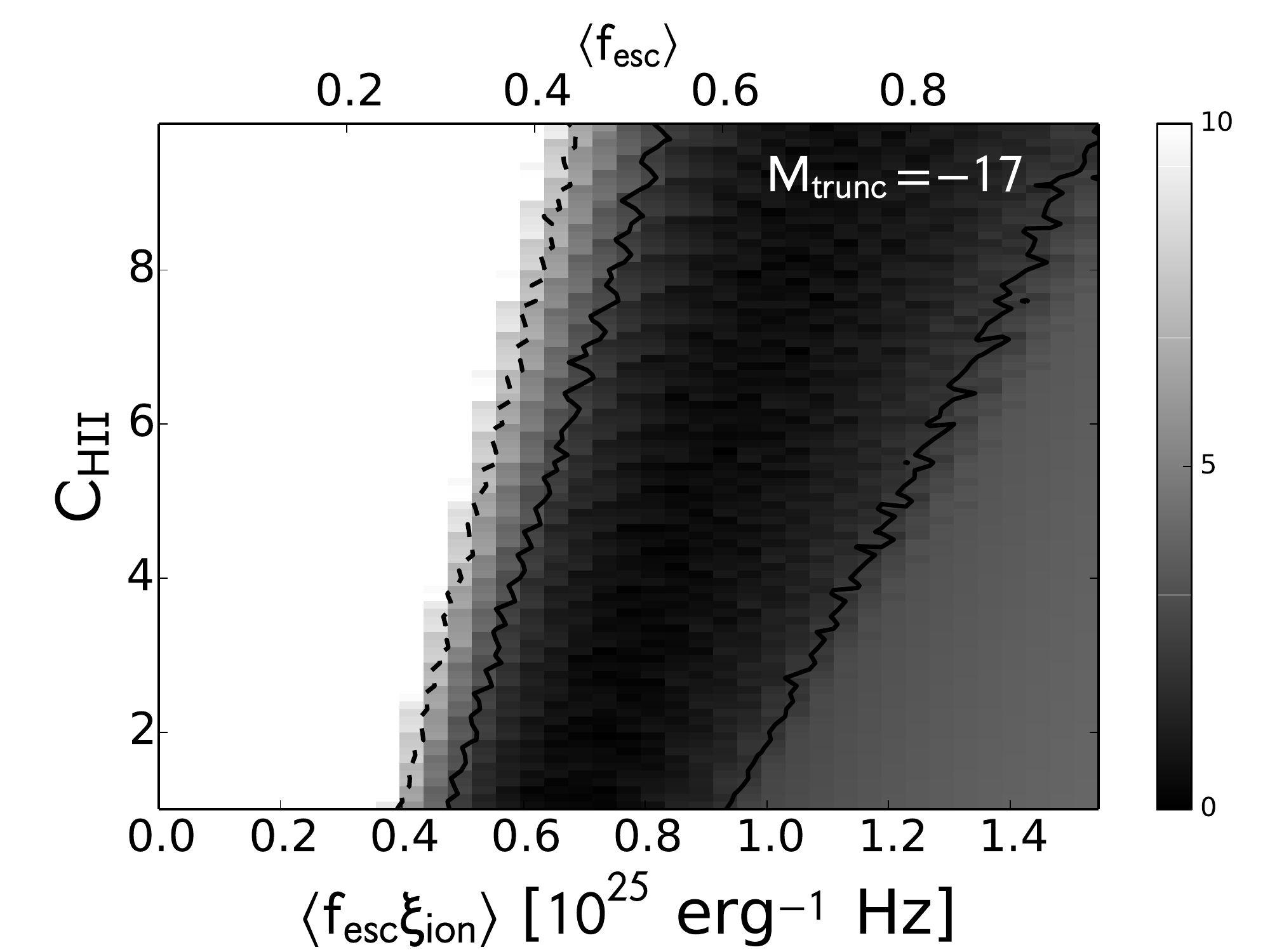}
	\plotone{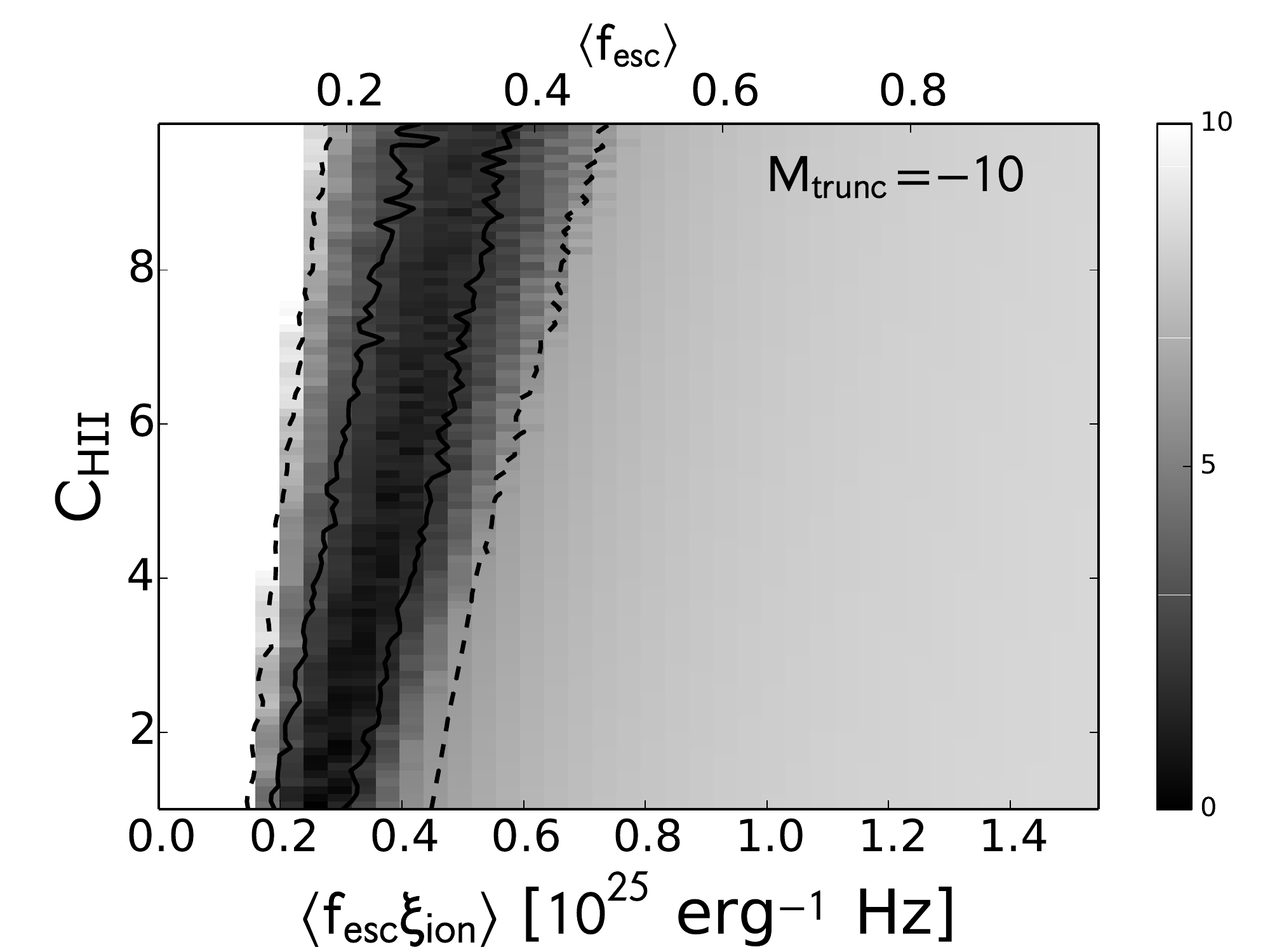}
	\caption{$\Delta \chi^2$ maps and confidence contours of $C_{\rm H_{II}}$ and $\left< f_{\rm esc}\xi_{\rm ion} \right>$ 
	by the SC method with $M_{\rm trunc} = -17$ (upper panel) and $M_{\rm trunc} = -10$ (bottom panel).
The upper axes indicate the average escape fraction $\left< f_{\rm esc} \right>$ under the assumptions of the constant value of $\xi_{\rm ion} = 10^{25.2}\ {\rm erg}^{-1}\ {\rm Hz}$, which is used in \citet{2013ApJ...768...71R}.
	The darker shade indicates the lower $\Delta \chi^2$.
				The solid and dashed lines show 68\% and 95\% likelihood contours, respectively.}
	\label{fig:C-fesc}
\end{figure}

From the $\chi^2$ minimization of the SC method, we find the best-fit parameters for $M_{\rm trunc} = -17$ and $-10$.
The best-fit parameters and the $\chi^2$ values are shown in Table \ref{table:rhoparam}.
Figure \ref{fig:C-fesc} presents $\Delta \chi^2$ values on the $\left<f_{\rm esc}\xi_{\rm ion}\right>$ versus $C_{\rm H_{II}}$ plane calculated by the SC method.
$\Delta \chi^2$ is determined by $\Delta \chi^2 \equiv \chi^2 - \chi^2_{\rm min}$, where 
$\chi^2_{\rm min}$ is the minimum $\chi^2$ value.
In Figure \ref{fig:plot_tile}, we show the best-fit functions of $\rho_{\rm UV}(z)$ and $\tau_e (z)$
for $M_{\rm trunc} = -17$ and $-10$.
Figure \ref{fig:plot_tile} indicates that
the best-fit $\rho_{{\rm UV}}(z)$ agrees with the data points,
but that the best-fit $\tau_e$ is significantly lower than the one of the CMB measurement.
The $\chi^2$ values and the degrees of freedom (dof) shown in Table \ref{table:rhoparam} suggest that 
the probabilities of these $\chi^2$ values occurring by chance are $0.4\%$ and $1.8\%$ with $M_{\rm trunc} = -17$ and $-10$, respectively.
It indicates that 
our analytic reionization models may not be good enough 
to explain the reionization history and sources of reionization.

The best-fit functions by the SC method are weighted by the $\rho_{\rm UV}$ data more strongly than the $\tau_e (z)$ data,
because the number of data points of $\rho_{\rm UV}$ is 8, while that of $\tau_e (z)$ is 1.
Here, we calculate $\chi^2$ values by another method which gives equal weight to $\rho_{\rm UV}$ and $\tau_e$ data sets.
In this method, we divide the $\chi^2$ values of the $\rho_{\rm UV}$ data by 8, that is the number of the data points. 
We refer to this method as an equally-weighted $\chi^2$ (EWC) method.
The best-fit parameters and functions by the EWC method are shown in Table \ref{table:rhoparam} and Figure \ref{fig:plot_tile}, respectively.
In the case of EWC, 
the best-fit $\tau_e$ falls in the error range of the CMB measurement as we expect.
However, there is a discrepancy between the best-fit $\rho_{{\rm UV}}(z)$ and the
observational $\rho_{{\rm UV}}$ data points especially at $z\gtrsim 9$ where the observational $\rho_{{\rm UV}}$ data
exhibit the rapid decrease towards high-$z$.

Even if we change the weights of the fitting and allow 
the large parameter space of $\left<f_{\rm esc}\xi_{\rm ion}\right>$ and $C_{\rm H_{II}}$,
we have found that no single model can reproduce both the $\tau_e$ and $\rho_{{\rm UV}}$ data points
from the observations.
This is because the data points of $\rho_{{\rm UV}}$ decrease too rapidly at $z>8$ 
to contribute to adding $\tau_e$. 
This conclusion is in contrast with the claims of the pioneering study of 
\citet{2013ApJ...768...71R} who find a parameter space of the similar
analytic models explaining
the observational $\tau_e$ and $\rho_{{\rm UV}}$ data available in 2013.
While the best measurement value of $\tau_e$ is almost unchanged since then,
the rapid decrease of $\rho_{{\rm UV}}$ at $z>8$ 
is clearly identified by the subsequent observational studies including our HFF work.
The strong constraints on the evolution of $\rho_{{\rm UV}}$ at $z>8$
probably allow us to find the discrepancy between the analytic models
and the observational data.

\setlength{\tabcolsep}{6pt}
\begin{deluxetable*}{lccccccc}
\tabletypesize{\scriptsize}
\tablecaption{Best-fit parameters and the $\chi^2$ values \label{table:rhoparam}}
\tablewidth{0pt}
\tablehead{
		\colhead{} & \colhead{$M_{\rm trunc}$}  & \colhead{$\log \left< f_{\rm esc}\xi_{\rm ion} \right>$  } & \colhead{$C_{\rm H_{II}}$} & \colhead{log $\rho_{{\rm UV}, z=8}$ } & \colhead{$a$} & \colhead{$b$} & \colhead{$\chi^2 / $ dof} \\
		\colhead{} & \colhead{} & \colhead{(log erg$^{-1}$ Hz)} & \colhead{} & \colhead{(log ${\rm ergs\ s^{-1}\ Hz^{-1}\ Mpc^{-3}}$)} & \colhead{} & \colhead{}
}
\startdata
SC method & $-17$ & $24.85$ & $1.9$ & $25.60$ & $0.13$ & $0.40$ & $15.53 / 4$\\
                & $-10$ & $24.38$ & $1.0$ & $26.11$ & $0.11$ & $0.09$ & $11.95 / 4$\\
EWC method& $-17$ & $24.50$ & $1.1$ & $25.97$ & $0.01$ & $0.01$ &  \nodata\\
                & $-10$ & $24.20$ & $1.1$ & $26.26$ & $0.004$ & $0.01$ &  \nodata
\enddata
\end{deluxetable*}

There are three possible explanations for the discrepancy between the models and the observational data of 
$\rho_{\rm UV}$ and $\tau_e$.
First, the decrease of $\rho_{\rm UV}$ at $z > 11$ may not be as rapid as that
found at $z=8-11$. Because there is no $\rho_{\rm UV}$ data at $z > 11$,
in our model we extrapolate the best-fit power-law $\rho_{{\rm UV}} (z)$ of $z=8-11$ towards $z=30$.
If the real $\rho_{\rm UV}$ values at $z > 11$ are larger than this extrapolation,
the $\tau_e$ value becomes larger, which eases the tension between
the model prediction and the observations.
The slow decrease of $\rho_{\rm UV}$ at $z>11$ may be made by
a $M_{\rm trunc}$ fainter than $-10$ mag and/or a luminosity function slope ($\alpha$)
steeper than the values found at $z\sim 6-8$ (Table \ref{lf_parameters}).
In other words, the discrepancy that we find
may suggest that faint galaxies dominate at $z>11$ even more
than at $z\sim 6-8$.
It is also possible that the rapid decrease would be weakened by the luminosity function slope steepening and/or $M_{\rm trunc}$ becoming fainter at $z \sim 9-10$.
Second, the evolution of $\left<f_{\rm esc}\xi_{\rm ion}\right>$ or $C_{\rm H_{\rm II}}$ can 
increase $Q_{\rm H_{II}}$, such suggested by \citet{2012MNRAS.423..862K}.
If the $\left<f_{\rm esc}\xi_{\rm ion}\right>$ value becomes large towards high-$z$,
$Q_{\rm H_{II}}$ (accordingly $\tau_e$)  could be boosted.
Similarly, the small $C_{\rm H_{\rm II}}$ would enhance $\tau_e$,
although $C_{\rm H_{\rm II}}$ can be as low as unity by definition.
Third, another source of ionizing photons besides massive stars of galaxies may exist, which
contributes to the cosmic reionization significantly. 
X-ray sources such as X-ray binaries and faint AGNs would not leave a clear signature in the $\rho_{\rm UV}$ measurements,
but provide a fraction of ionizing photons via X-ray necessary for the cosmic reionization 
\citep{2013ApJ...776L..31F,2004ApJ...604..484M,2013MNRAS.431..621M}.

The bottom right panel of Figure \ref{fig:plot_tile} shows $Q_{\rm H_{II}}$ as a function of redshift, 
reproduced by our best-fit models. In Figure \ref{fig:plot_tile}, we also plot $Q_{\rm H_{II}}$
estimated from observational results of 
the Ly$\alpha$ forest transmission \citep{2010MNRAS.407.1328M, 2011MNRAS.415.3237M}, 
Ly$\alpha$ near-zone sizes around high-redshift quasars \citep{2010ApJ...714..834C, 2011MNRAS.416L..70B}, 
Ly$\alpha$ damping wing absorption in a GRB spectrum \citep{2006PASJ...58..485T,2008MNRAS.388.1101M}, 
evolution of the Ly$\alpha$ luminosity function and Ly$\alpha$ emitter clustering \citep{2008ApJ...677...12O, 2010ApJ...723..869O, 2014arXiv1404.6066K}, 
and the Ly$\alpha$ emitting galaxy fraction evolution \citep{2011ApJ...743..132P, 2012ApJ...744..179S, 2012ApJ...744...83O}.
Because these measurements have uncertainties too large to constrain our model parameters,
we do not use these measurements for our model fitting. However, Figure \ref{fig:plot_tile}
illustrates that our best-fit models are in good agreement with most of the $Q_{\rm H_{II}}$ measurements.

\section{Summary}
\label{sec:Summary}

We conduct the comprehensive analyses of the full-depth HFF Abell 2744 cluster and parallel field data
whose observations completed in July 2014, and study faint dropout galaxies at $z \sim 5-10$.
We construct a mass model for Abell 2744 to evaluate the gravitational lensing effects of the cluster. 
Then we estimate number densities of our dropout candidates with realistic Monte-Carlo simulations
in the {\it image plane} including detection completeness, contamination, and all lensing effects 
such as magnification, distortion, and multiplication of images.

The major results of our study are as follows.
\begin{enumerate}
\item We identify $54$ dropout candidates at $z \sim 5-10$ with the $i$-, $Y$-, and $YJ$-dropout selection criteria.
The magnifications of our dropout candidates range from $1.03$ to $14$.
The intrinsic magnitudes of our dropout candidates reach $M_{\rm UV} \sim -17$ mag that is
comparable to survey limits of the deepest blank-field observations of
the HUDF.
\item The number densities of our dropout candidates are consistent with previous results of blank-field surveys. 
However, we find a slight excess of the number of our bright dropout candidates at $z \sim 8$ 
probably due to field-to-field variance.
\item We derive the UV luminosity functions at $z \sim 6-7$, $8$, and $9$ combining 
our HFF results with the previous blank-field surveys.
We confirm that the faint-end slopes of the luminosity functions ($\alpha$) are as steep as $-2$ both 
at $z \sim 6-7$ and $z \sim 8$. The number of dropout candidates at $z\sim 9$ increases significantly
by our HFF study, and strengthen the early claim of the rapid decrease from $z\sim 8$ to $\sim 10$
from the evolution of $\rho_{\rm UV}$.
\item We use the simple analytic reionization models to explain 
the observational results of the $\rho_{\rm UV}$ evolution and the CMB's $\tau_e$.
None of our models can reproduce both of these observational measurements,
due to the rapid decrease of $\rho_{\rm UV}$ and the large $\tau_e$ value,
even if we allow a large parameter space of $M_{\rm trunc}$, $\left< f_{\rm esc}\xi_{\rm ion} \right>$, and $C_{\rm H_{II}}$.
This problem could be resolved by the slow decrease of $\rho_{\rm UV}$ at $z>11$, 
the evolution of $\left< f_{\rm esc}\xi_{\rm ion} \right>$ and/or $C_{\rm H_{II}}$,
or another source of reionization such as X-ray bright populations of X-ray binaries and faint AGN.
\end{enumerate}

The HFF program will provide a significantly large sample of high-redshift galaxies 
when the observations of the planned six clusters are completed.
In the Abell 2744 cluster field, we find 3 dropout candidates whose magnifications are $\gtrsim 10$.
A simple scaling suggests that the complete HFF observations 
will provide $\sim 20$ highly-magnified ($\mu \gtrsim 10$) systems at high redshift,
which 
will uncover the properties of the faint galaxies at the epoch of cosmic reionization
and greatly improve our understanding of sources of reionization up to $z\sim 12$.

\acknowledgments
We are grateful to Rychard Bouwens, Richard Ellis, Andrea Ferrara, Akio Inoue, Akira Konno, 
Jennifer Lotz, Kentaro Nagamine, Brant Robertson, Tomoki Saito, Takatoshi Shibuya, Dan Stark,
and Masayuki Umemura for useful information, comments, and discussions.
We particularly thank Hakim Atek and Rychard Bouwens for providing their data tables.
This work is based on observations made with the NASA/ESA {\it Hubble Space Telescope}, 
obtained at the Space Telescope Science Institute (STScI),
which is operated by the Association of Universities for Research in Astronomy, Inc., 
under NASA contract NAS 5-26555.
The {\it HST} image mosaics were produced by the Frontier Fields Science Data Products Team at STScI.
This work utilizes gravitational lensing models produced by 
P.I.s Brada\v{c}, Kneib \& Natarajan, Merten \& Zitrin, Sharon, and Williams,
funded as part of the {\it HST} Frontier Fields program conducted by STScI.
We thank these teams for their invaluable help.
We are grateful to Dan Coe for the help in posting our mass model on the website.
This work was supported by KAKENHI (23244025) Grant-in-Aid for Scientific Research (A) 
through Japan Society for the Promotion of Science (JSPS).
This work was supported in part by World Premier International Research
Center Initiative (WPI Initiative), MEXT, Japan, and Grant-in-Aid for
Scientific Research from the JSPS (26800093).
The work of M.I is partly supported by an Advanced Leading Graduate Course for Photon Science grant.

{\it Facilities:} \facility{{\sl HST} (WFC3, ACS)}


\bibliographystyle{apj}
\bibliography{apj-jour,ishigaki2014}

\end{document}